\documentclass{article}
\usepackage[utf8]{inputenc}
\usepackage[numbers]{natbib}
\usepackage{manfnt} 

\usepackage{amsthm}

\newtheorem{thrm}{Theorem}
\newtheorem{lmm}{Lemma}
\newtheorem{prpstn}{Proposition}

\newtheorem{dfntn}{Definition}
\usepackage{amsmath}			% Permet de taper des formules mathématiques
\usepackage{amssymb}			% Permet d'utiliser des symboles mathématiques
\usepackage{amsfonts}			% Permet d'utiliser des polices mathématiques
\usepackage{nicefrac}			% Fractions 'inline'

\usepackage{multirow}
\usepackage{booktabs}
\usepackage{colortbl}
\usepackage{tabularx}
\usepackage{multirow}
\usepackage{threeparttable}
\usepackage{etoolbox}
	\appto\TPTnoteSettings{\footnotesize}

\usepackage{dsfont}

\usepackage{longtable}

\usepackage{tikz}
\usetikzlibrary{%
  arrows,%
  calc,%
  mindmap,%
  trees,%
  tikzmark,%
  shapes.geometric,%
  shapes.misc,%
  shapes.symbols,%
  shapes.arrows,%
  automata,%
  through,%
  positioning,%
  scopes,%
  decorations.shapes,%
  decorations.text,%
  decorations.pathmorphing,%
  shadows}
  \usetikzlibrary{mindmap,trees}

\usepackage{algorithm,algorithmic,refcount}
\usepackage{wrapfig}
\usepackage{tikz}
\usepackage{makecell}

\usetikzlibrary{positioning}
\usepackage{mathtools}

\usepackage{manfnt} 
\usepackage{xcolor}
\usepackage[most]{tcolorbox}
\definecolor{light-gray}{gray}{0.89}
\usepackage{array}
\newcolumntype{C}[1]{>{\arraybackslash}m{#1}}

\usepackage{subcaption}

\usepackage{enumitem}

\usepackage{framed}
\definecolor{shadecolor}{gray}{0.9}
\usepackage{hyperref}
\title{Reliable Time Prediction in the Markov Stochastic Block Model}

\usepackage[margin=1in]{geometry}

\date{}
\author{%
  Quentin Duchemin\thanks{This work was supported by grants from Région Ile-de-France.} \\
 LAMA, Univ Gustave Eiffel, CNRS, Marne-la-Vallée, France.\\
  \texttt{quentin.duchemin@univ-eiffel.fr} 
}

\begin{document}

\maketitle

\begin{abstract}%
We introduce the Markov Stochastic Block Model (MSBM): a growth model for community based networks where node attributes are assigned through a Markovian dynamic. We rely on HMMs' literature to design prediction methods that are robust to local clustering errors. We focus specifically on the link prediction and collaborative filtering problems and we introduce a new model selection procedure to infer the number of hidden clusters in the network. Our approaches for reliable prediction in MSBMs are not algorithm-dependent in the sense that they can be applied using your favourite clustering tool.  

In this paper, we use a recent SDP method to infer the hidden communities and we provide theoretical guarantees. In particular, we identify the relevant signal-to-noise ratio (SNR) in our framework and we prove that the misclassification error decays exponentially fast with respect to this SNR.
\end{abstract}

\maketitle

\section{Introduction}

Large random graphs have been very popular in the last decade since they are powerful tools to model complex phenomena like interactions on social networks \cite{TYS11} or the spread of a disease \cite{ahmad2017effects}. The relevance of random graph models can be quantified by understanding their ability $i)$ to reproduce properties observed in most real world networks and $ii)$ to make predictions regarding missing information or future evolution of the system.  

\medskip

{\bf Random graph models and characteristics of real world networks.}
Among important features found in most of real world networks, one can mention the so-called scale-free property which states that degree distribution follows a power-law, or the small-world phenomenon in social networks. A large span of random graph models that exhibit such behaviours have been proposed such as the scale-free network model of Barabasi and Albert \cite{barabasi09} or the small-world networks of Watts and Strogatz \cite{watts98}. Another important characteristic of networks in applications is the existence of groups of nodes that are more densely connected internally than with the rest of the network. To deal with such cases, latent space models for network data emerged (see \cite{smith19}). One of the most studied latent model is the Stochastic Block Model (SBM): a community based random graph model where each node is assumed to belong to one community while the connection probabilities between two nodes in the graph depend on their respective membership. The SBM gives a relevant framework to detect communities of well connected nodes in a graph and a large number of algorithms have been proposed to recover the hidden groups in SBMs from the observation of the edges. One can mention  belief propagation algorithms \cite{Abbe16}, spectral methods \cite{CRV15}, neural networks \cite{SG19}, Bayesian approaches \cite{TYS11} or Maximum Likelihood estimation \cite{CDP12}. Another powerful and popular tool is Semi-Definite Programming (SDP) which is known to have interesting robustness features \cite{PW15}, \cite{FC17}. Recently, \cite{Verzelen} proposed a SDP method to address community detection by solving a relaxed version of $K$-means. They prove that the proportion of clustering errors made by their algorithm decays exponentially fast with respect to a well-chosen Signal to Noise Ratio (SNR). Such result is known as a {\it partial recovery bound} in the literature and we refer to the survey \cite{Abbe} for further details regarding  recent developments for community detection in the SBMs.

\begin{figure}[!ht]
\begin{center}
\scalebox{0.9}{
\begin{tikzpicture}[>=stealth,yscale=2,xscale=1.4]  
\node(pred) at (0,-0.275)[rectangle,draw,text width=3cm,text centered] {Link prediction};
\node(dyn) at (3,0.65)[rectangle,draw,text width=3cm,text centered] {Dynamic network};
\node(stat) at (3,-1.3)[rectangle,draw,text width=3cm,text centered] {Static network};
\node(temp) at (6,1.15)[rectangle,draw,text width=3cm,text centered] {Temporal network};
\node(gro) at (6,0.15)[rectangle,draw,text width=3cm,text centered] {Growth model};
\node(mrgg) at (10,0.3)[rectangle,draw,text width=5cm,text centered] {Random Geometric Graph: \cite{DCD20}};
\node(us) at (8.935,0)[rectangle,draw,text width=2cm,text centered] {This paper};

\node(nn) at (10,1)[rectangle,draw,text width=5cm,text centered] {Neural-Network: \cite{li2014deep}};
\node(jordan) at (10,0.7)[rectangle,draw,text width=5cm,text centered] {Non-parametric approach: \cite{sarkar2012nonparametric}};
\node(mf) at (10,1.3)[rectangle,draw,text width=5cm,text centered] {Matrix factorization: \cite{ma2018graph}};
\node(markov) at (10,1.6)[rectangle,draw,text width=5cm,text centered] {Markovian model: \cite{SARUKKAI2000377,7996909}};

\node(proba) at (6,-0.7)[rectangle,draw,text width=3cm,text centered] {Probabilistic approaches};

\node(simi) at (6,-1.3)[rectangle,draw,text width=3cm,text centered] {Similarity-based};

\node(algo) at (6,-1.75)[rectangle,draw,text width=3cm,text centered] {Algorithmic methods};

\node(baye) at (10,-0.4)[rectangle,draw,text width=5cm,text centered] {Bayesian approach: \cite{Guimer2009,ZHANG2014553}};
\node(hierarchical) at (10,-0.7)[rectangle,draw,text width=5cm,text centered] {Hierarchical model: \cite{Clauset_2008}};
\node(logit) at (10,-1)[rectangle,draw,text width=5cm,text centered] {Logistic model: \cite{baldin2018optimal}};

\node(simicit) at (10,-1.3)[rectangle,draw,text width=5cm,text centered] {\cite{newman2001clustering,zhou2009predicting,ADAMIC2003211}};
\node(statfactor) at (10,-1.6)[rectangle,draw,text width=5cm,text centered] {Matrix factorization: \cite{wu2016link}};
\node(kernel) at (10,-1.9)[rectangle,draw,text width=5cm,text centered] {Kernel method: \cite{YUAN20191}};

\draw[->] (pred.east) -- (dyn.west);
\draw[->] (pred.east) -- (stat.west);
\draw[->] (dyn.east) -- (temp.west);
\draw[->] (dyn.east) -- (gro.west);
\draw[->] (gro.east) -- (mrgg.west);
\draw[->] (gro.east) -- (us.west);
\draw[->] (temp.east) -- (nn.west);
\draw[->] (temp.east) -- (jordan.west);
\draw[->] (temp.east) -- (mf.west);
\draw[->] (temp.east) -- (markov.west);

\draw[->] (stat.east) -- (proba.west);
\draw[->] (stat.east) -- (simi.west);
\draw[->] (stat.east) -- (algo.west);
\draw[->] (algo.east) -- (statfactor.west);
\draw[->] (algo.east) -- (kernel.west);
\draw[->] (simi) -- (simicit);
\draw[->] (proba.east) -- (logit.west);
\draw[->] (proba.east) -- (hierarchical.west);
\draw[->] (proba.east) -- (baye.west);

\end{tikzpicture}
}
\end{center}
\caption{Classification of several methods proposed for link prediction in random graphs.}
\label{fig:linkpredreview}
\end{figure}
\medskip

{\bf Link prediction in random graphs.}  Random graph models are not only used to answer questions about the properties of the studied network, but their goal is also to make predictions about missing information or future events (cf.\cite{armengol2015evaluating}). In the last
decade, a lot of work has been done on the link prediction problem and several review articles have been written to synthesize this abundant literature (cf.\cite{reviewsurvey} or \cite{KUMAR2020124289}). Papers tackling this question can be classified into three groups. In the first category, a network is partially observed and one aims at predicting the missing links. The simplest approaches are based on similarity-metrics~\cite{newman2001clustering,zhou2009predicting,ADAMIC2003211} where for each pair of nodes a similarity score is calculated. Among the non-observed edges, the pair of nodes having a higher score represents the predicted link. Another important line of research is based on probabilistic models. One can mention for example~\cite{baldin2018optimal} where the authors consider that covariates are observed for each node and propose a generative model formulated as a matrix logistic regression. Other approaches include matrix factorization techniques (cf.~\cite{wu2016link}) or the use of kernel methods (cf.~\cite{YUAN20191}). 

Papers that we classify in the second category tackle the link prediction problem in {\it temporal networks}: they observe several snapshots of a network with a fixed number of nodes where edges can appear or disappear over time. Based on the sequence of snapshots, the goal is to predict the connections in the network in the future. The difference between various temporal link prediction algorithms lies in how they capture the temporal or dynamic nature of the networks and also how they define the network property to be preserved. Several probabilistic approaches has been deployed to solve link prediction in temporal networks. In \cite{7996909}, the authors present a stochastic Markov model over a time-varying graph. The temporal analysis in this model considers link (local structural) evolution over fine-grained time scale and cluster (semi-global structural) evolution over coarse grained time scale. Some of the probabilistic temporal link prediction algorithms follow non-parametric approaches such as in \cite{sarkar2012nonparametric} where the authors model the out-edges of a node $i$ at time $t$ as a function of the local neighbourhood of $i$ over a moving time window. We refer to \cite{divakaran2020temporal} for a more detailed description of the different methods proposed to tackle link prediction in temporal networks.

In the third and last category, the goal is to tackle link prediction in {\it growth models}, i.e. in networks where at each time step new nodes are joining the graph. Given the observation of the graph up to time $t$, the goal is to predict how a node that joins the network at time $T > t$ will connect to nodes already existing in the network. Growth models aim at mimicking the way the nature generates heterogeneous networks. Each of these models has its own growth dynamics or generating process, proposed as an hypothesis explaining the emergence of a target feature. Famous growth models include the preferential attachment model (cf.\cite{newman2001clustering}), the team-based Yule model (cf.\cite{grow7}) or the copying mechanism from~\cite{grow5} where at each time step a new node enters the network and copies a number of links from a “prototype” node that is selected randomly from the existing nodes whereas choosing the remaining neighbors is random. In \cite{DCD20}, the authors introduced a new growth model based on the Random Geometric Graph on the euclidean sphere $\mathbb S^{d-1}$. A latent attribute $X_i \in \mathbb S^{d-1}$ is associated to each node $i\in[n]$ and they assume that the process $(X_i)_{i\geq 1}$ is a Markov chain. Two nodes $i$ and $j$ are connected with a probability that is a function of the euclidean distance between the latent representations $X_i$ and $X_j$. Using non-parametric methods, the authors prove that they can estimate the probability of connection between the upcoming node $n+1$ and any node $i\in[n]$ that already exists in the graph. Figure~\ref{fig:linkpredreview} gives a synthetic presentation of the different link prediction approaches discussed so far.

 \medskip
 
Among all the above mentioned link prediction techniques, only a small fraction of them discuss the reliability of the proposed algorithms in the presence of spurious links or when information is missing. 
Some works have investigated the reliability of link prediction for temporal networks using Bayesian approaches (cf.~\cite{Guimer2009,feng2012link}), but this question remains so far understudied.

\bigskip

In this paper, we propose a reliable link prediction method in a new growth model for community-based networks which is a dynamic extension of the standard SBM.

\medskip

{\bf Dynamic community-based networks.}  
Several time evolving SBMs have been recently introduced. In \cite{MM15}, a Stochastic Block Temporal Model is considered where the temporal evolution is modeled through a discrete hidden Markov chain on the nodes membership and where the connection probabilities also evolve through time.
In \cite{PZ17}, connection probabilities between nodes are functions of time, considering a maximum number of nodes that can switch their communities between two consecutive time steps. Following the work of \cite{KBN11}, \cite{LR15} study the Degree Corrected Stochastic Block Model where the degree of the nodes can vary within the same community. They show that for the relatively sparse case (i.e. when the maximum expected node degree is of order $\log(n)$ or higher), the proportion of misclassified nodes tends to~$0$ with a probability that goes to $1$ when the number of nodes $n$ increases using spectral clustering. This result inspired the recent paper \cite{KV20} which considers a Dynamic Stochastic Block Model where the communities can change with time. They provide direct connection between the density of the graph and its smoothness (which measures how much the graph changes with time). Several other dynamic variants of the SBM have been proposed so far like in \cite{X14} where the presence of an edge at the time step $t+1$ directly depends on its presence or absence at time $t$.

The above mentioned works are mainly considering temporal networks where membership of nodes or edges can evolve with time, but only few papers are interested in growth model for SBMs (meaning that the size of the graph increases over time) and we aim at filling this gap.

\medskip

\textbf{Model and Motivations.} 
While previous works mainly consider a fixed number of nodes with an evolving graph where communities or connection probabilities can evolve, the Markov Stochastic Block Model (MSBM) is a growth model where a new node enters the graph at each time step and its community is drawn from a distribution depending only on the community of its predecessor. 
Our model could find interesting applications as in the study of bird migrations (see Section~\ref{sec:appli}) where animals have regular seasonal movement between breeding and wintering grounds. Another possible application of our model is for recommendation systems or the analysis of tumor growth that we describe in Sections~\ref{sec:recommendation} and~\ref{sec:tumor} of the Appendix. We provide a reliable link prediction method in the MSBM and we also propose an algorithm to solve {\it collaborative filtering problems}.

\begin{figure}
\centering
\includegraphics[scale=0.4]{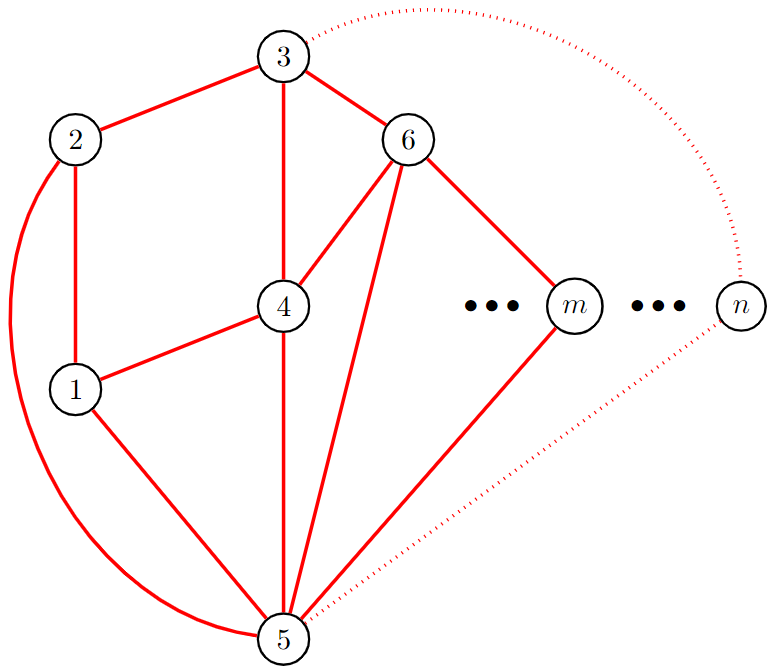}
\caption{Collaborative filtering in a growth model: the graph is fully observed until time $m$ and one aims at inferring the community of node $n$ based on a partial information on the connections between node $n$ and nodes in $[m]$.}
\label{fig:collabointro}
\end{figure}

Collaborative filtering is mainly studied in recommender systems and refers to the ability to exploit the relationships between users to recommend items to the active user according to the ratings of his/her neighbors (cf. \cite{collabo1}). In this paper, we consider the more general sense of collaborative filtering defined as the process of searching for information using strategies that involve several agents. More precisely, our goal is to infer the community of some node $n$ when we have only partial information about how node $n$ is connected to the other nodes of the graph. Typically, we consider a growth model where we fully observe the graph at time $m$ while a poor information transmission occurs from time $m$ making available only a small number of edges between node $n$ and the nodes in $[m]$. The goal is to infer the hidden community of node $n$. Figure~\ref{fig:collabointro} gives a visualization of the collaborative filtering problem tackled in this paper.

\vspace{0.4cm}
\textbf{Contributions.}
We show that the MSBM gives a convenient framework to study community-based growing networks with a Markovian underlying dynamic. Our goal is to predict future information based
on historical data and we tackle specifically the problems of link prediction and collaborative filtering.
We show that the standard plug-in method is highly sensitive to clustering errors. This is the reason why we propose a new approach that is more robust to possible errors in the estimated communities. To do so, we borrow ideas from mean field approximation (see e.g. \cite{meanfield}) by considering that the joint distribution of the communities of nodes ({\it hidden} states) and the communities returned by the clustering algorithm ({\it observed} states) factorizes according to the graph of a homogeneous Hidden Markov Model (HMM). Using the Baum Welch algorithm, we learn the emission probabilities of this HMM, namely the probability that the clustering algorithm predicts community $l$ for a node belonging to community $k$ for any $k,l \in[K]$. Based on these quantities, we also propose a model selection procedure. In our simulations, we use the algorithm from \cite{Verzelen} to recover communities and as far as we know, we are the first to provide an implementation of their algorithm. From a theoretical point of view, we show that the misclassification error decays exponentially fast with respect to the SNR and we provide regimes where we can estimate consistently the parameters of our model.

\medskip
\textbf{Outline.} 
In Section \ref{model}, we formally define SBMs and we introduce MSBMs. In Section \ref{results}, we establish a partial recovery bound and we show that we can consistently estimate the parameters of our model. Sections~\ref{sec:link-pred} and \ref{sec:colab} are dedicated to our main contributions: we address the link prediction (cf. Section~\ref{sec:link-pred}) and collaborative filtering problems~(cf. Section~\ref{sec:colab}), and we give heuristics to be robust to potential local clustering errors of the algorithm. In Section~\ref{simulations}, we propose a model selection procedure to estimate the number of communities in our model and we apply our methods on real data.

In the Appendix, we provide the proofs of the theoretical results presented in Section~\ref{results} and we present additional experiments. Let us mention that Appendix~\ref{apdx:notations} contains a list of the different notations used in this paper.

\medskip

{\bf Notations.} For any $n\in \mathbb N$ with $n\geq1$, we denote by $[n]$ the set $\{1,\dots,n\}$.  Given any matrix $A =(A_{i,j})_{i \in [n], j \in [d]}\in \mathbb R^{n\times d}$, we denote by $\|A\|_{\infty} : = \max_{i,j} |A_{i,j}|$ the max norm of the matrix $A$, while $\|A\|_2:= \sqrt{\sum_{i\in[n] , j\in[d]}A_{i,j}^2}$ is the Frobenius norm of the matrix $A$. Given two sequences~$(a_n)_{n \in \mathbb N}$ and~$(b_n)_{n\in \mathbb N}$ of real numbers such that for some~$N \in \mathbb N$, $b_n\neq 0$ for all $n\geq N$, we write~$b_n\underset{n\to +\infty}{=} \omega(a_n)$ if the sequence~$(a_n/b_n)_{n\geq N}$ goes to zero as $n$ tends to $+\infty$.

\section{Model}
\label{model}

An undirected graph is defined by a set of nodes $V$ and a set of edges $E\subset V \times V$. For an undirected graph with $n$ nodes, we define the adjacency matrix of this graph $X \in \{0,1\}^{n\times n }$ such that for all $i,j \in [n]$,
\begin{equation*}
X_{i,j} = \left\{
    \begin{array}{ll}
        1& \mbox{if } \{i,j\} \in E \\
        0 & \mbox{otherwise.}
    \end{array}
\right.
\end{equation*}
\textbf{Stochastic Block Model.} Let us consider $K \geq 2$ communities and a set of $n$ nodes $V=[n]$. The communities $(c_i)_{i \in [n]}\in K^n$ are assigned independently to each node according to a probability distribution $\nu \in [0,1]^{K}$, $\sum_{k \in [K]} \nu_k =1$. Stated otherwise, the community $c_i$ of node $i \in [n]$ is randomly sampled from the distribution $\nu$. Considering the symmetric connectivity matrix $Q \in [0,1]^{K \times K}$, the adjacency matrix of the graph $X \in \{0,1\}^{n \times n}$ related to the assignment of the communities $(c_i)_{i \in [n]}$ is defined by
\begin{equation*}
X_{i,j} \sim \mathrm{Ber}(Q_{c_i,c_j}),
\end{equation*}
where $\rm{Ber}(p)$ indicates a Bernoulli random variable with parameter $p \in [0,1]$. In the standard SBM, $X$ is observed while the latent variables $(c_i)_{i\in [n]}$ are unknown.

For a parameter $\alpha_n \in (0,1)$ varying with the number of nodes $n$, we will be focused on connectivity matrix of the form $$Q:= \alpha_n Q_0,$$ where $Q_0 \in [0,1] ^{K \times K}$ is a matrix independent of~$n$. As highlighted for example in \cite{Abbe2015}, the rate of $\alpha_n$ as $n \to \infty$ is a key property to study random graphs sampled from SBMs. Typical regimes are $\alpha_n \sim 1$ (dense regime), $\alpha_n \sim \frac{\log(n)}{n}$ (relatively sparse regime) and $\alpha_n \sim \frac{1}{n}$ (sparse regime).

\vspace{0.4cm}

\textbf{Markovian assignment of communities in the SBM.} We introduce in this paper the Markov Stochastic Block Model (MSBM) which assigns a community to each node using a Markovian dynamic. We start by ordering the $n$ nodes in $V$ and without loss of generality, we consider the increasing order of the integers $1,2, \dots,n$. For all $i \in [n]$, we denote $C_i \in [K]$ the random variable representing the community of the node $i$ and we consider that they satisfy the following assumption.

\medskip

\underline{Assumption A1.} \hypertarget{A1}{}
$(C_i)_{i  \in [n]}$ is a positive recurrent Markov chain on the finite space $[K]$ with stationary measure $\pi$, with transition matrix $P \in \mathbb{R}^{K \times K}$ and initial distribution $\pi$. $K$ is independent of~$n$.

\smallskip
The community of the first node $C_1$ is drawn from the probability distribution~$\pi$. Then for any $i\geq2$, the community of the $i$-th node $C_i$ is sampled from the probability distribution $P(C_{i-1},\cdot)$. In the following, we will denote by \[ G_k:=\{i\in [n]\; |\; C_i=k\},\] the set of nodes belonging to some community~$k\in[K]$. Once the community of each node is assigned, we draw an edge between the nodes $i$ and $j$ with probability $Q_{C_i,C_j}$,
$$X_{i,j} \sim \text{Ber}(Q_{C_i,C_j}) \quad \text{ with }\quad Q:=\alpha_n Q_0.$$

Here, $Q_0 \in [0,1]^{K \times K}$ is independent of~$n$ and $\alpha_n \in (0,1)$ is varying with~$n$. Figure \ref{fig:model} presents a graphical representation of our model. We observe the adjacency matrix $X$ but the latent variables $(C_i)_{i \in [n]}$ are unknown.

\begin{figure}[h!]
\centering
\scalebox{0.36}{
\begin{tikzpicture}[->,>=stealth',shorten >=1pt,auto,node distance=2.9cm,
                    semithick]
  \tikzstyle{every state}=[fill=red,draw=none,text=white]

  \node[state,scale=1.3] (c1)                 {$\mathbf{C_1}$};
  \node[state,scale=1.3] (c2) [right of=c1]   {$\mathbf{C_2}$};
  \node[state,scale=1.3] (c3) [right of=c2]   {$\mathbf{C_3}$};
  \node[state,scale=1.3] (c4) [right of=c3]   {$\mathbf{C_4}$}; 
  \node[state, fill=white,text=black]   (dots) [right of=c4]   {$\dots$};
  \node[state,scale=1.3] (ci) [right of=dots]   {$\mathbf{C_i}$};
  \node[state,scale=1.3,fill=white,text=black]  (dots2) [right of=ci]   {$\dots$}; 
    \node[state,scale=1.3]  (cn) [right of=dots2]   {$\mathbf{C_n}$}; 
  \node[state,scale=1.3,fill=white,text=black,draw]         (X2) [below of=c2]       {$X_{2,1}$};
  \node[state,scale=1.3,fill=white,text=black,draw]         (X3) [below of=c3]       {$X_{3,1} \; X_{3,2}$};
  \node[state,scale=1.3,fill=white,text=black,draw]         (X4) [below of=c4]       {$X_{4,1} \; X_{4,2} \; X_{4.3}$};
\node[state,scale=1.3,fill=white,text=black,draw]         (Xi) [below of=ci]       {$\left(X_{i,j}\right)_{1 \leq j\leq i-1} $};
\node[state,scale=1.3,fill=white,text=black,draw]         (Xn) [below of=cn]       {$\left(X_{n,j}\right)_{1 \leq j\leq n-1} $};
  \path (c1) edge              node {} (c2)
        (c2) edge              node  {} (c3)
        (c3) edge              node {} (c4)
        (c4) edge              node {} (dots)
        (dots) edge            node {} (ci)
        (ci) edge              node {} (dots2)
        (dots2) edge           node {} (cn)
        (c2) edge              node {} (X2)
        (c3) edge              node {} (X3)
        (c4) edge              node {} (X4)
        (ci) edge              node {} (Xi)
        (cn) edge              node {} (Xn);
    \path[opacity=0.2]
        (c1) edge              node {} (X2)
        (c1) edge              node {} (X3)
        (c1) edge              node {} (X4)
        (c1) edge              node {} (Xi)
        (c1) edge              node {} (Xn)
        (c2) edge              node {} (X3)
        (c2) edge              node {} (X4)
        (c2) edge              node {} (Xi)
        (c2) edge              node {} (Xn)
        (c3) edge              node {} (X4)
        (c3) edge              node {} (Xi)
        (c3) edge              node {} (Xn)
        (c4) edge              node {} (Xi)
        (c4) edge              node {} (Xn)
        (ci) edge              node {} (Xn);
\end{tikzpicture}
}
\caption{Graphical model presenting the SBM with Markovian assignment of the communities.}
\label{fig:model}
\end{figure}
The following quantities are independent of~$n$ and will be crucial in the definition of the SNR
\[ L:=\|Q_0\|_{\infty},   \quad \pi_m := \min_{c \in [K]}\; \pi(c),\quad D^2:= \min_{l \neq k} \; \| (Q_0)_{:,k}-(Q_0)_{:,l}\|_2^2.\]

\textbf{Identifiability.} 
Let us consider some integer $K\geq 1$ and $\mathcal P$ (resp. $ \mathcal Q$) a subset of the set of Markov kernels (resp. of connectivity matrices) of dimension $K\times K$. For any $(P,Q) \in \mathcal P \times \mathcal Q$, let us denote $\mathbb P_{n,(P,Q)}$ the distribution of the adjacency matrix of a graph of size $n$ sampled from the MSBM with parameters $(P,Q)$. Following~\cite{celisse}, we consider the number of clusters $K$ as known and we say that MSBM parameters are identifiable in the parameter space $\mathcal P\times \mathcal Q$ if there exists $N\in \mathbb N$ such that

\begin{tabular}{ccc}\multirow{2}{*}{$\forall P,P'\in \mathcal P, \; \forall Q,Q' \in \mathcal Q, \quad \Bigg[ \, \big(\forall n \geq N, \quad \mathbb P_{n,(P,Q)}=\mathbb P_{n,(P',Q')} \big)\quad \implies \quad$} & $(P,Q)\; \text{and} \; (P',Q') \;$ are equal& \multirow{2}{*}{$\Bigg]$}\\
& up to label switching\footnotemark .& \end{tabular}
\footnotetext{$(P,Q)$ and $(P',Q')$ are equal up to label switching if there exists some permutation $\sigma$ of $[K]$ such that $\forall i,j\in[K], \quad P_{i,j} = P'_{\sigma(i),\sigma(j)}$ and $ Q_{i,j} = Q'_{\sigma(i),\sigma(j)}$.}
\medskip

We consider the following additional assumption.
\smallskip

\underline{Assumption A2.} \hypertarget{A2}{} $D^2=\min_{l \neq k} \; \| (Q_0)_{:,k}-(Q_0)_{:,l}\|_2^2>0$.
\smallskip

\noindent Provided that Assumptions \hyperlink{A1}{A1} and \hyperlink{A2}{A2} hold, Theorems~\ref{Qrecovery} and~\ref{kernelrecovery} (cf. Section~\ref{estimation}) prove that we are able to get consistent estimation (for the max norm) of the parameters $P$, $\pi$ and $Q$ of our model when \begin{equation}\label{eq:alphan} \alpha_n \log(n) \leq 1/L \quad \text{and} \quad \alpha_n \underset{n\to +\infty}{=}\omega\left(\frac{\log(n)}{n}\right).\end{equation}
In particular, denoting $\mathcal P$ (resp. $\mathcal Q$) the set of $K\times K$ Markov kernels (resp. of connectivity matrices) satisfying Assumption~\hyperlink{A1}{A1} (resp. satisfying Assumption~\hyperlink{A2}{A2}), the MSBM parameters are identifiable in $\mathcal P\times \mathcal Q$ provided that the conditions in Eq.\eqref{eq:alphan} are satisfied. The first condition in Eq.\eqref{eq:alphan} is a technical assumption related to the clustering algorithm considered in this paper (cf.\cite{Verzelen}) while the second condition ensures that for $n$ large enough, the clustering algorithm recovers exactly the hidden communities with high probability. Note that the paper~\cite{HMMaliased} suggests that the condition~$D^2>0$ may not be necessary for identifiability since in classical Hidden Markov Models, the additional temporal structure allows for identifiability even, say, when some states have exactly the same output distributions.

\vspace{0.4cm}

\textbf{Error measure.} Given two partitions
$\hat{G}= (\hat{G}_1, \dots , \hat{G}_K)$ and $G = (G_1, \dots ,G_K)$ of $[n]$ into $K$ non-void groups, we define the
proportion of non-matching points
\[\text{err}(\hat{G},G) = \min_{\sigma \in \mathcal{S}_K} \frac{1}{2n} \sum_{k=1}^K \left|
\hat{G}_k \; \Delta \; G_{\sigma(k)} \right| ,\]
where $A\;\Delta \; B=\left(A\setminus B\right)\cup \left(B\setminus A\right)$ represents the symmetric difference between the two sets $A$ and $B$, $|A|$ is the cardinality of the set $A$ and $\mathcal{S}_K$
represents the set of permutations on $\{1, \dots ,K\}$. When $\hat{G}$ is a partition estimating $G$, we
refer to $\text{err}(\hat{G},G)$ as the misclassification proportion (or error) of the clustering.
\medskip

In this paper, $\hat G=(\hat G_1,\dots,\hat G_K)$ refers to the partition in $K$ groups of the $n$ nodes of the studied graph using the clustering algorithm presented in \cite{Verzelen}. This clustering algorithm is presented in Appendix \ref{algo} and uses as input only the adjacency matrix of the graph and a predefined number of clusters $K$.

\section{Estimation procedures and theoretical results}
\label{results}

\subsection{Partial recovery bound for the MSBM}
Using the clustering algorithm from \cite{Verzelen} to infer the hidden communities, we provide a partial recovery bound in the Stochastic Block Model when the communities are assigned through a Markovian dynamic. In the following, $(\hat{C_i})_{1\leq i \leq n}$ and $(\hat{G}_k)_{k \in [K]}$ denote respectively the estimators of $(C_i)_{1\leq i \leq n}$ and $(G_k)_{k \in [K]}$ provided by the Algorithm 1 from \cite{Verzelen} which is described in Section~\ref{algo} of the Appendix.

We define the signal-to-noise ratio as \[S^2 := \frac{n \alpha_n\pi_m D^2 }{L}, \]
 reminding that $\pi_m = \min_{c \in [K]} \; \pi(c) $, $\|Q_0\|_{\infty}\leq L$ and $D^2= \min_{l \neq k} \; \| (Q_0)_{:,k}-(Q_0)_{:,l}\|^2_2 $. The SNR should be understood as the ratio between $i)$ the {\it signal} $\alpha_n^2 n\pi_m D^2$, which is an asymptotic lower bound on the minimal distance between two distinct centers $\Delta^2$ defined as
 \[\Delta^2:= \underset{k\neq j}{\min}  \sum_l |G_l| (Q_{k,l} - Q_{j,l})^2=\alpha_n^2 \sum_l |G_l| ((Q_0)_{k,l} - (Q_0)_{j,l})^2,\]
 and $ii)$ the {\it noise} $\alpha_n L$. We shed light on the fact that this quantity matches asymptotically the SNR from Theorem~\ref{thmVerzelen} (cf. Section~\ref{algo} in the Appendix) when $\pi$ is the uniform distribution over $[K]$ and when the communities are assigned independently to each node according to the probability distribution $\pi$. Moreover, $\pi_m$ can be related to standard quantities that measure how fast the chain converges to its stationary distribution $\pi$ (see Section~\ref{gap}-Proposition~\ref{pi-mixing} in the Appendix). The smaller $\pi_m$, the slower the convergence of the chain towards $\pi$ and the smaller the SNR. Similarly to Theorem~\ref{thmVerzelen}, we prove with Theorem \ref{theorem} that the misclassification error decays exponentially fast with respect to the SNR $S^2$.

\begin{thrm}
Let us recall that $\hat G=(\hat G_1,\dots, \hat G_K)$ is the partition in $K$ groups of the $n$ nodes of the graph obtained using the clustering algorithm from \cite{Verzelen}. We assume that $\alpha_n \log(n)\leq 1/L$ and we consider that assumptions A1 and A2 of Section~\ref{model} are satisfied. Then there exist three constants $a,b,c >0$ such that for any $n$ satisfying \begin{equation}\label{eq:condition-theorem}n\alpha_n>a,\end{equation} it holds with probability at least $1-b/n^2$,
\[ \mathrm{err}(\hat{G},G) \leq e^{-cS^2}.\]
In particular, it holds with probability at least $1-b/n^2$,
\[ -\log \left( \mathrm{err}(\hat{G},G) \right) = \Omega(n\alpha_n).  \]
The constant $a$ only depends on the parameters $\pi$ and $Q_0$, the constant $b$ only depends on $\pi$, $P$ and $K$ while $c$ is a universal constant.
\label{theorem}
\end{thrm}

The proof of Theorem~\ref{theorem} and the explicit expressions of the constants $a$ and $b$ are provided in Section~\ref{proofs} of the Appendix. Theorem \ref{theorem} states that in the relatively sparse regime (i.e. when $\alpha_n \sim \log(n)/n$), we achieve a polynomial decay of the misclassification error with order $\pi_m D^2/L$. The greater the quantity $\pi_m D^2/L$ is, the faster the misclassification error decays. In particular, for $n\alpha_n/\log(n)\geq L/(c\pi_mD^2)$ it holds with high probability $\mathrm{err}(\hat G,G)<1/n$ which means that $\hat G=G$. The condition on the sparsity parameter $\alpha_n$ indicates that Theorem~\ref{theorem} can still be informative in the sparse regime (i.e. when $\alpha_n \sim 1/n$). Typically if $\lim_{n \to \infty} \alpha_n n > A$ for some $A>a$, then Theorem~\ref{theorem} ensures that for $n$ large enough it holds with high probability, $\mathrm{err}(\hat G, G) \leq e^{-c A \pi_m D^2/L}$.

\subsection{Consistent parameter estimation}
\label{estimation}

Note that Theorem~\ref{theorem} is a straightforward consequence of the work of \cite{Verzelen}. Our methods from Sections~\ref{sec:link-pred} and~\ref{sec:colab} could be easily applied using your favorite clustering algorithm and we have decided to work with this recent SDP method for our simulations. In this section, we give estimates $\hat \pi$, $\hat P$ and $\hat Q$ of the parameters of our model, namely $\pi$, $P$ and $Q$. We prove that they are consistent for the max norm when the average degree is of order $\log n$ or higher.

In Theorems \ref{Qrecovery}, \ref{pirecovery} and \ref{kernelrecovery}, we only specify on which parameters of the model the constants $a,b,b'$ depend. In the Appendix, Lemmas~\ref{lemma:4},~\ref{lemma:3} and~\ref{lemma:2} provide the explicit expressions of those constants with respect to the parameters of the model.
 
In Theorems \ref{Qrecovery}, \ref{pirecovery} and \ref{kernelrecovery}, the condition on the sparsity parameter $\alpha_n$ indicates that we get consistent estimation (for the max norm) respectively of the transition matrix, the stationary measure and the connectivity matrix in the relatively sparse regime (i.e. when $\alpha_n \sim \log(n)/n$) for $n$ large enough provided that $\underset{n \to \infty}{\lim} \; n \alpha_n/\log(n) >a$.

\subsubsection{The connectivity matrix}
\label{sec:connectivity-matrix}
In the relatively sparse setting (i.e. when $\alpha_n \sim \log(n)/n$), Theorem~\ref{theorem} ensures that for $n$ large enough it holds with high probability $\mathrm{err}(\hat G,G)<1/n$ which implies that the partition of the nodes is correctly recovered. In this case, a natural estimator for $Q_{k,l}$ (for $k,l \in [K]^2$) consists in computing the ratio between the number of edges between nodes with communities $k$ and $l$ and the maximum number of edges between nodes with communities $k$ and $l$. For any $  k,l \in [K]^2,$
\[ \hat{Q}_{k,l}:= \left\{
    \begin{array}{ll} \displaystyle
        \frac{1}{|\hat{G}_k|.|\hat{G}_l|}\sum_{i \in \hat{G}_k} \sum_{j \in \hat{G}_l} X_{i,j}& \mbox{if } k \neq l \\
        \displaystyle \frac{1}{|\hat{G}_k|.(|\hat{G}_k|-1)}\sum_{i,j \in \hat{G}_k}  X_{i,j}& \mbox{if } k = l 
    \end{array}
\right..\]
At first glance, it would be tempting to state that $\hat Q_{k,l}$ is a sum of i.i.d. Bernoulli random variables. Actually, this is not true since the random variables $(\hat G_k)_{k\in [K]}$ depend on the random variables $(X_{i,j})_{i,j\in[n]}$. Taking this lack of independence carefully into account, one can show Theorem~\ref{Qrecovery} which ensures the consistency of our estimate of the connectivity matrix for the norm $\|\cdot\|_{\infty}$. 
\begin{thrm}
Let us consider $\gamma >0$. We assume that $\alpha_n \log(n)\leq 1/L$ and we consider that assumptions A1 and A2 of Section~\ref{model} are satisfied. Then there exist three constants $a,b,b'>0$ such that for any $n$ satisfying 
\begin{equation}\label{eq:n-condition-Qrecovery} \frac{n \alpha_n}{\log(n) } \geq a \quad \text{ and } \quad n > \left( \frac{\gamma+1}{\pi_m}\right)^2,\end{equation}
it holds with probability at least $1-b(1/n^2 \vee \exp(-b'\gamma^2))$,
\[\|\hat{Q}-Q\|_{\infty} \leq \frac{\gamma}{\sqrt{n}}.\]
The constant $a$ only depends on the parameters $\pi$ and $Q_0$, the constant $b$ only depends on $K$ while $b'$ depends on $\pi$ and $P$.
\label{Qrecovery}
\end{thrm}
Theorem~\ref{Qrecovery} is proved in details in Section~\ref{sec:Qrecovery} and we only provide here the main arguments. To cope with the non-standard dependence structure of the entries of the matrix~$\hat Q$ previously highlighted, we consider in the proof of Theorem~\ref{Qrecovery} the matrix $\widetilde Q$ defined by
\[ \forall k,l\in[K],\quad \widetilde{Q}_{k,l}:= \left\{
    \begin{array}{ll} \displaystyle
        \frac{1}{|{G}_k|.|{G}_l|}\sum_{i \in {G}_k} \sum_{j \in {G}_l} X_{i,j}& \mbox{if } k \neq l \\
        \displaystyle \frac{1}{|{G}_k|.(|{G}_k|-1)}\sum_{i,j \in {G}_k}  X_{i,j}& \mbox{if } k = l 
    \end{array}
\right.,\]
where the sum in $\widetilde Q_{k,l}$ has binomial distribution as it is a sum of i.i.d. random variables with mean $Q_{k,l}$. The proof goes as follows.  
\begin{itemize}
\item[$i)$] {\bf Bounding $\|\hat Q-\widetilde Q\|_{\infty}$.} First, we ensure that the clustering algorithm is recovering correctly the hidden communities with high probability in which case $\hat Q=\widetilde Q$. This result is a direct consequence of Theorem~\ref{theorem} for $\alpha_n$ satisfying the first inequality in Eq.\eqref{eq:n-condition-Qrecovery}. The proof of Theorem~\ref{Qrecovery} is then reduced to bound $\|\widetilde Q-Q\|_{\infty}$.
\item[$ii)$] {\bf Bounding $\|\widetilde Q-Q\|_{\infty}$.} Then, we aim at proving that $\widetilde Q_{k,l}$ is close to $Q_{k,l}=\mathbb E[X_{i,j} \; |\; i\in C_k,j \in C_l]$ with high probability. For this, we need to choose $n$ large enough so that the Markov chain $(C_i)_{i\in [n]}$ {\it mixed enough}, meaning that for any $k\in [K]$, $|G_k|$ is close to $\mathbb E[ \sum_{i=1}^n \mathds 1_{C_i=k}]=\pi(k)n$. This is ensured for $n$ satisfying the second inequality in Eq.\eqref{eq:n-condition-Qrecovery}.
\end{itemize}

\subsubsection{The stationary distribution of the Markov chain}

Thanks to the ergodic theorem, we know that the average number of visits in each state of the chain converges toward the stationary probability of the chain at this particular state. Stated otherwise, for all community $k \in  [K]$, the average number of nodes with community $k$ in the graph converges toward $\pi(k)$ as $n$ tends to $+ \infty$. Therefore we propose to estimate the stationary measure of the chain $(C_i)_{i\geq 1}$ with $\hat{\pi}$ defined by
\[ \forall k \in [K], \quad   \hat{\pi}_{k}:= \frac{1}{n} \sum_{i=1}^n \mathds 1_{\hat{C}_i=k}.\]
Theorem~\ref{pirecovery} ensures the consistency of our estimate $\hat \pi$. Its proof can be found in Section~\ref{sec:proofmsbmpirecovery}.
\begin{thrm}
Let us consider $\gamma>0.$ We assume that $\alpha_n \log(n)\leq 1/L$ and we consider that assumptions A1 and A2 of Section~\ref{model} are satisfied. Then there exist three constants $a,b,b'>0$ such that for any $n$ satisfying  \begin{equation}\label{eq:pireco}\frac{n\alpha_n}{ \log(n)} \geq a,\end{equation}
it holds with probability at least $1-b(1/n^2 \vee \exp(-b' \gamma^2))$,
\[\|\hat{\pi}-\pi \|_{\infty} \leq \frac{\gamma}{\sqrt{n}}.\]
The constant $a$ only depends on the parameters $\pi$ and $Q_0$, the constant $b$ only depends on $\pi$, $P$ and $K$ whereas $c$ is a universal constant.
\label{pirecovery}
\end{thrm}
Similarly to Theorem~\ref{Qrecovery}, the condition on $\alpha_n$ given by Eq.\eqref{eq:pireco} ensures that the clustering algorithm recovers the correct hidden communities with high probability (cf. Theorem~\ref{theorem}). Under this assumption, Theorem~\ref{pirecovery} is a direct consequence of the uniform ergodicity of the chain $(C_i)_{i\in[n]}$ using concentration inequality for Markov chains (cf. Appendix~\ref{gap} or \cite[Theorem 2]{Jiang}).

\subsubsection{The transition matrix of the Markov chain}
\label{defY}

We define $(Y_i)_{i \geq 1}$ a Markov Chain on $[K]^2$ by setting $Y_i = (C_i, C_{i+1})$. We define naturally the sequence $\left( \hat{Y}_i\right)_{i \geq 1}$ by $\hat{Y}_i=(\hat{C}_i,\hat{C}_{i+1})$. The transition kernel of the Markov Chain $(Y_i)_{i\geq1}$ is $\mathcal{P}_{(k,l),(k',l')} = \mathds 1_{l=k'} P_{l,l'}$ and its stationary measure is given by $\mu$ such that  $\forall k,l$, $\mu(k,l) = \pi(k)P_{k,l}$. We propose to estimate each entry of the transition matrix $P$ of the Markov chain $(C_i)_{i\geq 1}$ with 
\[\displaystyle \forall k,l \in [K]^2, \quad \hat{P}_{k,l}:=\frac{n}{n-1} \frac{\sum_{i=1}^{n-1} \mathds 1_{\hat{Y}_i=(k,l)}}{\sum_{i=1}^{n}\mathds 1_{\hat{C}_i=k}}.\]

\begin{thrm}
\label{kernelrecovery}

Let us consider $\gamma> \frac{5K}{2 \pi_m^2}$. We assume that $\alpha_n \log(n)\leq 1/L$ and we consider that assumptions A1 and A2 of Section~\ref{model} are satisfied. Then there exist three constants $a,b,b'>0$ such that for any $n$ satisfying $$\frac{n\alpha_n}{ \log(n)}\geq a \quad \text{ and } \quad n \alpha_n \geq \frac{a}{\gamma^2},$$ it holds with probability at least $1-b\left[1/n^2 \vee \exp\left(-b' (\gamma-\frac{5K}{2\pi_m^2})^2\right)\right],$
\[\| \hat{P} - P \|_{\infty} \leq \frac{\gamma}{\sqrt{n}} .\]
The constant $a$ only depends on the parameters $\pi$ and $Q_0$, $b$ depends only on $K$ while the constant $b'$ depends on $\pi$, $P$ and $K$.
\end{thrm}

To prove Theorem~\ref{kernelrecovery}, we consider the Markov chain $(Y_i)_{i \geq 1}$ built considering two consecutive states of the Markov chain $(C_i)_{i \geq 1}$. Stated otherwise, the state number $i$ of the Markov chain used is formed by the couple of the communities of the nodes number $i$ and number $i+1$.

\section{Link Prediction}
\label{sec:link-pred}

\subsection{The plug-in approach}
\label{sec:plugin}

 In this section, we use the underlying dynamic structure of the MSBM to solve link prediction problems. More precisely, considering a graph of size $n$ with adjacency matrix $X$, we want to find an algorithm that predicts the absence or presence of an edge between nodes $n+1$ and $i$ for any $i\in[n]$. Such algorithm can be understood as a binary classifier and can be represented by a random vector valued in $\{0,1\}^n$. As in binary classification, the risk of a classifier $\mathbf g=(g_i)_{i\in[n]}$ is given by
  \begin{align}\mathcal R(\mathbf g):=& \frac{1}{n} \sum_{i=1}^n \mathbb P\left(  g_i \neq X_{i,n+1}\; | \; \mathbf C_{1:n}\right)\label{BLP:risk}\\
=&\frac{1}{n} \sum_{i=1}^n  (1-\eta_i(\mathbf C_{1:n}))  \mathbb E\left\{ \mathds 1_{g_i =1}\;|\; \mathbf C_{1:n} \right\}+ \eta_i(\mathbf C_{1:n})  \mathbb E\left\{ \mathds 1_{g_i =0}\;|\; \mathbf C_{1:n} \right\}, \notag
\end{align}
 where $\mathbf C_{1:n} = (C_i)_{i\in[n]}$ and for all $i\in[n]$, $\eta_i$ is the posterior probability function and is defined by
 \begin{equation}\label{def:posterior-proba}\eta_i(\mathbf c_{1:n}) = \mathbb P\left( X_{i,n+1}=1 \;|\; \mathbf C_{1:n}=\mathbf c_{1:n}   \right) = \sum_{k\in[K]} Q_{c_i,k}P_{c_n,k} .\end{equation}

 Pushing further this analogy, we can define the classification error of some classifier $\mathbf g$ by $L(\mathbf g) = \mathbb E \left[\mathcal R(\mathbf g, \mathbf C_{1:n})\right]$. Proposition \ref{prop:bayes-optimal} shows that the Bayes classifier - introduced in Definition \ref{def:bayes-estimator} - is optimal for the risk defined in Eq.\eqref{BLP:risk}. 

 \begin{dfntn} \label{def:bayes-estimator} (Bayes classifier) \\
The Bayes classifier $\mathbf g^*=(g^*_i)_{i\in [n]}$ is defined by
\[ \forall i \in [n], \quad g^*_i = \left\{
    \begin{array}{ll}
        1 & \mbox{if } \eta_i(\mathbf C_{1:n}) \geq \frac12 \\
        0 & \mbox{otherwise.}
    \end{array}
\right.
 \] 
\end{dfntn}

\begin{prpstn} \label{prop:bayes-optimal} (Optimality of the Bayes classifier for the risk $\mathcal R$)\\
For any classifier $\mathbf g$ which is $\sigma(\mathbf C_{1:n},(X_{i,j})_{i,j\in[n]})$-measurable, it holds for any $i \in [n]$,
\begin{align*} &\mathbb P\left( g_i \neq X_{i,n+1}  \; |\; \mathbf C_{1:n}\right) - \mathbb P\left( g_i^* \neq X_{i,n+1}  \; |\;  \mathbf C_{1:n}\right)\\
&= 2\left| \eta_i(\mathbf C_{1:n})-\frac12\right|  \mathbb E\left\{ \mathds 1_{g_i \neq g_i^*}\;|\; \mathbf C_{1:n} \right\},\end{align*}
which immediately implies that \[\mathcal R(\mathbf g,\mathbf C_{1:n}) \geq  \mathcal R(\mathbf g^*,\mathbf C_{1:n}) \text{  and therefore  } L(\mathbf g) \geq L(\mathbf g^*).\] 
\end{prpstn}
The computation of the Bayes classifier is intractable since we only observe the adjacency matrix $X$ while the communities $\mathbf C_{1:n}$ and the model parameters remain unknown. A reasonable approximation of the Bayes classifier is the MSBM classifier (cf. Definition~\ref{def:mrgg-classifier}).

\begin{dfntn} (MSBM classifier)\label{def:mrgg-classifier}
The MSBM classifier $\mathbf g^{MSBM}$ is defined by 
\[ \forall i \in [n], \quad g^{MSBM}_i = \left\{
    \begin{array}{ll}
        1 & \mbox{if } \hat \eta_i(\hat {\mathbf C}_{1:n}) \geq \frac12 \\
        0 & \mbox{otherwise}
    \end{array}
\right.,
 \] 
 where for all $\mathbf c_{1:n} \in [K]^n$ and for all $i\in [n]$, \begin{equation}\label{eq:link-esti}\hat \eta_i({\mathbf c}_{1:n})=\sum_{k \in [K]} \hat Q_{{ c}_i,k} \hat P_{{ c}_{n},k} .\end{equation}
In Eq.\eqref{eq:link-esti}, $\hat Q$ (resp. $\hat P$) corresponds to the estimate of the connectivity matrix (resp. of the transition matrix) presented in Section~\ref{estimation}.
\end{dfntn}
Proposition~\ref{prop:consistance-BLP} shows that the MSBM classifier is consistent, meaning that given a training set, the probability of correct classification approaches - as the size of the training set increases - the best probability theoretically possible if the population distributions were fully known. Proposition~\ref{prop:consistance-BLP} is proved in Section~\ref{apdx:prop:consistance-BLP} of the Appendix.
\begin{prpstn} (Consistency of the MSBM classifier) \label{prop:consistance-BLP}
Let us consider $\gamma> \frac{5K}{2 \pi_m^2}$. Assume that $\alpha_n \log(n)\leq 1/L$. Then there exist three constants $a,b,b'>0$ such that for any $n$ satisfying 
\begin{equation} \label{conditions-MSBM} \frac{n \alpha_n}{\log(n) } \geq a, \; n \alpha_n \geq \frac{a}{\gamma^2} \; \text{ and } \; n > \left( \frac{\gamma+1}{\pi_m}\right)^2, \end{equation}
it holds with probability at least 
 $1-b\left[1/n \vee n\exp\left(-b' (\gamma-\frac{5K}{2\pi_m^2})^2\right)\right],$ \begin{equation} \label{eq:consistency}\forall i \in [n], \quad \left| \eta_i(\mathbf C_{1:n})  - \hat \eta_i(\hat{\mathbf C}_{1:n}) \right|  \leq \frac{\gamma}{\sqrt n}\left( \alpha_n K L+1\right).\end{equation}
 Let us finally mention that the constant $a$ only depends on the parameters $\pi$ and $Q_0$, $b$ depends only on $K$ while the constant $b'$ depends on $\pi$, $P$ and $K$.

\end{prpstn}

Obtaining a non-trivial result from Proposition~\ref{prop:consistance-BLP} may require to choose $\gamma$ as function of $n$. Typically, choosing $\gamma= n^{1/4}$, we obtain that for $n$ large enough, it holds with probability at least $1-b/n$,
\begin{equation*}\forall i \in [n], \quad \left| \eta_i(\mathbf C_{1:n})  - \hat \eta_i(\hat{\mathbf C}_{1:n}) \right| \leq n^{-1/4}\left( \alpha_n K L+1\right).\end{equation*}

\begin{figure}[ht!]
  \begin{minipage}[c]{0.4\textwidth}
    \includegraphics[width=\textwidth]{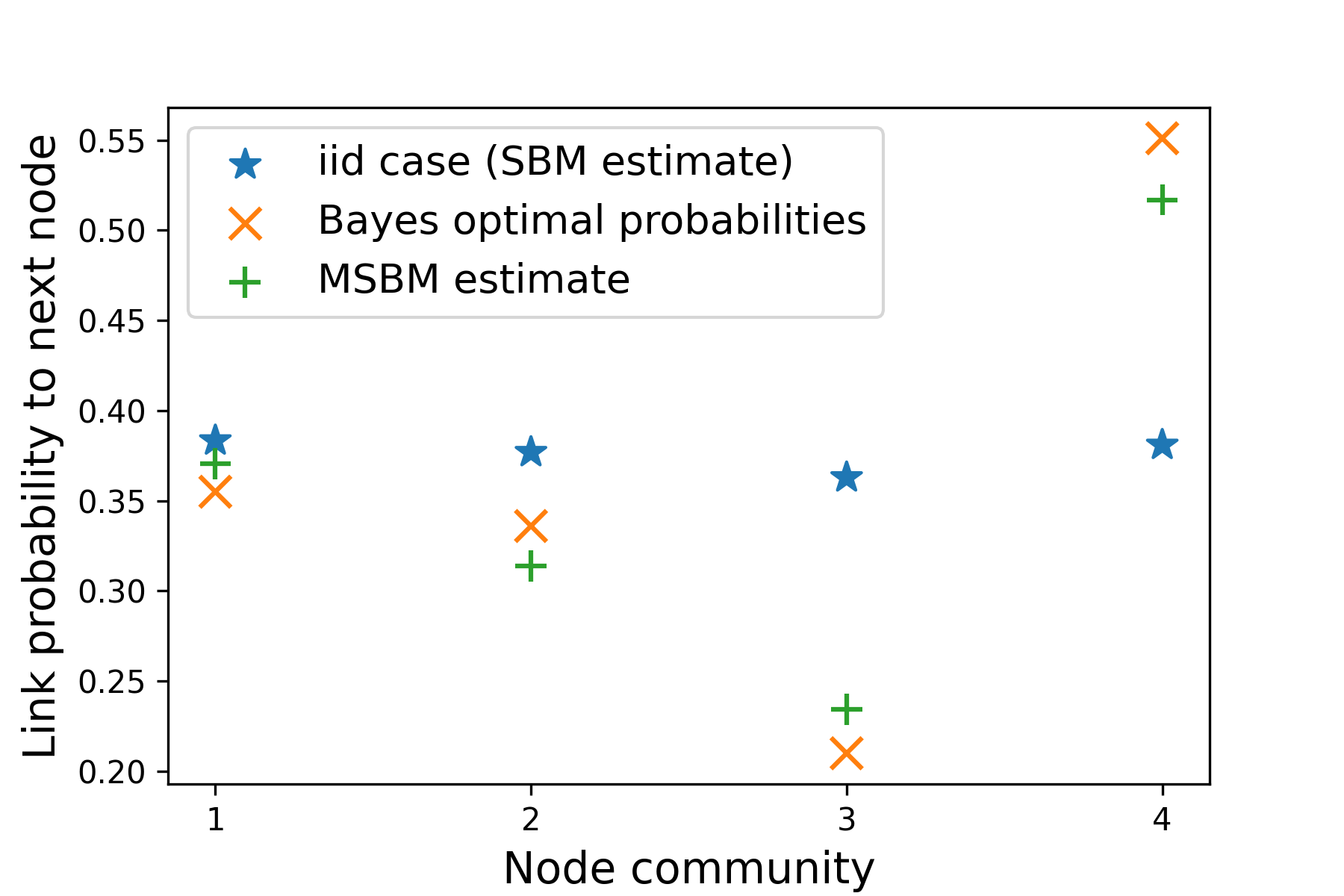}
  \end{minipage}
  \begin{minipage}[c]{0.59\textwidth}
   \caption{We plot the posterior probability functions considering that $i)$ communities are assigned independently according to $\pi$ (blue stars), or that $ii)$ communities are assigned with a Markovian dynamic (orange crosses). The green plus represent the estimates of the orange crosses given by $\hat \eta_k(\hat { C}_{n})$'s for $k \in [K]$.}
        \label{fig:link-prediction}
  \end{minipage}
\end{figure}

Since $\eta_i(\mathbf c_{1:n})$ only depends on $c_i$ and $c_n$, we simply denote $\eta_i(\mathbf c_{1:n})$ by $\eta_{c_i}(c_n)$. We use an analogous slight abuse of notation for the $\hat \eta_i$'s and we denote $\hat \eta_i( {\mathbf c}_{1:n})$ by $\hat \eta_{ c_i}( c_n)$ for any sequence $\mathbf c_{1:n} \in [K]^n$. Let us illustrate Proposition~\ref{prop:consistance-BLP} with a specific numerical example. We consider $K=4$ communities and we sample a random graph of size $n=180$ from the MSBM using the matrices $P$ and $Q$ defined by Eq.\eqref{tmatrixK5} in the Appendix. We denote by $\pi$ the stationary measure of the Markov kernel $P$. 
With Figure~\ref{fig:link-prediction}, we aim at comparing the posterior probabilities $(\eta_k(C_{n}))_{k\in[K]}$ (orange crosses) with $i)$ the posterior probabilities obtained considering that communities have been assigned independently to each node using the probability measure $\pi$ (blue stars), and $ii)$ the estimates of the $\eta_k(C_{n})$'s given by the $\hat \eta_k(\hat { C}_{n})$'s (green plus). More precisely:
 \begin{itemize}
  \item[$\textcolor{blue}{\mathbf \star}$] {\bf i.i.d. case.} For each $k\in [K]$, the corresponding blue star represents $\eta_k^{\mathrm{iid}}:=\sum_{l\in[K]} \pi_l Q_{k,l}$.
  \item[$\textcolor{orange}{\mathbf \times}$] {\bf Bayes optimal probabilities.} For each $k\in [K]$, the corresponding orange cross represents $\eta_k(C_{n})=\sum_{l\in[K]} P_{C_n,l} Q_{k,l}$.
  \item[$\textcolor{green}{\mathbf +}$] {\bf MSBM estimate.} For each $k\in [K]$,
  the corresponding green plus represents $\hat{\eta}_k(\hat{C}_{n}).$
  \end{itemize}

Comparing the blue stars and the orange crosses on Figure~\ref{fig:link-prediction}, we see that considering an independent assignment of the communities lead to bad estimates of the Bayes optimal probabilities. The latter quantities are unknown in practice and Figure~\ref{fig:link-prediction} shows that they can be correctly estimated by the $\hat \eta_k(\hat { C}_{n})$'s, which confirms that the MSBM classifier from Definition~\ref{def:mrgg-classifier} is consistent.

\subsection{Reliable link prediction}
\label{sec:baumwelch}
Proposition~\ref{prop:consistance-BLP} is an asymptotic result and its proof relies on the ability of the clustering algorithm to recover with high probability the correct partition of the nodes when $n$ is large enough and satisfies the conditions from Eq.\eqref{conditions-MSBM}. In practice, we observe a graph with a finite number of nodes and the partition recovered by the algorithm is likely to contain some errors. In this case, using the plug-in approach of the previous section may give bad predictions. Indeed, the MSBM classifier only uses the two estimated communities $\hat C_i$ and $\hat C_n$ to compute $\hat \eta_i(\hat {\mathbf C}_{1:n})$ without taking advantage of the complete sequence of recovered communities $\hat {\mathbf C}_{1:n}$. This makes the MSBM classifier heavily sensitive to local clustering errors.

To cope with this issue, we adopt a new perspective on our problem borrowed from the field of Hidden Markov Model (HMM). This section is divided into three parts: $i)$ first we give a brief introduction to HMMs, then $ii)$ we present the Baum-Welch algorithm which allows to conduct inference in HMMs, and finally $iii)$ we explain how this algorithm can be used in our context to design a {\it reliable link prediction} method. By {\it reliable}, we mean that our method takes advantage of the complete sequence of estimated clusters $\hat{\mathbf C}_{1:n}$ and is less sensitive to local clustering errors compared to the plug-in approach from Section~\ref{sec:plugin}.

\subsubsection{A brief introduction to HMMs for finite state spaces} \label{subsub:1}
Originally introduced by Baum and colleagues at the Institute for Defense Analyses in Princeton (cf. \cite{baum1966statistical}), HMMs have found applications in different areas such as speech recognition \cite{juang1991hidden}, neurophysiology
\cite{fredkin1987correlation} or biology \cite{churchill1989stochastic}. In the simpler version of HMMs, the system being modeled is assumed to be a homogeneous Markov chain $\mathbf Z_{1:T} = (Z_1, Z_1, \dots , Z_T )$ taking values in a finite space $[K]$ (for some $K\in \mathbb N$) with transition matrix $P$ and initial probability distribution $\mu$. The so-called hidden states $(Z_1,\dots,Z_T)$ are not observed and we have only access to another sequence of random variables $\mathbf Y_{1:T} = (Y_1, \dots , Y_T )$ taking values in some finite set $ \mathcal Y$ called the observation space. The goal is to learn about $\mathbf Z_{1:T}$ by observing only $\mathbf Y_{1:T}$. To do so, we need to specify the way the hidden states influence the distribution of the observed states. We assume that the random
variables $(Y_1,\dots , Y_T )$ are independent conditional on the state sequence $\mathbf Z_{1:T}$, meaning that the probability mass function of $\mathbf Y_{1:T}$ conditional to $\mathbf Z_{1:T}$ is given by
\[\mathbb Q(\mathbf Y_{1:T}\; |\, \mathbf Z_{1:T}  ) = \prod_{t=1}^T \mathbb Q(Y_t \, |\, Z_t).\]
We consider that the HMM is homogeneous, which means that the conditional probability $\mathbb Q(Y_t\,|\,Z_t )$ does not depend on~$t$. We write for $z \in [K]$ and $y \in \mathcal Y$, $O_{z,y} := \mathbb Q(Y_t = y\, |\, Z_t = z)$ where the $O_{z,y}$'s are called the emission probabilities. In Table~\ref{table:hmms}, we give some classical inference problems in HMMs and we refer to \cite{cappe2009inference} for further details.
\begin{table}[!ht]
\begin{tabular}{c||c|c|c|c}
Inference problem& Filtering & Smoothing & Prediction & Likelihood\\\hline
Quantity we aim at computing &  $\mathbb Q(Z_t \, |\, \mathbf Y_{1:t})$ & $\mathbb Q(Z_t \, |\, \mathbf Y_{1:s})$, $s>t$ & $\mathbb Q(Z_t \, |\, \mathbf Y_{1:s})$, $s<t$ & $\mathbb Q( \mathbf Y_{1:T})$
\end{tabular}
\caption{Classical inference problems in HMMs.}
\label{table:hmms}
\end{table}
In this paper, we are particularly interested in the smoothing problem. In the next section, we present the general methodology to solve this inference task.
\medskip

{\bf Notation}: Considering $\mu$ a probability distribution on $[K]$ and $P,O$ two Markov kernels on $[K]$, we will denote by $\mathbb Q_{\theta}$ with $\theta=(\mu,P,O)$ the probability under which the joint distribution of $(\mathbf Y_{1:T},\mathbf Z_{1:T})$ factorizes according to the graph of a homogeneous HMM where 
\begin{itemize}
\item $\mathbf Z_{1:T}$ (resp. $\mathbf Y_{1:T}$) are the hidden states (resp. the observed states), 
\item the emission probabilities are $O_{z,y}=\mathbb Q(Y_t=y \, |\,Z_t=z)$ for any $y\in\mathcal Y$, $z\in [K]$ and any $t\in[T]$,
\item $\mathbf Z_{1:T}$ is a Markov chain with initial distribution $\mu$ and transition kernel $P$.
\end{itemize}

\subsubsection{The Baum-Welch algorithm}
\label{subsub:2}

We consider $\theta^*=(\mu^*,P^*,O^*)$ with $\mu^*$ a probability distribution on $[K]$ and $P^*,O^*$ two Markov kernels on $[K]$. We consider that the joint distribution of $(\mathbf Y_{1:T},\mathbf Z_{1:T})$ is $\mathbb Q_{\theta^*}$ that we simply denote by $\mathbb Q$.
We focus on a smoothing problem and we aim at computing
\begin{align}
\label{eq:smooth} \mathbb Q(Z_t \, |\, \mathbf Y_{1:T}) &= \frac{\mathbb Q(Z_t,\mathbf Y_{1:t}) \mathbb Q(\mathbf Y_{t+1:T}\; |\; Z_t)}{\mathbb Q(\mathbf Y_{1:T})}\propto \underbrace{\mathbb Q(Z_t,\mathbf Y_{1:t})}_{=:\alpha_{Z_t}(t)} \underbrace{\mathbb Q(\mathbf Y_{t+1:T}\; |\; Z_t)}_{=:\beta_{Z_t}(t)}, \end{align}
where the symbol $\propto$ means that both sides are equal up to some factor that is a function of $\mathbf Y_{1:T}$. Hence, the probability distribution $\mathbb Q(\cdot \, |\, \mathbf Y_{1:T})$ can be obtained by normalizing $\mathbb Q(Z_t,\mathbf Y_{1:t}) \mathbb Q(\mathbf Y_{t+1:T}\; |\; Z_t)$. The first term $\alpha_{Z_t}(t)$ can be obtained using a forward filtering while the second term $\beta_{Z_t}(t)$ is computed using a backward recursion. We detail both procedures in what follows.

\medskip

$\bullet$ {\bf Forward filtering.} In the filtering problem, we are interested in the conditional probability mass function $\mathbb Q(Z_t\, |\, \mathbf Y_{1:t})$ of the state $Z_t$ conditional to the data observed
up to time $t$. Note that, by Bayes rule, $\mathbb Q(Z_t\, |\, \mathbf Y_{1:t})$ can be obtained by normalizing $\mathbb Q(Z_t, \mathbf Y_{1:t})=\alpha_{Z_t}(t)$. One can easily prove the following recursion formula
\[\alpha_{Z_t}(t) = \mathbb Q(Y_t\, |\, Z_t)\sum_{z_{t-1} \in [K]} \mathbb Q(Z_t \, |\,Z_{t-1}= z_{t-1}) \alpha_{z_{t-1}}(t-1)= O_{Z_t,Y_t}\sum_{z_{t-1} \in [K]} P_{z_{t-1},Z_t}\alpha_{z_{t-1}}(t-1).\]
Hence, one can compute $\alpha_z(t)$ for all $z\in [K]$ and all $t\in[T]$ using the forward iterative procedure presented in Algorithm~\ref{algo:forward}.

\medskip

$\bullet$ {\bf Backward procedure.} We are now interested in computing $\beta_{Z_t}(t)=\mathbb Q(\mathbf Y_{t+1:T}\, |\, Z_t)$. One can easily show that we have the following recursion
\[\beta_{Z_t}(t) =  \sum_{z_{t+1}\in[K]} \mathbb Q(Y_{t+1}\, |\, Z_{t+1}=z_{t+1}) \mathbb Q(Z_{t+1}=z_{t+1}\,|\;Z_t) \beta_{z_{t+1}}(t+1). \]
We deduce that the backward recursion presented in Algorithm~\ref{algo:back} allows to compute $\beta_{z}(t)$ for all $z \in [K]$ and all $t\in [T]$.

\begin{minipage}{0.46\textwidth}
\begin{algorithm}[H]
\centering
 \caption{Forward filtering.}
 \label{algo:forward}
	\begin{algorithmic}[1]
 \STATE \textbf{for} $k =1,\dots,K$ \textbf{set} $\alpha_k(1)= \mu_kO_{k,Y_1}$.\;
  \STATE \textbf{for} {$t=2,\dots,T$}
  \STATE \hskip1em \textbf{for} {$k=1,\dots,K$}
 \STATE \hskip2em $\alpha_k(t)=    O_{k,Y_{t}} \sum_{l \in [K]}  P_{l,k}  \alpha_l(t-1)$.
 \end{algorithmic}
\end{algorithm}
\end{minipage}\hfill
\begin{minipage}{0.46\textwidth}
\begin{algorithm}[H]
\centering
 \caption{Backward procedure.}
 \label{algo:back}
	\begin{algorithmic}[1]
 \STATE \textbf{for} $k =1,\dots,K$ \textbf{set} $\beta_k(T)= 1$.\;
  \STATE \textbf{for} {$t=T,\dots,2$}
  \STATE \hskip1em \textbf{for} {$k=1,\dots,K$}
 \STATE \hskip2em $\beta_k(t-1)=    \sum_{l \in [K]}  O_{l,Y_t}P_{k,l}  \beta_l(t)$.
 \end{algorithmic}
\end{algorithm}
\end{minipage}

\medskip

$\bullet$ {\bf The Baum-Welch algorithm.}

We have shown how to solve smoothing problems when the parameters of the HMMs given by $\theta:=(\mu, P, O)$ are known. In practice, $\theta$ is unknown and a standard approach to bypass this difficulty consists in estimating $\theta$ using a maximum likelihood approach:
\[\theta^{\mathrm{MLE}} = (\mu^{\mathrm{MLE}},P^{\mathrm{MLE}},O^{\mathrm{MLE}}) \in \underset{\theta}{\arg \max}  \left\{\log \mathbb Q_{\theta}(\mathbf Y_{1:T})\right\}= \underset{\theta}{\arg \max} \left\{\log  \sum_{\mathbf z_{1:T}} \mathbb Q_{\theta}(\mathbf Y_{1:T},\mathbf Z_{1:T}=\mathbf z_{1:T}) \right\}   .\]
However, this quantity is intractable since the distribution of $\mathbf Z_{1:T}$ is unknown. The Expectation-Maximization (EM) algorithm has been designed specifically for such situation where we aim at finding maximum likelihood estimates of parameters in statistical models, where the model depends on unobserved latent variables. Considering some parameters $\theta=(\mu,P,O)$ and $\theta'=(\mu',P',O')$ (with $\mu,\mu'$ probability distributions on $[K]$ and $P,O,P',O'$ Markov kernels on $[K]$), we have
\begin{align*}
&\log \sum_{\mathbf z_{1:T}} \mathbb Q_{\theta}(\mathbf Y_{1:T},\mathbf Z_{1:T}=\mathbf z_{1:T})=\log \sum_{\mathbf z_{1:T}} \mathbb Q_{\theta}(\mathbf Y_{1:T},\mathbf Z_{1:T}=\mathbf z_{1:T}) \frac{\mathbb Q_{\theta'}(\mathbf Z_{1:T}=\mathbf z_{1:T} \, |\, \mathbf Y_{1:T})}{\mathbb Q_{\theta'}(\mathbf Z_{1:T}=\mathbf z_{1:T}\, |\, \mathbf Y_{1:T})} \\
\geq& \sum_{\mathbf z_{1:T}}\log \left(\frac{\mathbb Q_{\theta}(\mathbf Y_{1:T},\mathbf Z_{1:T}=\mathbf z_{1:T})}{\mathbb Q_{\theta'}(\mathbf Z_{1:T}=\mathbf z_{1:T}\, |\, \mathbf Y_{1:T})}\right)\mathbb Q_{\theta'}(\mathbf Z_{1:T}=\mathbf z_{1:T} \, |\, \mathbf Y_{1:T}) =Q(\theta\, |\, \theta') + H(\theta'),
\end{align*}
where we used Jensen's inequality and where
\[Q(\theta\, |\, \theta') := \mathbb E_{\mathbf Z_{1:T} \sim \mathbb Q_{\theta'}(\cdot \, |\, \mathbf Y_{1:T})} \left[ \log \mathbb Q_{\theta}(\mathbf Y_{1:T},\mathbf Z_{1:T}) \right]\, \text{  and  } \, H(\theta'):=-\mathbb E_{\mathbf Z_{1:T} \sim \mathbb Q_{\theta'}(\cdot \, |\, \mathbf Y_{1:T})} \left[ \log \mathbb Q_{\theta'}(\mathbf Z_{1:T} \, |\, \mathbf Y_{1:T}) \right].\]
From Jensen's inequality, we get that $H(\theta')\geq 0$ so that $Q(\theta \, |\, \theta')$ is a lower-bound of the log-likelihood $\log \mathbb Q_{\theta}(\mathbf Y_{1:T})$ for any $\theta'$. Moreover, if we choose $\theta':=\theta^{\mathrm{MLE}}$, one can easily check that $\theta^{\mathrm{MLE}} \in \arg \max_{\theta} Q(\theta \, |\, \theta')$. Based on these observations, the EM algorithm seeks to find the MLE of the marginal likelihood by starting from an initial value of the parameters $\theta^{(0)}=(\mu^{(0)},P^{(0)},O^{(0)})$ and by iteratively applying these two steps:
\begin{itemize}
\item Expectation step (E step): Given the current value of the estimate of the parameters

$\theta^{(m)}=(\mu^{(m)},P^{(m)},O^{(m)})$, we compute the posterior probability distribution considering that the parameters of the HMM are given by $\theta^{(m)}$, namely $\mathbb Q_{\theta^{(m)}}(\cdot \, |\, \mathbf Y_{1:T})$.
\item Maximization step (M step): We find the parameter $\theta$ that maximizes $Q(\theta \, |\, \theta^{(m)})$ and we call it $\theta^{(m+1)}$.
\end{itemize}

When applied to HMMs, the EM algorithm is also called the Baum-Welch algorithm. In our framework, the M-step of the algorithm has a closed-form expression given by:
\begin{align}
\label{bw:1}\forall k \in [K], \quad  \mu^{(m+1)}_k &= \gamma_k^{(m)}(1) \\
\label{bw:2}\forall k,l \in [K], \quad P^{(m+1)}_{k,l} &= \frac{\sum_{t=1}^{T-1} \xi_{k,l}^{(m)}(t)}{\sum_{i=1}^{T-1} \gamma_k^{(m)}(t)} \quad \text{and} \quad  O^{(m+1)}_{k,l} = \frac{\sum_{t=1}^T \mathds 1_{Y_t = l} \gamma_k^{(m)}(t)}{\sum_{t=1}^T \gamma_k^{(m)}(t)},
\end{align}
where for all $k,l\in[K]$ and all $t\in[T]$,
\begin{align}
\label{bw:gamma}\gamma_k^{(m)}(t) &:= \mathbb Q_{\theta^{(m)}}(Z_t=k | \mathbf Y_{1:T}) = \frac{\alpha^{(m)}_k(t) \beta^{(m)}_k(t)}{\sum_{b \in [K]} \alpha^{(m)}_b(t) \beta^{(m)}_b(t) },\\
\notag \xi^{(m)}_{k,l}(t) &:= \mathbb Q_{\theta^{(m)}}(Z_{t}=k, Z_{t+1}=l \; | \; \mathbf Y_{1:T}) = \frac{\mathbb Q_{\theta^{(m)}}(Z_t=k,Z_{t+1}=l,\mathbf Y_{1:T})}{\mathbb Q_{\theta^{(m)}}(\mathbf Y_{1:T})}\\
\label{bw:beta} &=\frac{\alpha^{(m)}_k(t)  P^{(m)}_{k,l}\beta^{(m)}_l(t+1)O^{(m)}_{l,Y_{t+1}}}{\sum_{b,c \in [K]} \alpha^{(m)}_c(t)  P^{(m)}_{c,b}\beta^{(m)}_b(t+1)O^{(m)}_{b,Y_{t+1}}} .
\end{align}
In Eqs.\eqref{bw:gamma} and~\eqref{bw:beta}, the quantities $\alpha^{(m)}_k(t)$ and $\beta_k^{(m)}(t)$ for $k\in[K]$ and $t\in [T]$ are the counterparts of the $\alpha_k(t)$'s and $\beta_k(t)$'s of the beginning of this section but computed using the probability distribution $\mathbb Q_{\theta^{(m)}}$. The expressions for the M-step given by Eqs.\eqref{bw:1} and~\eqref{bw:2} show that given $\theta^{(m)}$, the E-step of the Baum-Welch algorithm only requires to perform the forward and backward procedures (cf. Algorithms~\ref{algo:forward} and \ref{algo:back}) to compute the $\alpha_k^{(m)}(t)$'s and the $\beta_k^{(m)}(t)$'s. We summarize the Baum-Welch algorithm with Algorithm~\ref{algo:bw}.

\begin{algorithm}[H]
    \centering
    \caption{Baum-Welch algorithm.}
    \label{algo:bw}
    \begin{algorithmic}[1]
    \STATE \textbf{Initialization:}  $\theta^{(0)}=(\mu^{(0)},P^{(0)},O^{(0)})$, $m_{\max}\geq1$.
    \STATE $m\leftarrow 0$.
        \FOR {$m=0\dots m_{\max}$}
        \STATE 1. Forward-Backward calculations:
        \STATE \hskip1em Compute $\alpha^{(m)}_k(t)$ for all $k\in [K]$ and all $t\in [T]$ using Algorithm~\ref{algo:forward} with $\theta^{(m)}$.
        \STATE \hskip1em Compute $\beta^{(m)}_k(t)$ for all $k\in [K]$ and all $t\in [T]$ using Algorithm~\ref{algo:back} with $\theta^{(m)}$.
        \STATE 2. E-step: 
        \STATE \hskip1em Compute $\gamma^{(m)}_k (t)$ for all $k\in[K]$ and all $t\in [T]$ using Eq.\eqref{bw:gamma}.
        \STATE \hskip1em Compute $\xi^{(m)}_{k,l} (t)$ for all $k,l\in[K]$ and all $t\in [T]$ using Eq.\eqref{bw:beta}.
        \STATE 3. M-step: \\
        \STATE \hskip1em Compute $\mu^{(m+1)}_k$ for all $k\in [K]$ using Eq.\eqref{bw:1}.
        \STATE \hskip1em Compute $P^{(m+1)}_{k,l}$ and $O^{(m+1)}_{k,l}$ for all $k,l\in [K]$ using Eq.\eqref{bw:2}.
        \STATE  \hskip1em $\theta^{(m+1)} \leftarrow (\mu^{(m+1)},P^{(m+1)},O^{(m+1)})$.
        \ENDFOR
    \end{algorithmic}
\end{algorithm}

We showed how the Baum-Welch algorithm allows at the same time to provide estimates of the parameters of the model and of $\alpha_k(t)$ and $\beta_k(t)$ (for $k\in[K]$ and $t\in [T]$) that are particularly useful to solve smoothing problems (cf. Eq.\eqref{eq:smooth}). Note that the convergence of the algorithm is ensured but it might stop at a local maximum of the log-likelihood. We refer to \cite[Chapters 3 and 10]{cappe2009inference} for further details on the Baum-Welch and the EM algorithms.

\subsubsection{From the Baum-Welch algorithm to reliable link prediction}

\label{sec:hmmbmlink}

\begin{figure}[!ht]
\vskip 0.2in
\begin{center}
\includegraphics[width=0.8\linewidth]{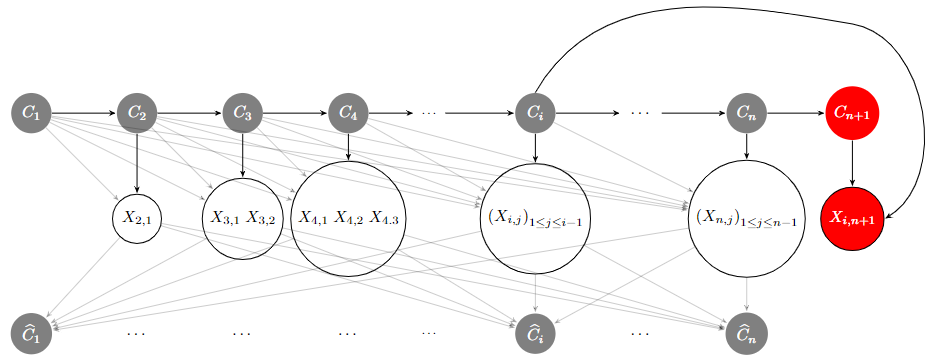}
\caption{Graphical model presenting the conditional dependencies between random variables.}
\label{fig:graphicalmodel}
\end{center}
\vskip -0.2in
\end{figure}

\begin{figure}[!ht]
\vskip 0.2in
\begin{center}
\includegraphics[width=0.8\linewidth]{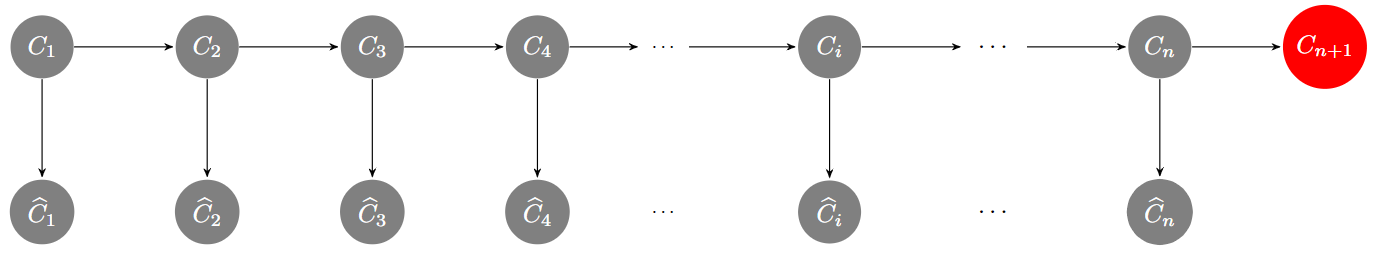}
\caption{We assume that the joint distribution of $(\mathbf C_{1:n},\hat {\mathbf C}_{1:n})$ factorizes according to the graph of a homogeneous HMM.}
\label{fig:gm-hmm}
\end{center}
\vskip -0.2in
\end{figure}

As previously mentioned, the plug-in approach for link prediction described in Section~\ref{sec:plugin} is highly sensitive to local clustering errors. We would like to use the complete sequence of estimated communities given by the clustering algorithm to estimate the probability that some node $i$ in the observed graph of size $n$ will be connected to the node $n+1$ (i.e. the upcoming node). Therefore, we aim at computing the following conditional probability $ \mathbb P(X_{i,n+1}=1 \, |\, \hat{ \mathbf C}_{1:n})$. Figure~\ref{fig:graphicalmodel} presents the graphical model expressing the conditional dependencies between the different random variables.
Thanks to Figure~\ref{fig:graphicalmodel}, we get that:
\begin{itemize}
\item[$\textcolor{red}{(*_1)}$] $X_{i,n+1}$ and $\hat {\mathbf C}_{1:n}$ are independent conditional to $(C_i,C_{n+1})$,
\item[$\textcolor{blue}{(*_2)}$] $C_{n+1}$ is independent of $(C_i,\hat {\mathbf C}_{1:n})$ conditional to $C_n$,
\end{itemize}
which implies that
\begingroup
\allowdisplaybreaks
\begin{align}& \notag\mathbb P(X_{i,n+1}=1 \, |\, \hat{ \mathbf C}_{1:n})\\
\notag= & \sum_{c_i,c_{n+1} \in [K]}\mathbb P(C_i=c_i,C_{n+1}=c_{n+1},X_{i,n+1}=1 \, |\, \hat{ \mathbf C}_{1:n})\\
\notag= & \sum_{c_i,c_{n+1} \in [K]}\frac{ \mathbb P(C_i=c_i,C_{n+1}=c_{n+1} ) }{\mathbb P ( \hat{ \mathbf C}_{1:n})}\mathbb P(X_{i,n+1}=1, \hat{ \mathbf C}_{1:n} \, |\, C_i=c_i,C_{n+1}=c_{n+1})\\
\notag \underset{cf.\textcolor{red}{(*_1)}}{=}&  \sum_{c_i,c_{n+1} \in [K]}\frac{ \mathbb P(C_i=c_i,C_{n+1}=c_{n+1} ) }{\mathbb P (  \hat{ \mathbf C}_{1:n})}  \underbrace{\mathbb P(X_{i,n+1}=1\, |\, C_i=c_i,C_{n+1}=c_{n+1})}_{=Q_{c_i,c_{n+1}}}\mathbb P( \hat{ \mathbf C}_{1:n} \, |\, C_i=c_i,C_{n+1}=c_{n+1})\\
\notag= & \sum_{c_i,c_{n+1} \in [K]} Q_{c_i,c_{n+1}}\mathbb P(  C_i=c_i,C_{n+1}=c_{n+1}\, |\, \hat{ \mathbf C}_{1:n})\\
\notag= & \sum_{c_i,c_n,c_{n+1} \in [K]} Q_{c_i,c_{n+1}}\mathbb P(  C_i=c_i,C_n=c_n,C_{n+1}=c_{n+1}\, |\, \hat{ \mathbf C}_{1:n})\\
\notag= & \sum_{c_i,c_n,c_{n+1} \in [K]} Q_{c_i,c_{n+1}}\mathbb P(  C_i=c_i,C_n=c_n\, |\, \hat{ \mathbf C}_{1:n})\mathbb P( C_{n+1}=c_{n+1}\, |\, C_i=c_i,C_n=c_n, \hat{ \mathbf C}_{1:n})\\
\label{eq:motiv} \underset{cf.\textcolor{blue}{(*_2)}}{=} & \sum_{c_i,c_n,c_{n+1} \in [K]} Q_{c_i,c_{n+1}}\mathbb P( C_i= c_i,C_n=c_n\, |\, \hat{ \mathbf C}_{1:n})P_{c_n,c_{n+1}}.
\end{align}
\endgroup
In Section~\ref{estimation}, we proposed estimates $\hat Q$ and $\hat P$ of the model parameters $Q$ and $P$ with theoretical guarantees. Hence the challenge lies in the estimation of the probability mass function $\mathbb P(  C_i,C_n\, |\, \hat{ \mathbf C}_{1:n})$ to get an approximation of \eqref{eq:motiv}. The computation of $\mathbb P(  C_i,C_n\, |\, \hat{ \mathbf C}_{1:n})$ looks like a smoothing problem where the random variables $\hat {\mathbf C}_{1:n}$ are observed and the random variables $\mathbf C_{1:n}$ are hidden. The important difference with our discussion from Sections~\ref{subsub:1} and~\ref{subsub:2} is that the emission probabilities are not independent, i.e. $\hat C_i$ is not independent from all other random variables (namely $(C_j,\hat C_j)_{j\neq i}$) conditional to $C_i$ (cf. Figure~\ref{fig:graphicalmodel}). Inspired by mean field approximation techniques (see for example~\cite{meanfield}), we propose to simplify the complex dependence structure of our problem depicted in Figure~\ref{fig:graphicalmodel} by assuming that the emission probabilities are independent and time invariant, meaning that the joint distribution of $(\mathbf C_{1:n},\hat {\mathbf C}_{1:n})$ factorizes according to the graph of a homogeneous HMM (cf. Figure~\ref{fig:gm-hmm}).
Denoting by $\mathbb Q$ the probability distribution with model parameter $(\mu,P,Q)$ and where we assume the HMM structure presented in Figure~\ref{fig:gm-hmm}, it holds
\begin{equation}\label{eq:QQ}\mathbb Q(X_{i,n+1}=1 \, |\, \hat {\mathbf C}_{1:n}) = \sum_{c_i,c_n,c_{n+1}\in [k]} Q_{c_i,c_{n+1}}P_{c_n,c_{n+1}} \mathbb Q(C_i=c_i,C_n=c_n \, |\, \hat {\mathbf C}_{1:n}),\end{equation}
where $\quad \mathbb Q(C_i=c_i,C_j=c_j \, |\, \hat {\mathbf C}_{1:n})\propto\mathbb Q(C_i=c_i,C_j=c_j, \hat {\mathbf C}_{1:n})
=  \alpha_{c_i}(i)  \chi^{(i,j)}_{c_i,c_j} \beta_{c_j}(j)\quad$ with \[ \alpha_{c_i}(i) := \mathbb Q(C_i=c_i, \hat {\mathbf C}_{1:i}), \quad   \chi^{(i,j)}_{c_i,c_j}:=\mathbb Q(C_j=c_j,\hat {\mathbf C}_{i+1:j}\, |\, C_i=c_i) \quad  \text{and}\quad  \beta_{c_j}(j):=\mathbb Q(\hat{\mathbf C}_{j+1:n} \, |\, C_j=c_j),\]
and where the symbol $\propto$ means that both side are equal up to some factor that is a function of $\hat {\mathbf C}_{1:n}$. Using the Baum-Welch algorithm, we obtain estimates $\hat \alpha_{c_i}(i)$'s of the $\alpha_{c_i}(i)$'s, estimates $\hat \beta_{c_j}(j)$'s of the $ \beta_{c_j}(j)$'s and estimates $\hat O_{k,l}$'s of the $O_{k,l}$'s where for all $k,l\in [K]$, $O_{k,l} = \mathbb Q(\hat C_t =l\, |\, C_t=k)$ (where $t\in [n]$ is arbitrary since we assumed that the HMM is homogeneous). We can also estimate $\chi_{c_i,c_j}^{(i,j)}$ using a plug-in approach with
\begin{equation}\label{eq:chihat}\hat \chi^{(i,j)}_{c_i,c_j}:= \sum_{c_{i+1},\dots,c_{j-1}} \hat P_{c_i,c_{i+1}}\hat O_{c_{i+1},\hat C_{i+1}}\hat P_{c_{i+1},c_{i+2}}\hat O_{c_{i+2},\hat C_{i+2}}\dots \hat P_{c_{j-1},c_{j}}\hat O_{c_{j},\hat C_{j}},\end{equation}
where we recall that $\hat P$ is defined in Section~\ref{defY}\footnote{Let us mention that we could have used the estimate of the transition matrix $P$ provided by the Baum-Welch algorithm instead of the one given in Section~\ref{defY}. We prefer the latter since its computation does not rely on the approximation made by considering a homogeneous HMM.}. Recalling the definition of $\hat Q$ given in Section~\ref{sec:connectivity-matrix} and using Eq.\eqref{eq:QQ}, we can then estimate $ \eta_{i}(\mathbf C_{1:n})$ by 
\begin{equation} \label{eq:etaR}\hat \eta_i^R(\hat {\mathbf C}_{1:n}):=\sum_{c_i,c_n,c_{n+1}\in[K]} \hat Q_{c_i,c_{n+1}} \hat P_{c_n,c_{n+1}} \hat \alpha_{c_i}(i)\hat \chi^{(i,n)}_{c_i,c_n}\hat \beta_{c_n}(n).\end{equation}
Hence, the Reliable MSBM (RMSBM) classifier is defined by replacing $\hat \eta_i$ by $\hat \eta^R_i$ in the definition of the MSBM classifier (cf. Definition~\ref{def:mrgg-classifier}). Note that this approach is a heuristic because of the local optimum reached by the EM algorithm and also because of the dependence of the emission probabilities in our model (cf. Figure~\ref{fig:graphicalmodel}). Algorithm~\ref{algo:RMSBM} summarizes our method.
\begin{algorithm}[H]
    \centering
    \caption{Reliable link prediction.}
    \label{algo:RMSBM}
    \begin{algorithmic}[1]
    \STATE \textbf{Data:} Adjacency matrix $X\in \{0,1\}^{n\times n}$, $K\in \mathbb N$ with $K\geq 2$ and a number of iterations $m_{\max} \in \mathbb N$.
    \STATE Run the clustering algorithm on $X$ with $K$ clusters to get the estimated communities \colorbox{light-gray}{$\hat{\mathbf C}_{1:n}$}.
    \STATE Compute the estimate \colorbox{light-gray}{$\hat Q$} of the connectivity matrix from $\hat{\mathbf C}_{1:n}$ and $X$ using Section~\ref{sec:connectivity-matrix}.
    \STATE Compute the estimate \colorbox{light-gray}{$\hat P$} of the Markov kernel from $\hat{\mathbf C}_{1:n}$ using Section~\ref{defY}.
    \STATE {\bf Baum-Welch algorithm:}
    \STATE \hskip1em Initialization:
    \STATE \hskip2em  $  \mu^{(0)} : =   \mathbf 1_K^{\top}$,
    \STATE \hskip2em  $ P^{(0)} := \frac{1}{K} \mathbf 1_K  \mathbf 1_K^{\top}$,
    \STATE \hskip2em  $O^{(0)} := (1-\epsilon) \mathrm{Id}_K + \frac{\epsilon}{K-1} \left(  \mathbf 1_K  \mathbf 1_K^{\top} - \mathrm{Id}_K\right),$ where $\epsilon \in (0,1)$ (typically $\epsilon = 10^{-2}$).
    \STATE \hskip1em Working with the probability $\mathbb Q$ (i.e. considering the assumption of homogeneous HMM),\par
  \hskip1em run the Baum-Welch algorithm (cf. Algorithm~\ref{algo:bw}) with $\theta^{(0)}:=(\mu^{(0)},P^{(0)},O^{(0)})$ and a number of iterations $m_{\max}$.\vspace{-0.3cm}
    \STATE \hskip1em \colorbox{light-gray}{$\hat \alpha_k(i)$}$\leftarrow \alpha^{(m_{\max})}_k(i),\quad \forall k\in[K], \, \forall i \in [n]$.
    \STATE \hskip1em \colorbox{light-gray}{$\hat \beta_k(i)$}$\leftarrow \beta^{(m_{\max})}_k(i),\quad \forall k\in[K], \, \forall i \in [n]$.
    \STATE \hskip1em \colorbox{light-gray}{$\hat O_{k,l}$}$\leftarrow O^{(m_{\max})}_{k,l}, \, \forall k,l \in [K]$.
    \STATE Compute \colorbox{light-gray}{$\hat \chi^{(i,n)}_{k,l}$} for all $k,l\in [K]$ and all $i\in[n]$ using Eq.\eqref{eq:chihat}. 
    \STATE Compute \colorbox{light-gray}{$\hat \eta_i^R(\hat {\mathbf C}_{1:n})$} for all $i\in [K]$ using Eq.\eqref{eq:etaR}.
    \STATE {\bf Return}: $\{i\in[n] \, : \, \hat \eta_i^R(\hat {\mathbf C}_{1:n})\geq \frac12 \}$ as the set of nodes from $[n]$ predicted to be connected to the node $n+1$.
    \end{algorithmic}
\end{algorithm}
  
 \subsubsection{Numerical results} 
 \label{num-expe-bw}
 
We perform our numerical experiment using the matrices $P$ and $Q$ provided in Eq.\eqref{tmatrixK5} of the Appendix. With Figure~\ref{fig:robust-MSBM}, we aim at showing the superiority of our reliable link prediction method compared to the plug-in one from Section~\ref{sec:plugin} while with Figure~\ref{fig:emission-probas}, we aim at showing that the learned emission probabilities carry relevant information on the clustering algorithm used (and also on the studied network as we will see in Section~\ref{sec:model-selection}).

\medskip
In Figure~\ref{fig:robust-MSBM}, we consider a graph of size $70$ and we plot the sorted list of the $L_1$ errors $\big(\big|\hat \eta_i(\hat{\mathbf{C}}_{1:n})-\eta_i(\hat{\mathbf{C}}_{1:n})\big|\big)_{i\in[n]}$ (resp. $\big(\big|\hat \eta_i^R(\hat{\mathbf{C}}_{1:n})-\eta_i(\hat{\mathbf{C}}_{1:n})\big|\big)_{i\in[n]}$). Our reliable estimation of the posterior probabilities allows to get a significantly smaller variance compared to the plug-in approach. The RMSBM classifier gives smaller $L_1$ errors on the posterior probabilities. The difference is significant when the clustering algorithm fails to recover the complete partition of the nodes which leads to bad estimates for the plug-in approach. 
\medskip

In Figure~\ref{fig:emission-probas}, we work with a graph of size $120$ and we plot the learned emission probabilities $\hat O_{k,l}$, $k,l \in [K]$. The ergodic theorem ensures that the first cluster is (asymptotically) the smaller. Indeed, the stationary measure of $P$ from~\eqref{tmatrixK5} is approximately $[0.14 \;,\; 0.22 \;,\;  0.38 \;,\; 0.26]$. We observe that the errors made by the algorithm consist in assigning nodes from community $2,3$ or $4$ to cluster~$1$. This means that the clustering algorithm from \cite{Verzelen} tends to overestimate the size of small clusters. In the Appendix (cf. Section~\ref{apdx:overesti}), we conduct experiments in other settings that confirm this overestimation of the size of small clusters.
\begin{figure}[ht!]
  \begin{minipage}[c]{0.5\textwidth}
    \includegraphics[width=\textwidth]{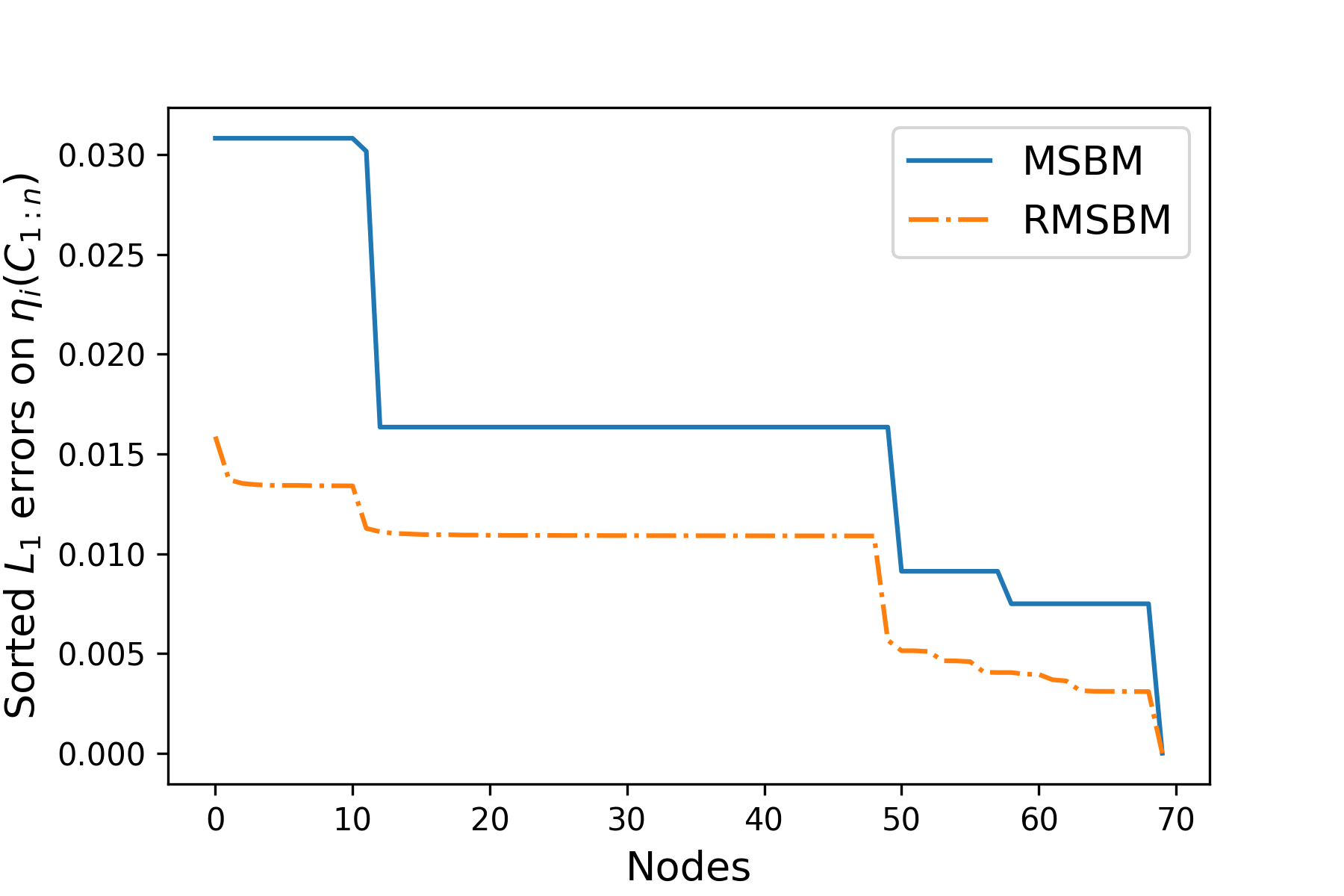}
    \caption{Sorted list of $L_1$ errors between $\hat \eta_i(\hat{\mathbf{C}}_{1:n})$ 
    (resp. $\hat \eta^R_i(\hat{\mathbf{C}}_{1:n})$) and $\eta_i(\hat{\mathbf{C}}_{1:n})$.} 
\label{fig:robust-MSBM}
  \end{minipage}
  \begin{minipage}[c]{0.48\textwidth}
      \includegraphics[width=\textwidth]{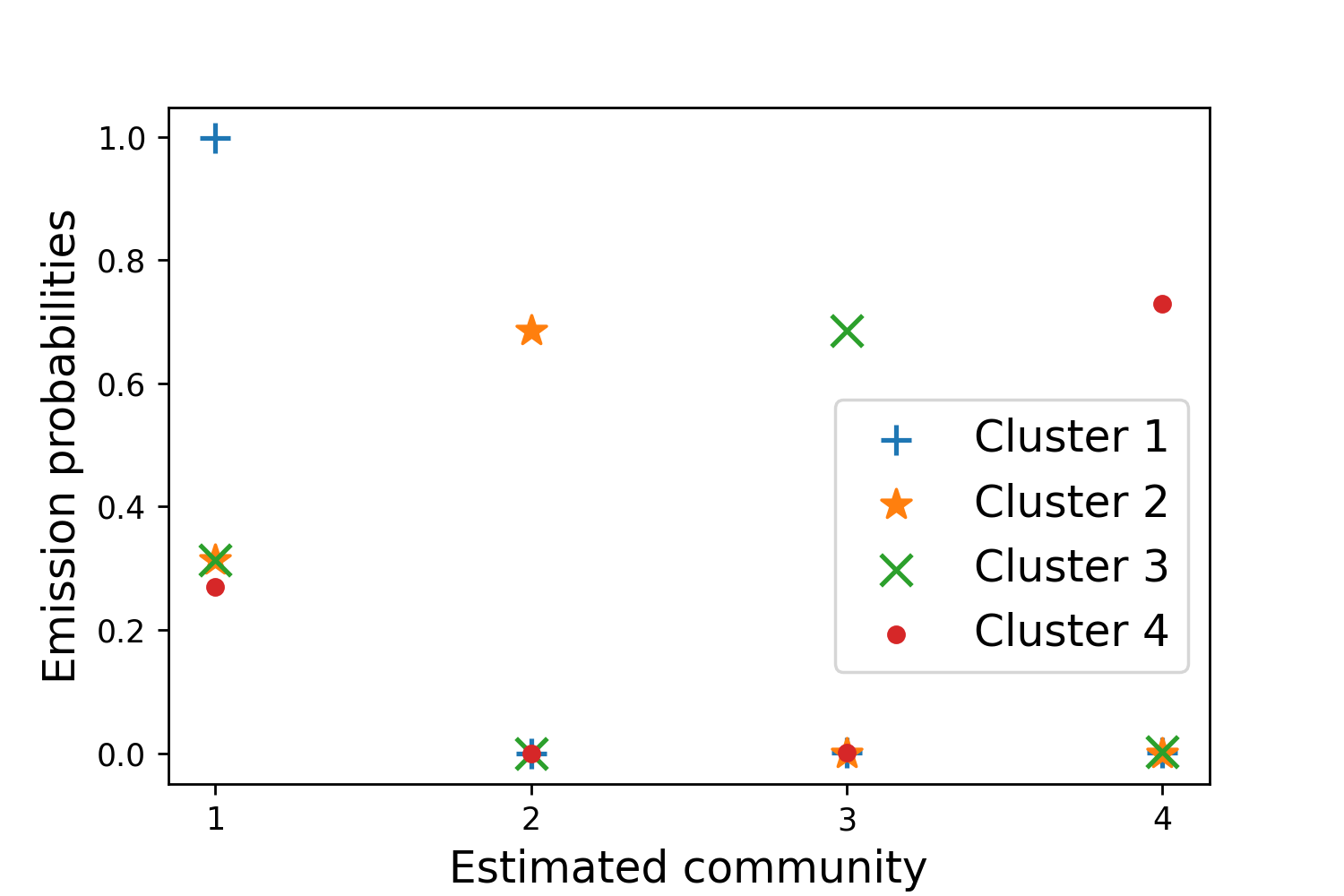}
 \caption{Plot of the learned emission probabilities $\hat O_{k,l}$, $k,l \in [K]$. Typically, on the first column we show $\hat O_{l,1}$ for $l\in [K]$.} 
 \label{fig:emission-probas}
  \end{minipage}
\end{figure}

\medskip

{\bf Notation:} In the following, we will denote by $\hat \mu$ (resp. $\hat O$) the probability measure $\mu^{(m_{\max})}$ (resp. the matrix $O^{(m_{\max})}$) returned by the Baum-Welch algorithm (cf. Algorithm~\ref{algo:bw}). Moreover, $\hat {\mathbb Q}$ will denote the probability measure under which 
\begin{itemize}
\item the Markov chain $(C_i)_{i \geq1}$ has initial distribution $\hat \mu$ and transition matrix $\hat P$ ($\hat P$ is defined in Section~\ref{estimation}), 
\item $X_{i,j} \sim \mathrm{Ber}(\hat Q_{C_i,C_j})$ for all $i, j \in [n], \; i\neq j$ ($\hat Q$ is defined in Section~\ref{estimation}),
\item we have the homogeneous HMM structure with emission probabilities the $\hat O_{k,l}$'s, i.e.
\[\hat {\mathbb Q}( \hat {\mathbf C}_{1:n}\, |\, \mathbf C_{1:n}) = \prod_{i=1}^n \hat {\mathbb Q}(\hat C_i  \, |\, C_i)=\prod_{i=1}^n \hat O_{C_i,\hat C_i}.\]
\end{itemize}

\section{Collaborative filtering}

 \label{sec:colab}
 \subsection{Reliable collaborative filtering}
 
 \label{sec:colabo}

Let us now consider another prediction task for dynamic networks: collaborative filtering. Solving a collaborative filtering problem consists in inferring the community of one node of the graph when we have only partial information about how this node is connected to the rest of the graph. More precisely, we observe fully the graph at time $m$ and for some $n>m$, we observe how the node $n$ is connected (or not) to a subset of nodes $\mathcal E \subset [m]$, i.e. we have access to $(X_{i,n})_{i \in \mathcal E}$. Our goal is then to predict the community of node $n$: $C_n$. We propose to use the maximum a posteriori (MAP) estimator to tackle this problem. In the following, we present three different strategies to solve the collaborative filtering problem using the maximum a posteriori (MAP) estimator. The practical implementation of the different methods are presented in the next section.

\begin{itemize}
\item {\bf The optimal MAP} has access to the hidden communities of the nodes in $[m]$ and to the model parameters. The community predicted for the node $n$  by the optimal MAP is
\[\hat C_n^{OPT} \in \underset{k \in [K]}{\arg \max} \; \mathbb P\left( C_n=k\;|\; (X_{i,n})_{i \in \mathcal E}, \mathbf{ C}_{1:m} \right).\]
\item {\bf The plug-in MAP} does not have access to the model parameters or to the hidden communities $\mathbf C_{1:n}$. It works with the probability $\mathbb P_{(\hat \mu,\hat P,\hat Q)}$ where $\hat \mu:=\mu^{(m_{\max})}$ is given by the Baum-Welch algorithm and where the Markov kernel $\hat P$ and the connectivity matrix $\hat Q$ are defined in Section~\ref{estimation}. $\mathbb P_{(\hat \mu,\hat P,\hat Q)}$ is the counterpart of $\mathbb P$ where the Markov chain $\mathbf C_{1:n}$ has initial distribution $\hat \mu$ and transition matrix $\hat P$, and where the connectivity matrix $Q$ is replaced by $\hat Q$. The plug-in MAP trusts the sequence of communities $\mathbf{\hat C}_{1:m}$ returned by the clustering algorithm and predicts the following community for the node $n$:
\[\hat C_n^{PI} \in \underset{k \in [K]}{\arg \max} \; \mathbb P_{(\hat \mu,\hat P,\hat Q)}\left( C_n=k\;|\; (X_{i,n})_{i \in \mathcal E}, \mathbf{ C}_{1:m}=\mathbf{\hat C}_{1:m} \right).\]
\item {\bf The Reliable MAP} does not know the model parameters or the hidden communities $\mathbf C_{1:n}$ either but it does not want to blindly rely on the estimated communities $\hat {\mathbf C}_{1:m} $ provided by the clustering algorithm. As a consequence, the Reliable MAP works with the probability $\hat {\mathbb Q}$ introduced in the previous section to be robust against possible clustering errors. The Reliable MAP predicts the following community for the node $n$:
\begin{align*} \hat C_n^R &\in \underset{k \in [K]}{\arg \max} \; \hat {\mathbb Q}\left( C_n=k\;|\; (X_{i,n})_{i \in \mathcal E} , \hat{\mathbf { C}}_{1:m}\right).
\end{align*}
\end{itemize}

\begin{figure}[!ht]
\centering
\includegraphics[scale=0.65]{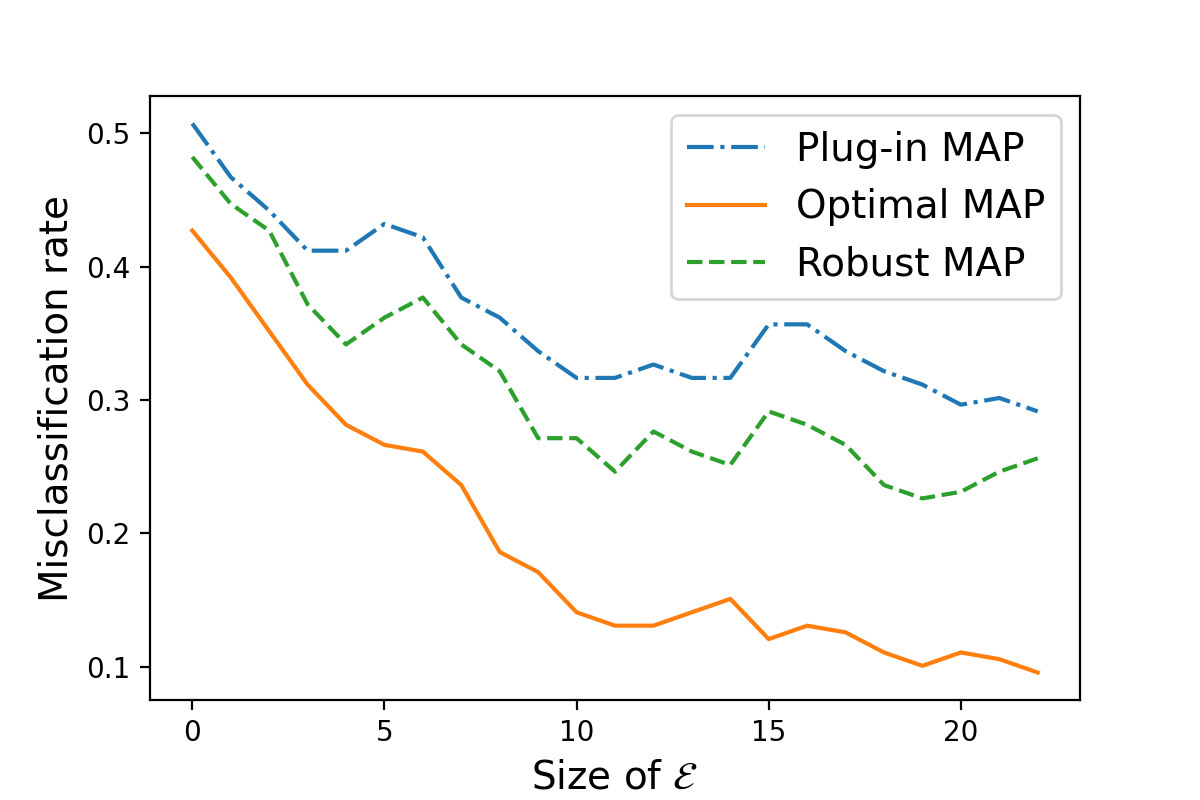}
\caption{Misclassification error rate for the optimal MAP, the plug-in MAP and the Reliable MAP.}
\label{fig:collaborative}
\end{figure}

Before describing in details how we compute $\hat C_n^{OPT}$, $\hat C_n^{PI}$ and $\hat C_n^{R}$, we present some numerical results. We consider a random graph drawn from the MSBM using the matrices given in \eqref{tmatrixK5}. We fully observe the graph until time $m=100$ and we observe how the node $n=120$ is connected to the nodes in $\mathcal E$ where $\mathcal E$ is equal to $\{m\}$, $\{m-1,m\}$, $\dots$ or $\{m-25, \dots, m\}$. For those different choices of $\mathcal E$, we plot in Figure~\ref{fig:collaborative} the average error on the clustering of the node $n$ using the optimal MAP, the plug-in MAP or the Reliable MAP. The plug-in MAP gives reasonable results but its misclassification rate is always lower bounded by the one of the Reliable MAP.

\subsection{Computing the optimal MAP, the plug-in MAP and the Reliable MAP}

In this section, we give the formula to compute the different estimates of the community of node $n$ in the collaborative filtering problem tackled in the previous section. We denote $\mathcal E = \{i_1, \dots, i_S\}$ with $1\leq i_1 < \dots < i_S \leq m$. In the following, the symbol $\propto$ will be used in the sense that the considered quantities are equal up to some normalization factor.
 
\begin{itemize}
 \item 
 The optimal MAP selects $\hat C_n^{OPT} \in \underset{k \in [K]}{\arg \max} \; \mathbb P\left( C_n=k\;|\; (X_{i,n})_{i \in \mathcal E}, \mathbf{ C}_{1:m} \right)$ with
 \begin{align*}
  \mathbb P\left( C_n=k\;|\; (X_{i,n})_{i \in \mathcal E}, \mathbf{ C}_{1:m} \right)& \propto   \mathbb P\left( C_n=k,  (X_{i,n})_{i \in \mathcal E}\;|\; \mathbf{ C}_{1:m} \right) \\
  &=  \mathbb P\left(  (X_{i,n})_{i \in \mathcal E}\;|\;C_n=k,  \mathbf{ C}_{1:m} \right)  \mathbb P\left( C_n=k \; | \;   \mathbf{ C}_{1:m} \right) \\
  &= \prod_{i \in \mathcal E} Q_{C_i,k}^{X_{i,n}}(1-Q_{C_i,k})^{X_{i,n}} \, .\,  \left( P^{n-m} \right)_{C_m,k}.
 \end{align*}
\item The plug-in MAP selects $\hat C_n^{PI} \in \underset{k \in [K]}{\arg \max} \;  \mathbb P_{(\hat \mu,\hat P,\hat Q)}\left( C_n=k\;|\; (X_{i,n})_{i \in \mathcal E}, \mathbf{ C}_{1:m}=\mathbf{\hat{ C}}_{1:m} \right)$ with
 \begin{align*}
  & \mathbb P_{(\hat \mu,\hat P,\hat Q)}\left( C_n=k\;|\; (X_{i,n})_{i \in \mathcal E}, \mathbf{ C}_{1:m}=\mathbf{\hat{ C}}_{1:m}  \right)\\
  &\propto    \mathbb P_{(\hat \mu,\hat P,\hat Q)}\left( C_n=k,  (X_{i,n})_{i \in \mathcal E}\;|\; \mathbf{ C}_{1:m}=\mathbf{\hat{ C}}_{1:m}  \right) \\
  &=   \mathbb P_{(\hat \mu,\hat P,\hat Q)}\left(  (X_{i,n})_{i \in \mathcal E}\;|\;C_n=k,  \mathbf{ C}_{1:m}=\mathbf{\hat{ C}}_{1:m}  \right)   \mathbb P_{(\hat \mu,\hat P,\hat Q)}\left( C_n=k \; | \; \mathbf{ C}_{1:m}=\mathbf{\hat{ C}}_{1:m} \right) \\
  &= \prod_{i \in \mathcal E} \hat Q_{\hat C_i,k}^{X_{i,n}}(1-\hat Q_{\hat C_i,k})^{X_{i,n}} \, . \,  \left( \hat P^{n-m} \right)_{\hat C_m,k}.
 \end{align*}
\item The Reliable MAP selects $\hat C_n^R \in \underset{k \in [K]}{\arg \max} \; \hat {\mathbb Q}\left( C_n=k\;|\; (X_{i,n})_{i \in \mathcal E} , \hat{\mathbf C}_{1:m}=\hat{\mathbf C}_{1:m}\right)$ with
\begin{align*}
\hat {\mathbb Q}\left( C_n=k\;|\; (X_{i,n})_{i \in \mathcal E} , \hat{\mathbf C}_{1:m}\right) &\propto \hat {\mathbb Q}\left( C_n=k, (X_{i,n})_{i \in \mathcal E} , \hat{\mathbf C}_{1:m}\right)\\
&=  \underbrace{\hat {\mathbb Q}\left(  (X_{i,n})_{i \in \mathcal E} , \hat{\mathbf C}_{1:m}\; |\; C_n=k\right)}_{=:\textcolor{red}{(\star)}}   \hat{\mathbb Q }\left( C_n=k\right).
\end{align*}
We have easily $  \hat{\mathbb Q }\left( C_n=k\right) =  \sum_{l \in [K]} \hat \mu_l \left(\hat P^n\right)_{l,k}$. Moreover, denoting
\[\forall i\in [n], \; \forall c,k \in [K], \quad \hat l_{i,n}(c,k) = \hat Q_{c,k}^{X_{i,n}}(1-\hat Q_{c,k})^{1-X_{i,n}},\]
we have
\begin{align*}
\textcolor{red}{(\star)}&=\hat {\mathbb Q}\left(  (X_{i,n})_{i \in \mathcal E} , \hat{\mathbf C}_{1:m}\; |\; C_n=k\right) \\
& = \hat {\mathbb Q}\left(  (X_{i_j,n})_{j \in [S]} , \hat{\mathbf C}_{1:m}\; |\; C_n=k\right)\\
&=  \sum_{c_{i_1}\in [K]} \hat {\mathbb Q}(C_{i_1}=c_{i_1},\hat {\mathbf C}_{1:i_1}) \hat Q_{c_{i_1},k}^{X_{i_1,n}}(1-\hat Q_{c_{i_1},k})^{X_{i_1,n}}   \hat {\mathbb Q}\left(  (X_{i_j,n})_{j \in \{2, \dots, S\}} , \hat{\mathbf C}_{i_1+1:m} \; |\; C_{i_1}=c_{i_1}, C_n=k\right).
\end{align*}
Replacing $\hat {\mathbb Q}(C_{i_1}=c_{i_1},\hat {\mathbf C}_{1:i_1})$ by its estimate $\hat \alpha_{c_{i_1}}(i_1)$ provided by the Baum-Welch algorithm, we have
\begin{align*}
\textcolor{red}{(\star)}&\approx  \sum_{c_{i_1}\in [K]} \hat \alpha_{c_{i_1}}(i_1) \hat Q_{c_{i_1},k}^{X_{i_1,n}}(1-\hat Q_{c_{i_1},k})^{X_{i_1,n}} \\
&\hspace{3cm}.\,  \hat {\mathbb Q}\left(  (X_{i_j,n})_{j \in \{2, \dots, S\}} , \hat{\mathbf C}_{i_1+1:m} \; |\; C_{i_1}=c_{i_1}, C_n=k\right)\\
&=  \sum_{c_{i_1},c_{i_2}\in [K]}\hat \alpha_{c_{i_1}}(i_1) \hat l_{i_1,n}(c_{i_1},k)  \hat {\mathbb Q}\left(  \hat{\mathbf C}_{i_1+1:i_2} ,C_{i_2}=c_{i_2}\; |\; C_{i_1}=c_{i_1}\right) \hat l_{i_2,n}(c_{i_2},k)\\
&\hspace{2cm}.\,  \hat {\mathbb Q}\left(  (X_{i_j,n})_{j \in \{3, \dots, S\}} , \hat{\mathbf C}_{i_2+1:m} \; |\; C_{i_2}=c_{i_2}, C_n=k\right) .
\end{align*}
Once again, we can replace $\hat {\mathbb Q}\left(  \hat{\mathbf C}_{i_1+1:i_2} ,C_{i_2}=c_{i_2}\; |\; C_{i_1}=c_{i_1}\right)$ by its estimate $ \hat \chi^{(i_1,i_2)}_{c_{i_1},c_{i_2}}$ given by the Baum-Welch algorithm (see Eq.\eqref{eq:chihat}). Iterating this procedure, one can derive the following approximation of $\textcolor{red}{(\star)}$:
\begin{align*}
\textcolor{red}{(\star)}&\approx   \sum_{c_{i_1},c_{i_2}\in [K]}\hat \alpha_{c_{i_1}}(i_1) \hat l_{i_1,n}(c_{i_1},k) \hat \chi^{(i_1,i_2)}_{c_{i_1},c_{i_2}} \hat l_{i_2,n}(c_{i_2},k) \hat {\mathbb Q}\left(  (X_{i_j,n})_{j \in \{3, \dots, S\}} , \hat{\mathbf C}_{i_2+1:m}  \; |\; C_{i_2}=c_{i_2}, C_n=k\right) \\
&= \dots \\
&\approx \sum_{\substack{c_{i_1},\dots,c_{i_S} \in [K]}} \hat \alpha_{c_{i_1}}(i_1) \prod_{j=1}^{S-1} \Bigg( \hat l_{i_j,n}(c_{i_j},k) 
\hat \chi^{(i_j,i_{j+1})}_{c_{i_j},c_{i_{j+1}}} \Bigg)\hat \beta_{c_{i_S}}(i_S) \hat l_{i_S,n}(c_{i_S},k).
\end{align*}

Thanks to the previous computations, we deduce that the estimated community for node $n$ provided by the Reliable MAP is
\[\underset{k \in [K]}{\arg \max} \; \sum_{l \in [K]} \hat \mu_l \left(\hat P^n\right)_{l,k} \; . \; \sum_{\substack{c_{i_1},\dots,c_{i_S} \in [K]}} \hat \alpha_{c_{i_1}}(i_1) \prod_{j=1}^{S-1} \Bigg( \hat l_{i_j,n}(c_{i_j},k) 
\hat \chi^{(i_j,i_{j+1})}_{c_{i_j},c_{i_{j+1}}} \Bigg)\hat \beta_{c_{i_S}}(i_S) \hat l_{i_S,n}(c_{i_S},k).\]
\end{itemize}

\section{Implementation and Experiments}
\label{simulations}

\subsection{Complexity and implementation}

The clustering algorithm used is a SDP method and, as a consequence, its time complexity scales with $n^3$, while the complexity of the Baum-Welch algorithm is of order $K^2n$. From here, computing $\hat \eta_i (\hat{\mathbf{c}}_{1:n})$ for all $i \in [n]$ requires $K^3 n^2 $ operations (see Sec.3.b of the notebook {\it experiments.ipynb}). Regarding the collaborative filtering task, the Reliable MAP estimator from Section~\ref{sec:colabo} has a time complexity of order $K^4 n^2$ (see method {\it collaborative\_filtering\_robustMAP} in the file {\it markovianSBM/BaumWelch.py}).

\subsection{Inferring the number of communities} \label{sec:model-selection}
In this section, we propose a heuristic based on the learned emission probabilities from the Baum-Welch algorithm to estimate the number of communities $K$ of our model. The proposed approach consists in running the Baum-Welch algorithm for a finite list of possible number of clusters $\{K_{min},\dots, K_{max}\}=\mathcal K \subset \mathbb N^*$. For each $K \in \mathcal K$, we denote $\hat O^{(K)}$ the matrix of emission probabilities learned by Algorithm~\ref{algo:RMSBM} when we consider that the number of communities is $K$. For any $K \in \mathcal K$, we define \[M^{(K)} := \max_{k,l \in [K], k\neq l}\left\{\hat  O^{(K)}_{l,k}+\hat O^{(K)}_{k,l} \right\}.\]
For any $K \in \mathcal K$ and any $k,l \in [K], \; k\neq l$, $\hat O^{(K)}_{l,k}+\hat O^{(K)}_{k,l} $ represents the probability that the clustering algorithm predicts community $k$ or $l$ if the true cluster is the other one.
\begin{itemize}
\item When $K $ is less than or equal to the true number of clusters,  $M^{(K)}$ stays small as soon as the graph is large enough and as the clustering algorithm used is efficient. \par 
\smallskip

\noindent Typically if the observed graph has $4$ hidden communities, we expect the clustering algorithm run with $K=3$ to return a partition of the nodes that is - in the ideal case - of the form $(G_1,G_2,G_3\cup G_4)$. In this ideal case, it holds $M^{(3)}=0$. Stated otherwise, we believe that the clustering algorithm will merge true clusters to output a partition of nodes with only $K$ groups while the true underlying structure of the graph contains more than $K$ clusters.
\item When $K $ becomes greater then the true number of clusters,  $M^{(K)}$ is larger compared to the previous case because at least one true cluster will be arbitrarily split in two different groups by the clustering algorithm. \par
\smallskip

\noindent For example, let us consider that the observed graph has $4$ hidden communities and that we run the clustering algorithm with $K=5$. In this case, a possible way to interpret the latent structure of the graph into $5$ groups is to consider the partition of the nodes given by $(G_1,G_2,G_3, G_4^{(1)}, G_4^{(2)})$ where $ G_4^{(1)} \cup G_4^{(2)} = G_4$. Stated otherwise, we have arbitrarily split one true cluster in two groups. Considering that the true labels of the nodes is given by $(G_1,G_2,G_3, G_4^{(1)}, G_4^{(2)})$, we understand that the clustering algorithm will have difficulties to obtain the label $4$ for nodes in $G_4^{(1)}$ and the label $5$ for nodes in $G_4^{(2)}$ because nodes belonging to $G_4=G_4^{(1)} \cup G_4^{(2)}$ have the same connection probabilities. Hence, we expect clustering errors leading to $\hat O_{4,5} + \hat O_{5,4}>0$ and thus $M^{(K)}>0$.
\end{itemize}
Based on this remark, we propose to estimate the number of communities by choosing the value $K \in \mathcal K$ leading to the larger positive jump of the function $K \mapsto M^{(K)}$ namely
\[\hat K \in \underset{K \in \{K_{min},\dots, K_{max}-1\}}{\arg \max} \left\{ M^{(K+1)}- M^{(K)}\right\}.\]

\begin{figure}[!ht]
  \centering
    \includegraphics[width=0.6\textwidth]%
    {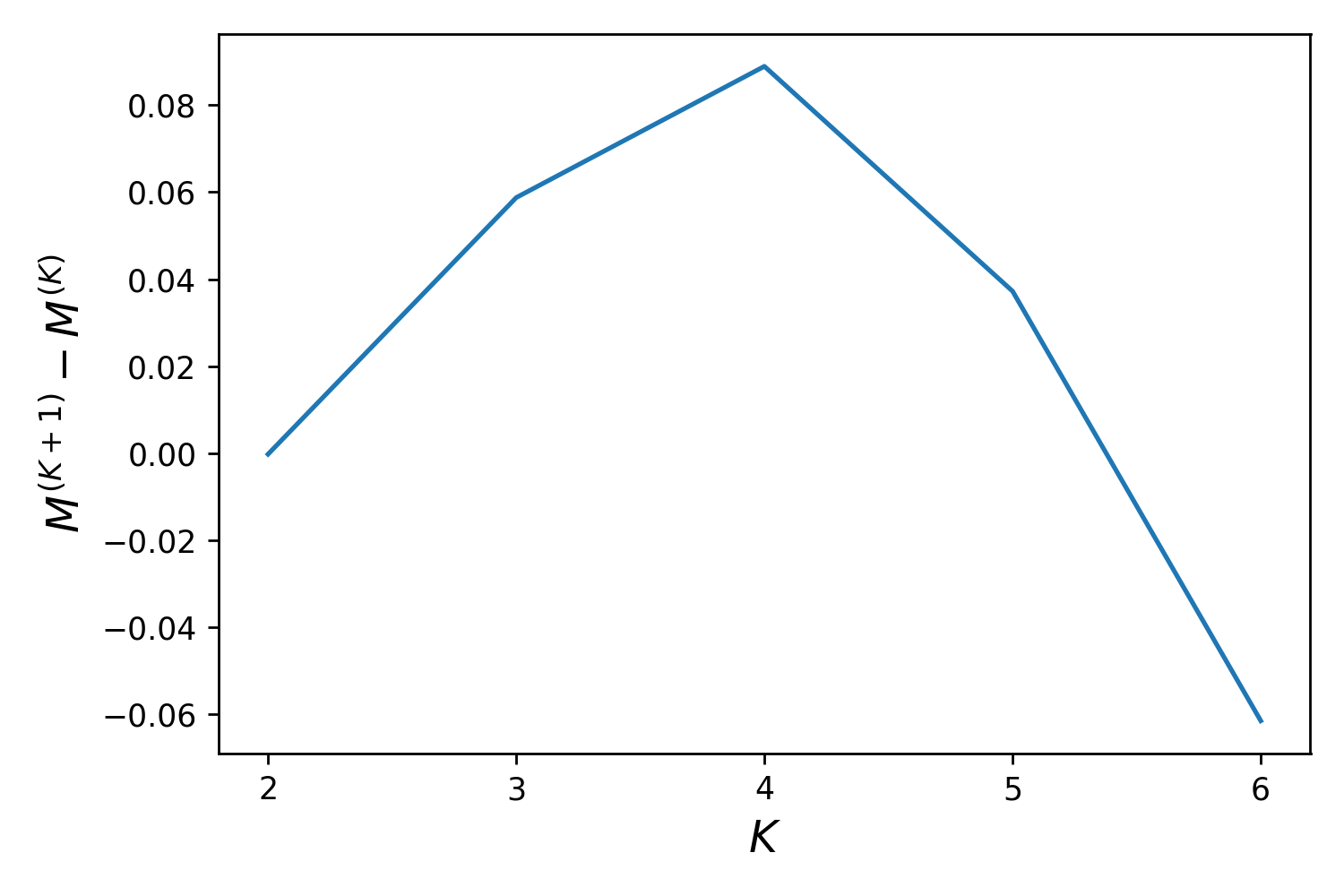}
  \caption[Heuristic to infer the number of clusters in the graph.]{$K \mapsto  M^{(K+1)}- M^{(K)}$ working with the connectivity matrix $Q$ from \eqref{tmatrixK5}.} 
\label{selec}
\end{figure}

First we test our method with a graph of size $n=110$ and with $K=4$ communities using the transition kernel $P$ and the connectivity matrix $Q$ defined by \eqref{tmatrixK5}. Figure~\ref{selec} shows that our approach allows to estimate the correct number of communities $K=4$.

\begin{figure}[!ht]
\vskip 0.2in
\begin{center}
\includegraphics[width=70mm]{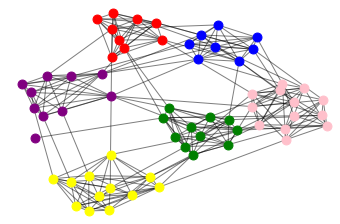}
\includegraphics[width=80mm]{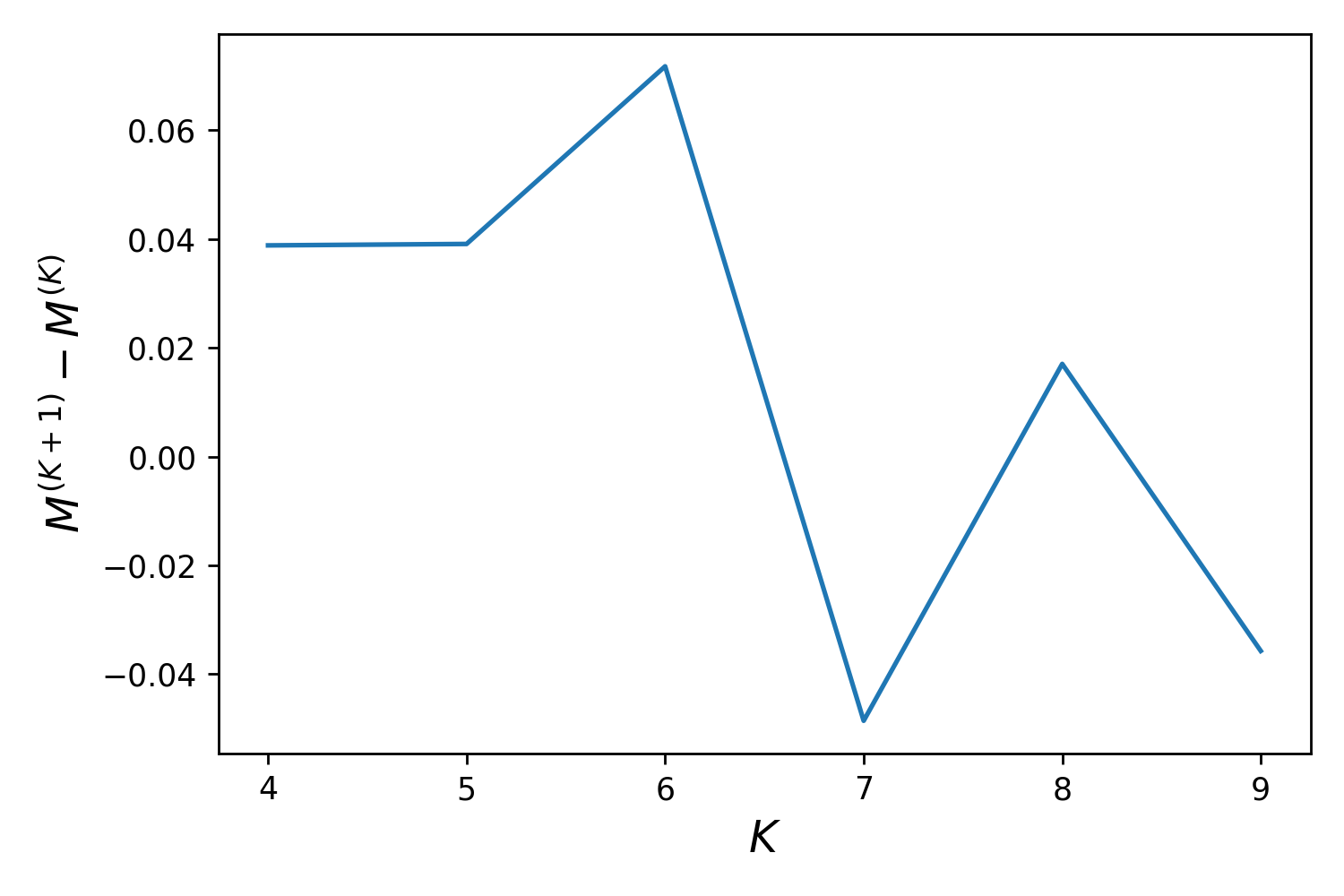}\\
$(a)$ Graph. \hspace{5cm} $(b)$ $K\mapsto M^{(K+1)}-M^{(K)}$.
\caption[Application of our model selection method to the football dataset from the Networkx python package.]{We test our model selection method on the \textit{football network} from the Networkx python package.}
\label{football}
\end{center}
\vskip -0.2in
\end{figure}

\medskip
{\bf Example of application.} 
We also test our procedure on a real network corresponding to American football games between Division IA colleges during regular season Fall $2000$. Two teams are connected if they played against each other. The nodes have values that indicate which conferences the corresponding team belongs to. We worked with $6$ different conferences \footnote{namely Atlantic Coast, Big East, Big Ten, Big Twelve, Conference USA, Mid-American.}. Figure~\ref{football} shows that our procedure infers the correct number of communities.

\medskip

{\bf Related literature.} Standard model selection methods such as the Bayesian Information Criterion (BIC) or
the Akaike Information Criterion (AIC) (cf.\cite{burnham2004multimodel}) are based on
the likelihood of the observed data, which is intractable in the MSBM (and in particular in the SBM). To tackle
this issue, \cite{mariadassou2010uncovering} and~\cite{daudin2008mixture} used the Integrated Completed
Likelihood (ICL) criterion in the SBM. The ICL method is based on an asymptotic approximation of the integrated complete-data
likelihood. The ICL was the first model-based criterion
developed for SBM. Let us mention that other approaches have been explored such as the Integrated Likelihood
Variational Bayes (cf.\cite{latouche2012variational}) where the authors aim at proposing a model selection method that relies on a non-asymptotic approximation of the marginal likelihood.

\medskip

\begin{figure}[!ht]
\vskip 0.2in
\begin{center}
\includegraphics[width=70mm]{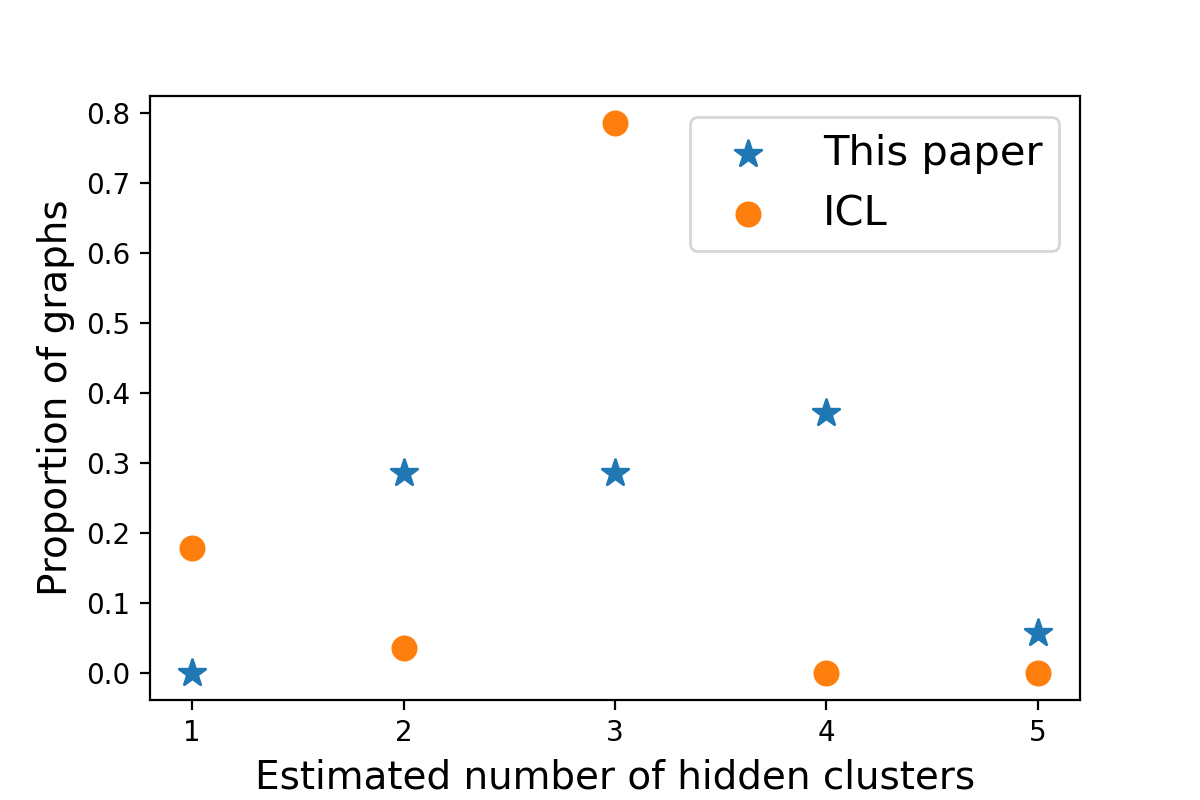}
\includegraphics[width=70mm]{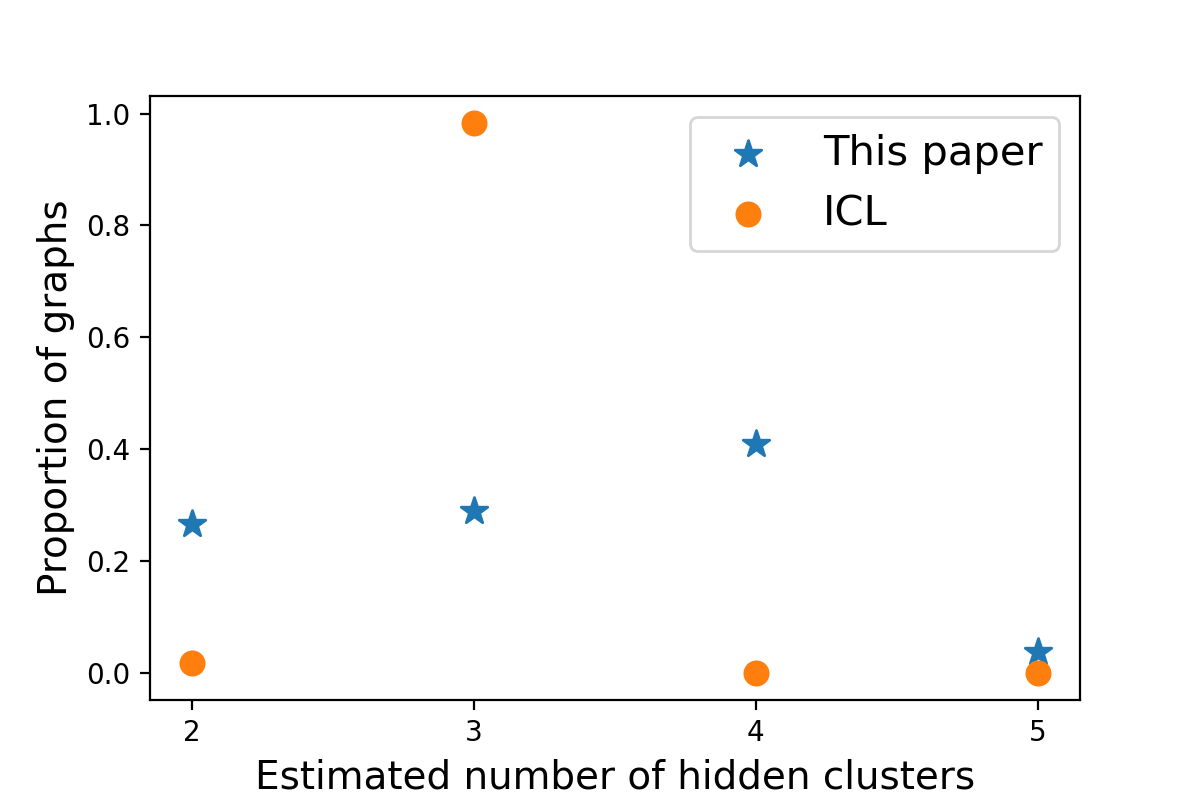}\\
$(a)$ $n=30$. \hspace{5cm} $(b)$ $n=40$.\\
\includegraphics[width=70mm]{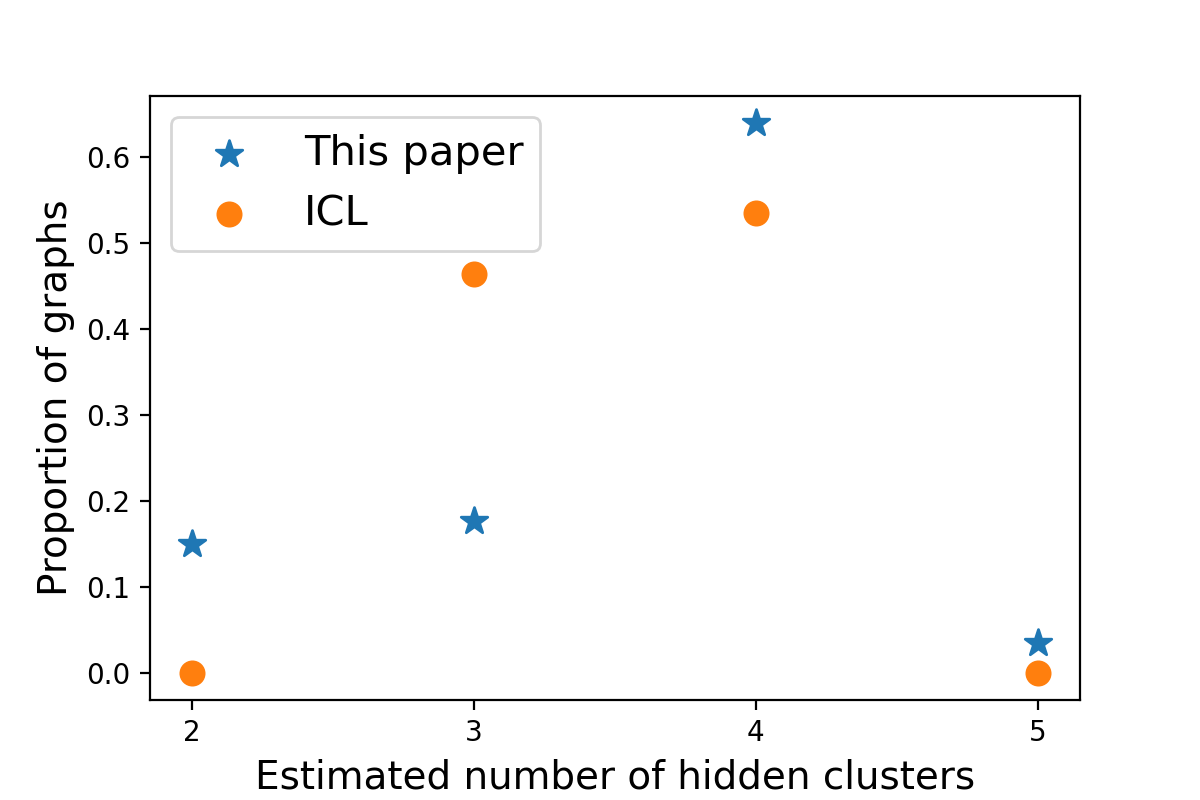}
\includegraphics[width=70mm]{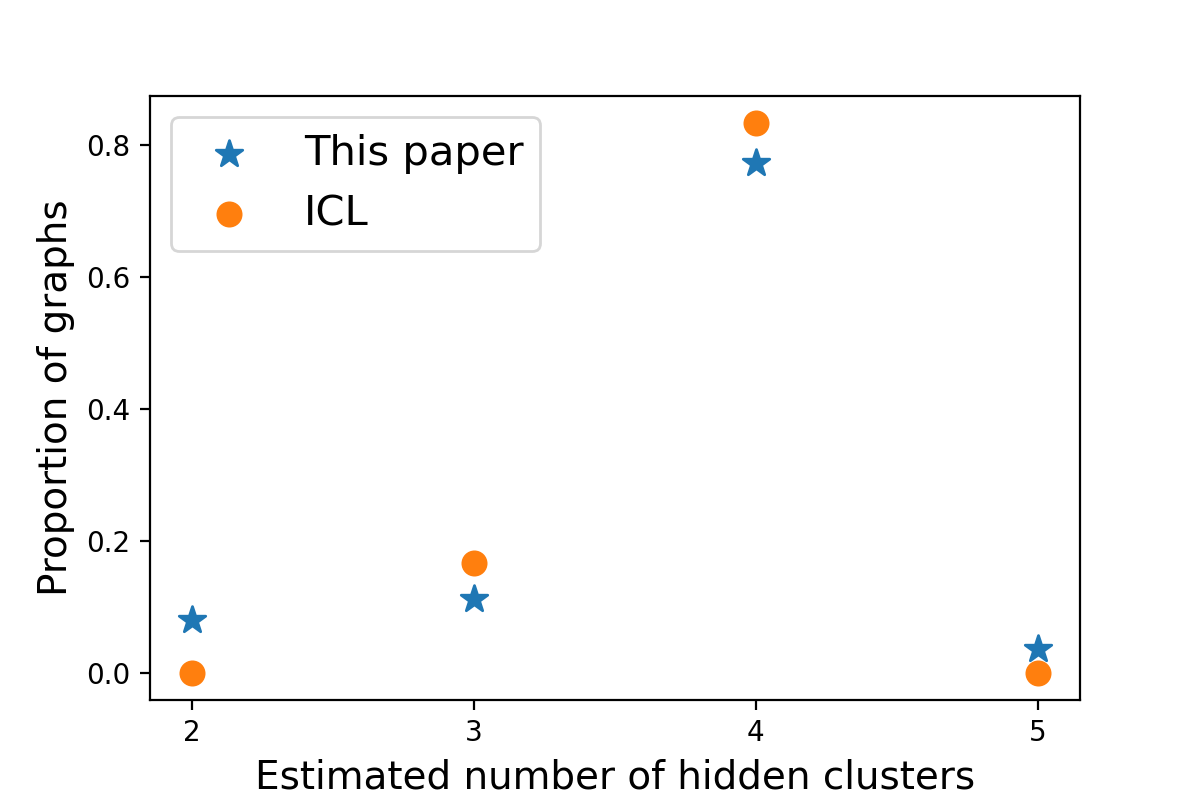}\\
$(c)$ $n=60$. \hspace{5cm} $(d)$ $n=80$.
\caption{Comparison of the model selection method proposed in this paper and the ICL criterion. We show the proportion of graphs leading to a specific estimate for the number of hidden communities. }
\label{modelsel}
\end{center}
\vskip -0.2in
\end{figure}

{\bf Comparison with the ICL criterion.} We consider $K=4$ communities and we sample graphs from the MSBM model using the transition kernel $P$ and the connectivity matrix $Q$ defined by \eqref{tmatrixK5}. For each $n \in \{30,40,60,80\}$, we sample randomly $100$ graphs of size $n$ and we use the estimated number of hidden clusters using our model selection method and using the ICL criterion. For each possible cluster size $K\in \mathbb N$, we compute of the proportion of graphs for which we estimate a number of hidden communities equal to $K$. Figure~\ref{modelsel} shows the result of these experiments. \cite{mariadassou2010uncovering} noticed that the ICL criterion tends to underestimate the number of classes when dealing with small networks and we recover similar results with our experiments as shown in Figure~\ref{modelsel}.$(a)$ and $(b)$. These experiments show that the model selection proposed in this paper should be prefered for small graphs. When the size of the graph is getting larger, the ICL criterion should be preferred to infer the number of hidden communities. We used the Python package {\it sparsebm} to obtain the results when applying the ICL criterion.

\subsection{Application on real data}
\label{sec:appli}
 
Migratory animals are essential components of the ecosystems that support all life on Earth. By acting as pollinators and seed distributors they contribute to ecosystem structure and function. They provide food for other animals and regulate the number of species in ecosystems. Migratory animals are potentially very effective indicators of environmental changes that affect us all. 

In \cite{real-data}, the authors proposed a periodic Markov model on a spatial migration network to formally describe the process of animal migration on the population level. They built their dataset using the Movebank data repository (see \cite{movebank}) that provides historic of animal movements. We propose to test our approach on this dataset. The data is publicly available \href{https://www.datarepository.movebank.org/handle/10255/move.747}{here}\footnotemark  \; and our experiments can be reproduced with the notebook \textit{experiments.ipynb}.

\paragraph{Description of the dataset.} The dataset presents the locations of several white-fronted gooses with timestamps. The animals have been tracked from 2006 to 2010. Each location can be associated with a class using classes defined from Argos User's Manual 2011. We refer to \cite{real-data} for details. \\
We focus on one specific white-fronted goose and we keep the list of its chronological locations between 2006 and 2010 for four location classes. Nodes correspond to the entries of the previous sequence of locations of the animal while communities are the classes associated to each location. In our network, we connect two nodes if the distance between the corresponding precise locations (given with latitude and longitude coordinates) is smaller than some specified threshold. 

\begin{figure}[!ht]
\vskip 0.2in
\begin{center}
\includegraphics[width=80mm]{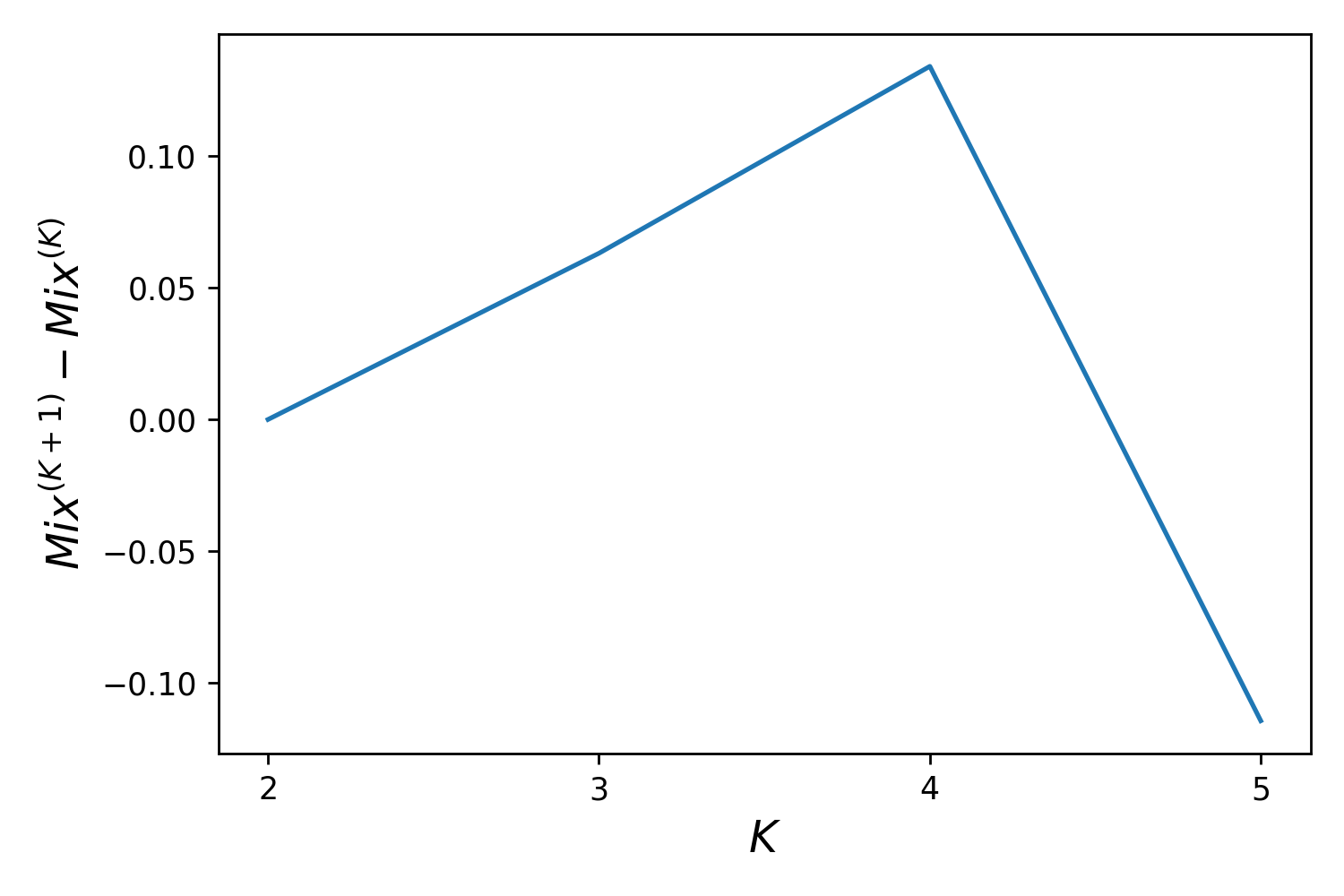}\hfill
\includegraphics[width=80mm]{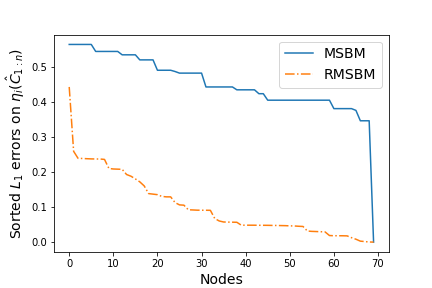}
\end{center}
\hspace{2cm}$(a)$ $K\mapsto M^{(K+1)}-M^{(K)}$. \hspace{2cm} $(b)$ $L^1$ errors between $\hat \eta_i(\hat{\mathbf{C}}_{1:n})$ (resp. $\hat \eta^R_i(\hat{\mathbf{C}}_{1:n})$) and $ \eta_i(\hat{\mathbf{C}}_{1:n})$.
\caption[Application of our methods to the bird migration dataset.]{We test our model on the bird migrations dataset from \cite{real-data}.}
\label{real}
\end{figure}

\paragraph{Results.} With Figure~\ref{real}$.(a)$, we show that the model selection method of Section~\ref{sec:model-selection} allows to retrieve the correct number of clusters on our dataset. In order to evaluate the performance of our reliable link prediction method, we compute the transition matrix $ P$ and the connection matrix $ Q$ associated with our network. More precisely, we define $\forall k,l \in [K],$
\[  Q_{k,l}:= \left\{
    \begin{array}{ll} \displaystyle
        \frac{1}{| G_k|.| G_l|}\sum_{i \in  G_k} \sum_{j \in  G_l} X_{i,j}& \mbox{if } k \neq l \\
        \displaystyle \frac{1}{| G_k|.(| G_k|-1)}\sum_{i,j \in  G_k}  X_{i,j}& \mbox{if } k = l 
    \end{array}
\right. \quad \text{and} \quad  P_{k,l}:=\frac{n}{n-1} \frac{\sum_{i=1}^{n-1} \mathds 1_{( C_i, C_{i+1})=(k,l)}}{\sum_{i=1}^{n}\mathds 1_{ C_i=k}},\]
where $X$ is the adjacency matrix, $ C_i$ is the community of node $i\in[n]$ and $ G_k$ is the set of nodes with label $k\in [K]$. We use these matrices to compute the posterior probabilities $\left(\eta_i(\hat{\mathbf C}_{1:n})\right)_{i\in [n]}$ (see Eq.~\eqref{def:posterior-proba}) and we can compare them with the estimations given by the plug-in approach and the reliable approach from Section~\ref{sec:link-pred}. Figure~\ref{real}$.(b)$ shows that the reliable approach allows to significantly improve the estimate of the posterior probabilities.

\footnotetext{https://www.datarepository.movebank.org/handle/10255/move.747}

\paragraph{Comments.} On simulated data with a small number of clusters, when $n$ gets larger, the
clustering algorithm will recover (almost) perfectly the true partition. In that case, it is clear that the reliable version cannot improve drastically the plug in method since this latter (almost) coincides with the Bayes classifier. However, real
datasets never fit a particular model and recovering the true partition is really unlikely even for very large graphs. In such
cases, our method is of great interest to provide reliable estimations for link prediction despite clustering errors.

\section{Conclusion}

We introduce a growth model for community-based random graphs and we addressed fundamental questions for practical applications in this setting. We propose a general approach to get reliable estimation of the probability of connection between a future node and the nodes already present in the graph. We show how the emission probabilities learned by our method can be used to derive interesting properties on the clustering algorithm and to estimate the number of communities. Our approach is shown to improve the results of the plug-in method for link prediction and collaborative filtering problems. Let us mention that we chose a recent SDP method to derive theoretical guarantees but all our work is directly transferable to other clustering algorithms. \\
Our approach could also be used to study the performance of the algorithm in the presence of spurious links in the graph by looking at the estimated matrix of emission probabilities $\hat O$. This could be an interesting direction of research to choose among the wide variety of clustering algorithms for practical cases where the graph may contain some fake links.

\clearpage

% Acknowledgments---Will not appear in anonymized 
{\bf Acknowledgement}

This work was supported by a grant from Région Ile-de-France.

\bibliography{sample}  

\clearpage

\begin{center}
{\LARGE \bf {Reliable prediction in the Markov Stochastic Block Model \\ \vspace{0.3cm}
Appendix}}
\end{center}

\appendix

\textbf{Guidelines for the Appendix}

\underline{Section~\ref{algo}: Algorithms}\\
In Section~\ref{algo}, we present the SDP algorithm used to recover the communities. We give theoretical elements allowing the reader to understand this approach. 

\underline{Section~\ref{secD}: Experiments}\\
We provide additional experiments. In particular, we present our Markovian hypothesis testing procedure and an application of our methods on real data. We also give details about a possible application on real data for items recommendation or for the analysis of tumor growth.

\underline{Sections~\ref{proofs} and~\ref{gap}: Proofs and Complements}\\
Section~\ref{proofs} contains the proofs of the main statements of our paper while Section~\ref{gap} is a complement recalling basic definitions and properties on Markov chains. 

\underline{Section~\ref{apdx:BW}: The Baum-Welch algorithm with information gap}\\
We provide further details on our reliable prediction methods when information is missing in the observed network.

\underline{Section~\ref{apdx:notations}: Notations}

\section{Clustering Algorithm}
\label{algo}

 \subsection{Partial recovery bound in SBMs with fixed assignment of the communities}
 
In~\cite{Verzelen}, the authors introduce a relaxed version of the $K$-means algorithms on the columns of the adjacency matrix. One specificity of their algorithm is the fact that they are working with the square of the adjacency matrix. This choice allows them to tackle problems outside of the assortative setting and with a wide set of possible connectivity matrices $Q$ contrary to previous works. 
 
 Theorem \ref{thmVerzelen} presents the result of Verzelen and Giraud in the SBM framework with a connectivity matrix $Q=\alpha_n Q_0$.

\begin{thrm}\cite[Theorem 2]{Verzelen}

Assume that $\|Q_0\|_{\infty}\leq L$. The size of the community $k \in [K]$ will be denoted $m_k$. The size of the smallest community will be denoted $m$. We define the signal-to-noise ratio $ s^2 = \Delta^2/(\alpha_nL)$, where $\Delta^2 = \underset{k\neq j}{\min} \; \Delta_{k,j}^2$ with $\Delta_{k,j}^2 = \sum_l m_l (Q_{k,l} - Q_{j,l})^2=\alpha_n^2 \sum_l m_l ((Q_0)_{k,l} - (Q_0)_{j,l})^2.$

Then, there exist three positive constants $c, c', c''$, such that for any $\displaystyle 1/m \leq  \alpha_n L \leq 1/\log(n) $,
$$ \frac{1}{m} \leq \beta \leq  \beta(\alpha_nL) :=
\frac{K^3}{n}e^{4n\alpha_nL}$$
and
$$s^2 \geq c''n/m,$$ 
with probability at least $1 - c/n^2$, 
$$ \mathrm{err}(\hat{G},G) \leq e^{-c's^2}.$$
\smallskip
In particular, since
$$s^2 = \frac{\alpha_n  \min_{k\neq j} \; \sum_{l \in [K]} \; m_l ((Q_0)_{k,l} - (Q_0)_{j,l})^2}{L} \geq \frac{\alpha_n m D^2}{L},$$
we get that with probability at least $1-c/n^2$, $$ -\log \left( \mathrm{err}(\hat{G},G) \right) = \Omega(m\alpha_n).$$
\label{thmVerzelen}
\end{thrm}

\subsection{Presentation of the SDP-based clustering algorithm}

In this Section, we present how we estimate the partition of the nodes $\hat{G}$ when communities are assigned using a Markovian dynamic. Our main result Theorem~\ref{theorem} shows that we are able to achieve \[-\log \; \mathrm{err}(\hat{G},G) = \Omega(n \alpha_n ).\]
Stated otherwise, we get a misclassification error that decays exponentially fast with respect to $n\alpha_n$. We recover the convergence rate recently proved in \cite{Verzelen} in the standard SBM\footnote{See Theorem~\ref{thmVerzelen}.} when the size of the smallest cluster scales linearly with $n$ like in our case. To reach this result, we use the SDP algorithm proposed by Giraud and Verzelen in \cite{Verzelen}. In the following, we expose how the method works.

Suppose the community of each node in the graph has been assigned. In all this subsection, all the communities are considered fixed. We denote $X$ the adjacency matrix of the graph and we refer to Theorem~\ref{thmVerzelen} for the definition of $(m_k)_k$ and $m.$ \cite{Verzelen} are interested in solving optimization problem similar to the following
\begin{align} &\underset{B \in  \mathcal{C}'}{\max} \; \langle X, B \rangle \; \text{ with } \label{SDP-MLE}\\
\mathcal{C}' &:= \{ B: \text{PSD}, \; B_{k,l} \geq 0, \; |B|_1 = \sum_{k} m_k^2 \}, \notag \end{align}
where PSD means that $B$ is positive semidefinite and where $|\cdot|_1$ is the element-wise $l_1$ norm, namely the sum of the absolute values of all entries of a given matrix.

We remind that, dealing with two communities, when the values of the probability matrix $Q$ are a constant~$p$ on the diagonal and another constant $q$ off the diagonal with $p>q$, we are in the assortative case. In the assortative setting, optimization problems like~\eqref{SDP-MLE} have been widely used to recover communities, see~\cite{CX16}, \cite{GV14}, \cite{PW15}, \cite{HWX16}, \cite{FC17}. Those SDP programs are trying to maximize the probability of connection between nodes belonging to the same community. Therefore, they cannot be used directly to solve community detection outside of the assortative framework. \\
\cite{PW07} showed that any partition $G$ of $[n]$ can be uniquely represented by a $n \times n$ matrix $B^* \in \mathbb{R}^{n\times n}$ defined by $\forall i,j \in [n],$
\begin{equation*} B_{i,j}^*=\left\{
    \begin{array}{ll} \displaystyle
        \frac{1}{m_k}& \mbox{if } i \text{ and } j \text{ belong to community } k\\
        0 & \text{ otherwise.}
    \end{array}
\right.\end{equation*}
The set of such matrices $B^*$ that can be built from a particular partition of $[n]$ in $K$ groups is defined by \begin{align*}\mathcal{S}=\{B\in \mathbb{R}^{n \times n} \; : \;  &B^{\top}=B, \; B^2=B, \; \text{Tr}(B) =K,\\
&B\textbf{1} =  \textbf{1}, \;  B\geq0\},\end{align*}
where $\textbf{1} \in \mathbb{R}^n$ is the $n$-dimensional vector with all entries equal to one and where $B \geq 0$ means that all entries of $B$ are nonnegative. \cite{PW07} proved that solving the $K$-means problem
\[\text{Crit}(G) = \sum_{k=1}^K \sum_{i \in G_k} \Bigg\| X_{:,i} - \frac{1}{|G_k|}\sum_{j \in G_k} X_{:,j}  \Bigg\|_2^2  ,\]
is equivalent to \begin{equation} \underset{B \in \mathcal{S}}{\max} \; \langle XX^{\top}, B \rangle. \label{Kmeans}\end{equation} 
Writing $B^*$ an optimal solution of \eqref{Kmeans}, an optimal solution for the $K$-means problem is obtained by gathering indices $i,j \in [n]$ such that $B^*_{i,j}\neq 0$. The set $\mathcal{S}$ is not convex and the authors of \cite{Verzelen} propose the following relaxation of problem \eqref{Kmeans}
\begin{align}\hat{B} \in \underset{B \in \mathcal{C}_{\beta}}{\arg \max} \; \langle XX^{\top},B \rangle &\text{ with } \label{relaxed-SDP}\\
\mathcal{C}_{\beta}:=\{B\in \mathbb{R}^{n \times n} \;: \;&  \text{ symmetric}, \; \text{Tr}(B) =K,\notag\\
& B\textbf{1} =  \textbf{1}, 0 \leq B \leq \beta\},\notag\end{align}

where $K/n \leq \beta \leq 1$. 
The constraint $B\leq \beta$ allows to deal with sparse graphs. Indeed, when $\alpha_n = o(\log(n)/n)$, solving $\eqref{relaxed-SDP}$ without this constraint will produce unbalanced partition.

At this step, we cannot ensure that $\hat{B}$ belongs to $\mathcal{S}$ and a final refinement is necessary to end up with a clustering of the nodes of the graph. This final rounding step is achieved by running a $K$-medoid algorithm on the rows of $\hat{B}$. Given a partition $\{G_1, \dots , G_k\}$ of the $n$ nodes of the graph into $K$ communities, we define the related membership matrix $A \in \mathbb{R}^{n \times K}$ where $A_{i,k} = \mathds 1_{i \in G_k}$. Working on the rows of $\hat{B}$, a $K$-medoid algorithm tries to find efficiently a pair $(\hat{A},\hat{M})$ with $\hat{A} \in \mathcal{A}_K$, $\hat{M} \in \mathbb{R}^{K \times n}$, $\text{Rows}(\hat{M}) \subset \text{Rows}(\hat{B})$ satisfying for some $\rho >0$
\begin{equation}|\hat{A}\hat{M}-\hat{B}|_1 \leq \rho \underset{A \in \mathcal{A}_K, \text{Rows}(M) \subset \text{Rows}(\hat{B})}{\min}\;|AM-\hat{B}|_1,\label{medoid}\end{equation}
where ${A}_K$ is the set of all possible membership matrices and  $\text{Rows}(\hat{B})$ the set of all rows of $\hat{B}$. The $K$-medoids algorithm proposed in \cite{CGTS02} gives in polynomial time a pair $(\hat{A},\hat{M})$ satisfying the inequality \eqref{medoid} with $\rho= 7$. From $\hat{A}$ we are able to define the final partition of the nodes of the graph by setting \[\forall k \in [K], \quad \hat{G}_k=\{i\in [n]\;:\; \hat{A}_{i,k}= 1\}.\]

\textbf{Remark.}

As highlighted in \cite{Verzelen}, the parameter $\beta$ can not be computed since $L$ is unknown. Verzelen and Giraud propose to set $\beta $ to value $\hat{\beta} = \frac{K^3}{n}e^{2nd_X} \wedge 1$, where $d_X$ denotes the density of the graph. We end up with the Algorithm \ref{ALGO} to estimate the communities in the SBM.

\begin{algorithm}[H]
\textbf{Data:} Adjacency matrix $X$ of a graph $G=(V,E)$, Number of communities $K$.\\

	\begin{algorithmic}[1]
 \STATE Compute the density of the graph $d_{X}=\frac{2|E|}{n(n-1)}$ and set $\hat{\beta} = \frac{K^3}{n}e^{2nd_X}\wedge 1$. \;
 \STATE Find $\hat{B} \in  \underset{B \in \mathcal{C}_{\hat{\beta}}}{\arg \max} \; \langle XX^{\top},B \rangle$ (using for example the interior-point method).\;
 \STATE Run the $K$-medoids algorithm from \cite{CGTS02} on the rows of $\hat{B}$. Note $\hat{A} \in \{0,1\}^{n \times K}$ the membership matrix obtained. \;
 \STATE Define $\forall k \in [K], \quad \hat{G}_k = \{ i\in [n] \; : \; \hat{A}_{i,k}=1 \}$ and $\forall i \in [n], \quad \hat{C}_i=k$ where $k \in [K]$ is such that $\hat{A}_{i,k}=1.$
 	\end{algorithmic}
 \caption{Algorithm to estimate the partition of the nodes of the graph.}
 \label{ALGO}
\end{algorithm}

\section{Additional Experiments}

\label{secD}

 \subsection{Markovian Dynamic Testing}
\label{markovian-testing}

We illustrate our model on a toy example with $K=4$ communities, with the transition matrix $P$ and the connectivity matrix $Q$ defined by
\begin{equation}
P = \begin{bmatrix} 0.1 & 0.3 & 0.5 & 0.1 \\ 
0.45 & 0.15 & 0.2 & 0.2 \\ 
0.15 & 0.3 & 0.1 & 0.45 \\ 
0.25 & 0.3 & 0.1  & 0.35\end{bmatrix}\quad \text{and} \quad 
Q =\begin{bmatrix} 0.22 & 0.48 & 0.29 & 0.44  \\ 

0.48 & 0.61 & 0.18 & 0.15 \\ 

0.29 & 0.18 & 0.08 & 0.87 \\ 

0.44 & 0.15 & 0.87 & 0.27    \end{bmatrix} .  \label{tmatrixK5}
\end{equation}

\begin{figure}[!ht]
\centering
\includegraphics[scale=0.51]{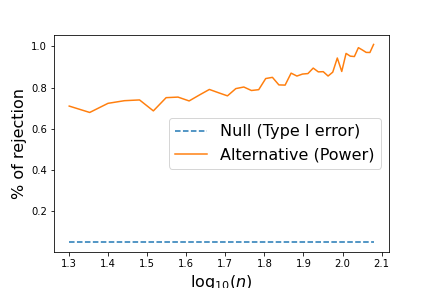}
\caption{Power of our hypothesis test with level $5 \%$. 
}
\label{fig:hypo-testing}
\end{figure}

We propose a hypothesis test to statistically distinguish between an independent assignment of the communities with the distribution $\pi$ and a Markovian assignment with a non-trivial dependence structure. More precisely, we consider the null $\mathbb H_0: $ \textit{communities are independently assigned with distribution $\pi$} where~$\pi$ denotes the stationary distribution of the transition matrix $P$ from \eqref{tmatrixK5}. Our test is based on estimate $\hat P$ of the transition matrix. The null can be rephrased as $\mathbb H_0:\  P= P^0$ where $P^0:=\begin{bmatrix} \pi \\ \vdots \\ \pi \end{bmatrix}$. One can use any {\it black-box goodness-of-fit test} comparing $\hat P$ to $P^0$. Figure~\ref{fig:hypo-testing} shows the power of this hypothesis test with level $5\%$ (Type I error)  and using the $\chi^2$-test described by \cite[Section 2.4]{bickenbach2001markov}. We choose alternative given by the matrices defined in \eqref{tmatrixK5}. Rejection region is calibrated (i.e., threshold of the $\chi^2$-test) by {\it Monte Carlo simulations under the null}. It allows us to control Type I error as depicted by dotted blue line. We run our algorithm to estimate the transition matrix from which we compute the $\chi^2$-test statistic namely
\[S_n:=\sum_{1\leq k,l\leq K}  \hat G_k \frac{\left(\hat P_{k,l}-\pi_l\right)^2}{\pi_l} \text{ with } \hat G_k = \sum_{i=1}^n \mathds 1_{\hat C_i =k}.\] $S_n$ is known to be asymptotically distributed as a $\chi^2$ random variable with $ K(K-1)$ degrees of freedom. Figure~\ref{fig:hypo-testing} shows that for graphs of size larger than $100$, the rejection rate is almost $1$ under the alternative (Type II error is almost zero), the test is very powerful.

\subsection{Experiments with 2 communities}

We test our algorithm on a toy example with $K=2$ communities, $\alpha_n=1$ and with the following matrices:

\begin{equation}
P = \begin{bmatrix} 0.2 & 0.8 \\ 0.6 & 0.4  \end{bmatrix} \text{  and  }Q_0 =\begin{bmatrix} 0.8 & 0.2 \\ 0.1 & 0.3 \end{bmatrix} .  \label{tmatrix}
\end{equation}

\begin{center}
\begin{figure}[!ht]
\centering
\includegraphics[scale=0.6]{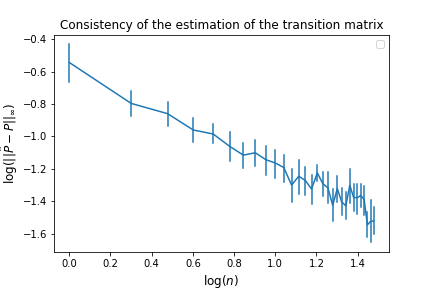}
\caption{We plot the $\log$ of the max norm of the difference between the true transition matrix $P$ and our estimate $\hat{P}$ according to the $\log$ of the number of nodes in the graph.}
\label{fig:K2}
\end{figure}
\end{center}

The Figure \ref{fig:K2} shows the evolution of the max norm of the difference between the true transition matrix $P$ and our estimate $\hat{P}$ when the size of the graph is increasing. For each point, the bar represents the standard deviation of the max norm error computed over thirty randomly generated graphs with the same number of nodes and using the matrices $P$ and $Q$ defined by \eqref{tmatrix}. Those numerical results are consistent with Theorem~\ref{kernelrecovery}: we recover the parametric convergence rate with our estimator of the transition matrix.

\begin{center}
\begin{figure}[!ht]
\centering
\includegraphics[scale=0.6]{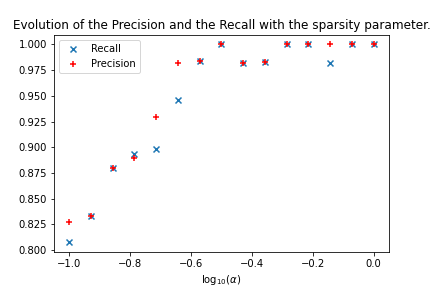}
\caption{Precision and recall of the studied binary classification problem when $K=2$ and $Q= \alpha Q_0.$
}
\label{fig:confusion}
\end{figure}
\end{center}

With Figure \ref{fig:confusion}, we shed light on the influence of the average degree of the nodes on the performance of our algorithm. We plot the recall and the precision of the output of our algorithm with a graph sampled from SBM with a Markovian assignment of the communities using $n=100$ nodes, a transition matrix $P$ defined in \eqref{tmatrix} and a connectivity matrix $Q=\alpha Q_0$ where $Q_0$ is defined in \eqref{tmatrix} and $\alpha$ varies on a log scale between $0.1$ and $1$. We show the recall and the precision with respect to the $\log_{10}$ of the sparsity parameter $\alpha$. We recall that in a binary classification problem, the precision is the ratio between the number of examples labeled $1$ that belong to class $1$ and the number of examples labeled $1$. The recall is the ratio between the number of examples labeled $1$ that belong to class $1$ and the number of examples that belong to class $1$. In our context, those definitions read as
\[\mathrm{precision}=\frac{\sum_{i=1}^n \mathds 1\{\hat{C}_i=1, \; C_i=1\}}{\sum_{i=1}^n \mathds 1\{\hat{C}_i=1\}} \quad \text{  and  } \quad  \mathrm{recall}=\frac{\sum_{i=1}^n \mathds 1\{\hat{C}_i=1, \; C_i=1\}}{\sum_{i=1}^n \mathds 1\{C_i=1\}}.\]

\subsection{Experiments with 5 communities}

\begin{figure}
\vskip 0.2in
\begin{center}
\includegraphics[width=80mm]{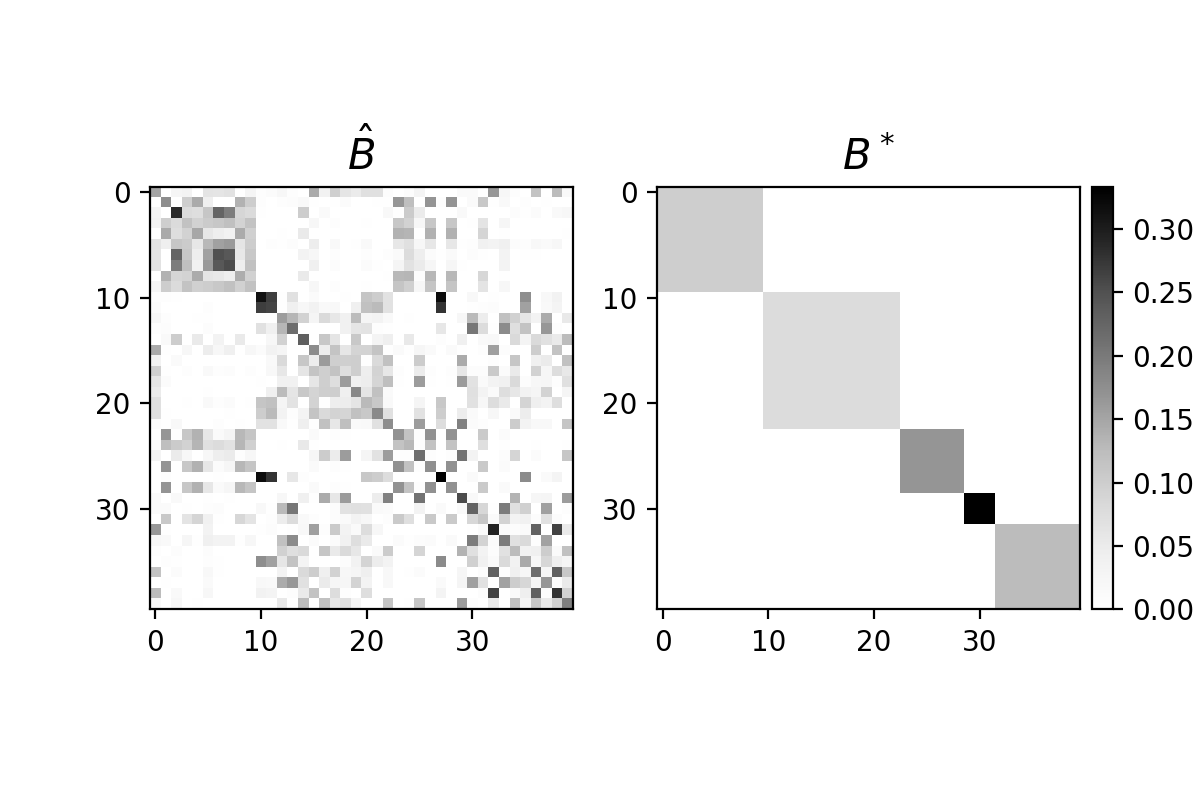}
\includegraphics[width=80mm]{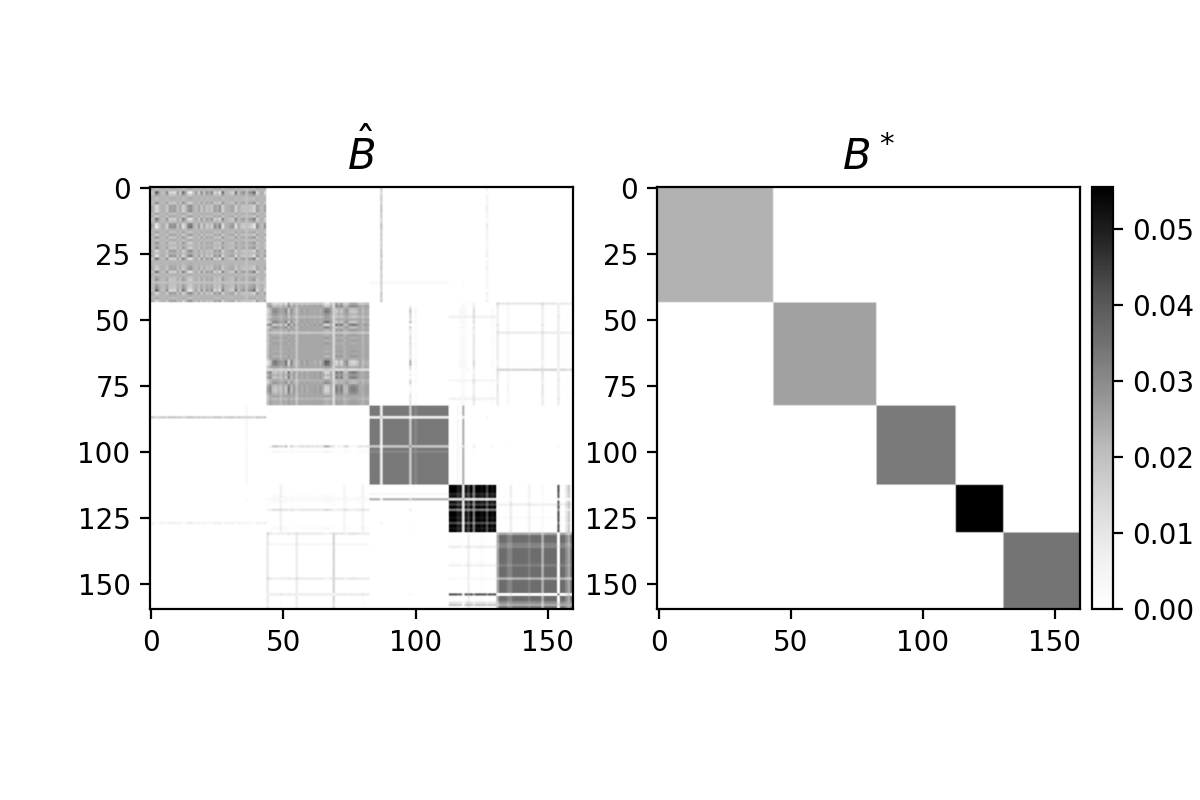}\\
\vspace{-0.8cm}$(a) \; n=40 $ \hspace{6.6cm}$(b) \; n=160$
\caption{Visualization of the matrix $B^*$ solution of Eq.\eqref{SDP-MLE} and its approximation $\hat{B}$ obtained by solving the SDP of Eq.\eqref{relaxed-SDP} for graphs with $40$ nodes (Figure (a)) or $160$ nodes (Figure (b)).}
\label{bmatrices}
\end{center}
\vskip -0.2in
\end{figure}

\begin{center}
\begin{figure}[!ht]
\centering
\includegraphics[scale=0.6]{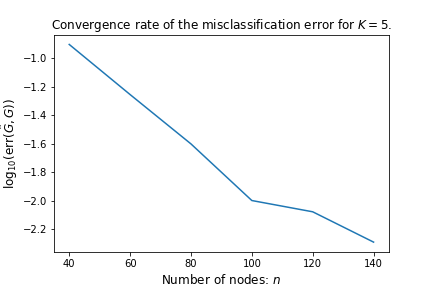}
\caption{$\log$ of the misclassification error as a function of the size of the graphs sampled.}
\label{fig:K5}
\end{figure}
\end{center}

We test our algorithm on a toy example with $K=5$ communities, with the transition matrix $P$ and the connectivity matrix $Q$ defined by
\begin{equation}
P = \begin{bmatrix} 0.1 & 0.3 & 0.5 & 0.01 & 0.09 \\ 

0.55 & 0.15 & 0.1 & 0.05 & 0.15 \\ 

0.15 & 0.3 & 0.1 & 0.2 & 0.25 \\ 

0.15 & 0.05 & 0.1 & 0.5 & 0.2 \\ 

0.2 & 0.3 & 0.1 & 0.05 & 0.35   \end{bmatrix} \text{  and  } Q =\begin{bmatrix} 0.6 & 0.1 & 0.15 & 0.1 & 0.2 \\ 

0.2 & 0.5 & 0.35 & 0.1 & 0.4 \\ 

0.4 & 0.15 & 0.6 & 0.25 & 0.05 \\ 

0.4 & 0.1 & 0.1 & 0.2 & 0.55 \\ 

0.3 & 0.35 & 0.2 & 0.1 & 0.7  \end{bmatrix} .  \label{tmatrixK5-supp}
\end{equation}
We order the nodes of the graph such that the true partition of the nodes is given by $G_1=\{1, \dots, m_1\}$, $G_2=\{m_1+1, \dots, m_1+m_2 \}$, \dots , $G_5=\{\sum_{j=1}^4m_j+1, \dots,n \}$. Figure~\ref{bmatrices} shows the matrix $B^*$ solution of Eq.\eqref{SDP-MLE} and its approximation $\hat{B}$ obtained by solving the SDP of Eq.\eqref{relaxed-SDP}. Thanks to the node ordering, the matrix $B^*$ has a block diagonal structure where each entry of one block is equal to the inverse of the size of the associated cluster. Figure~\ref{bmatrices}.(a) allows us to compare the matrices $B^*$ and $\hat{B}$ when the number of nodes in the graph is equal to $40$ while Figure~\ref{bmatrices}.(b) deals with a graph of size $160.$ 
\\
For a graph sampled with a size equal to $40$, Figure~\ref{bmatrices}.(a) shows us that the SDP algorithm defined in Algorithm 1 is able to capture relevant information about the clustering of the nodes in communities $1, \; 2$ and $5$. However,  we see that using a number of nodes equal to $40$ is not enough to distinguish nodes belonging to community $3$ or $4$. Figure~\ref{bmatrices}.(b) proves that increasing the size of the graph (i.e. for $n=160$) allows to solve this issue. One can easily guess that running a $K$-medoid algorithm on the rows of the matrix $\hat{B}$ plotted in Figure~\ref{bmatrices}.(b) will lead to an accurate clustering of the nodes of the graph. Figure \ref{fig:K5} shows that the $\log$ of the misclassification error decreases linearly with the size of the graph.

\subsection{Overestimation of the size of small clusters using the algorithm of Section \ref{algo}}\label{apdx:overesti}
\begin{figure}[!ht]
\includegraphics[scale=0.6]{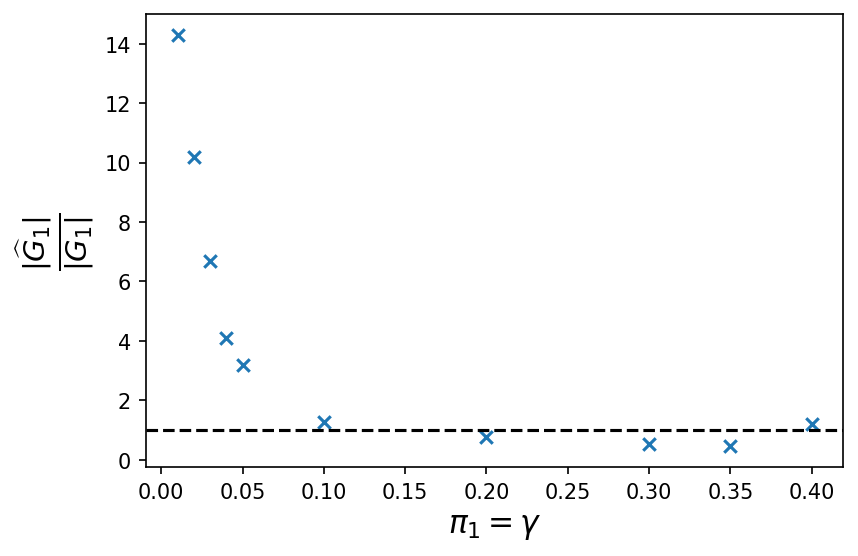}
\caption{Ratio between the size of the estimated first cluster and the true size of the first cluster. We vary the parameter $\gamma$ which is such that $n\gamma$ is the expected value of the size of the first cluster.}
\label{fig:size_first_cluster}
\end{figure}

In Figure~\ref{fig:emission-probas} of the main paper (cf. Section~\ref{num-expe-bw}), we showed the learned emission probabilities $\hat O_{k,l}$, $k,l\in[K]$ and we concluded that the size of the smaller cluster was overestimated by the clustering algorithm from~\cite{Verzelen}. In this section, we conduct additional experiments to check if we reach the same conclusion. We consider two communities and a probability vector $\pi=[\gamma,1-\gamma]$ with $\gamma \in (0,0.5).$ We consider a SBM with $100$ nodes with an independent assignment of the communities according to the vector $\pi$. The connectivity matrix is defined by $\displaystyle Q = \begin{bmatrix} 0.8 & 0.05 \\ 0.05 & 0.8  \end{bmatrix}$. Varying the value of $\gamma$ between $0.01$ and $0.4$, we compute the clustering  of the nodes (denoted $(\hat G_1,\hat G_2)$) obtained with the SDP algorithm from~\cite{Verzelen}. We compute the ratio between the size of the estimated first cluster (namely $|\hat G_1|$) and the size of the true first cluster (namely $|G_1|$). Figure~\ref{fig:size_first_cluster} gives the result we obtain. The horizontal dotted black line corresponds to the value $1$. Hence, crosses below that line show cases where the size of the first cluster is underestimated while crosses above that line show cases where the size of the first cluster is overestimated. These experiments go in the same direction as the one from Section~\ref{num-expe-bw}: the size of small clusters tend to be overestimated by the SDP algorithm from~\cite{Verzelen}.

\subsection{Other potential application on real data: The example of recommendation system}
\label{sec:recommendation}

In this section, we give more details on another possible application of our model for recommendation system as mentioned in the introduction. Let us remind the framework of our example. We suppose that we have access to the online purchases of different customers. For each of them, we know the dates and the product IDs of each of their purchases. Our goal is threefold: $i)$ learn the category of product sold by each url $ii)$ learn the purchasing behavior of each customer $iii)$ use this information to suggest relevant new products to each customer. For each customer $U$, we have a network where nodes are product IDs (ordered by timestamp of purchase). We connect two products $i$ and $j$ if the ratio $ W_{i,j}/ \sqrt{w_i. w_j}$ is larger than some threshold $\tau \in (0,1)$, where $W_{i,j}$ is the number of clients in the dataset who bought both the products $i$ and $j$, and $w_i$ is the number of clients who bought the product $i$. We can proceed as follows.
\begin{enumerate}
\item Running our algorithm, we can infer the number of different categories of products bought by $U$ using our heuristic from Section~\ref{sec:model-selection}.
\item Then, we can learn both the category of each product and the transition matrix $\hat P$ which gives the  purchasing behavior of $U$. 
\item To recommend a new product to the client $U$, one can use the purchasing behaviour of the client $V$ who shares the largest number of common purchased products with $U$. \end{enumerate}

Let us finally mention that the connection probabilities learned from client $V$ for categories unseen so far by $U$ could be used as an initialization of a stochastic bandit algorithm for recommendation of products for $U$.

\subsection{Other potential application on real data: The example of the tumor growth}
\label{sec:tumor}

To better characterize tumour heterogeneity and predict the
potential response of the constituent parts of a tumour to drugs, biologists often rely on single-cell RNA sequencing data (cf. \cite{costello2014community}). RNA sequencing allows the measurement of gene expression in thousands of cells in a single experiment. To identify the cell types and cell states present in a sample, unsupervised clustering are often used (cf. \cite{levine2015data}).

During the cell cycle, a cell increases in size, replicates its DNA and splits into daughter cells. Hence, a given sample of tumor is structured according to an underlying evolutionary history of cancer cells. It has been reported that incorporating the pathway information (i.e. the parent-child hierarchical structure) is key to reach better drug prediction (cf. \cite{yang2018linking}).

\medskip

In this context, an interesting application of the MSBM model would be to follow the ex-vivo evolution of a cell. Namely, starting from a specific cell, a biologist could choose a daughter of this cell, and a daughter of this one, etc. Stated otherwise, the biologist will follow a specific path in the genealogical tree where the root is the starting cell. Performing the RNA sequencing of each cell along the path, he could build a graph as follows:
\begin{itemize}
\item each node is a cell in the path selected in the genealogical tree,
\item the order of the nodes is simply given by the timestep of appearance of the cells,
\item the RNA sequencing gives for all cell $i$ a vector $x^{(i)}\in \mathbb R^d$ where each entry $x^{(i)}_s$ corresponds to the expression level of gene $s$ in cell $i$. One can then build a graph by simply putting an edge between cell $i$ and cell $j$ with probability equal to $\frac12\big(\mathrm{cor}(x^{(i)},x^{(j)})+1\big)$ where $\mathrm{cor}(x^{(i)},x^{(j)})$ is the correlation between vector $x^{(i)}$ and $x^{(j)}$. 
\end{itemize}
\medskip

{\bf Related literature.} \par 
Let us mention that this approach to tackle the problem of cell clustering has already been proposed in the literature such as in~\cite{Huizing2021} where the authors build a cell-cell similarity metric (based on the optimal transport cost between the probability distributions given by $\frac{x^{(i)}}{\|x^{(i)}\|_2}$ for $i \in [n]$). They use a spectral method to infer clusters considering a weighted graph whose adjacency matrix is given by $A=1-D$ where $D$ is a normalized version of the matrix gathering the optimal transport costs between the pairs of cells previously computed (the normalization simply ensures that the maximum entry of $D$ is $1$, so that weights in the graph are all non negative).
\medskip

{\bf Use of the MSBM in this context.} \par 
The biologist could:
\begin{itemize}
\item Use our model selection method from Section~\ref{sec:model-selection} to infer the number of hidden communities $K$ in the graph.
\item Run a clustering algorithm on the graph to obtain $K$ groups of cells.
\item Estimate the Markov kernel $P$ to understand the evolution process of cancer cells and answering questions such that: "What is the probability that the daughter of a cancer cell of type A is of type A, B, C, ...?"
\item By estimating the probability of connection between any node $i\in [n]$ and the upcoming node $n+1$ as explained in Section~\ref{sec:baumwelch}, the biologist would get the correlation between the gene expression profile of cell $i$ and the future cell $n+1$. 
\end{itemize}

\section{Proofs}
\label{proofs}

\subsection{Proof of Theorem~\ref{theorem}}

Lemma \ref{lemma:1} provides a more complete version of Theorem~\ref{theorem} by giving explicitly the constants.

\begin{lmm}
\label{lemma:1} Let us consider the three positive constants $c,\; c'$ and $c''$ involved in Theorem \ref{thmVerzelen}.

Assume that $\alpha_n \log(n)\leq 1/L$ and that $\displaystyle n \alpha_n > \max\left( \frac{4Lc''}{ \pi_m^2D^2}\;,\;  \frac{2}{ L\pi_m}\right)$. Then it holds

\begin{align*}
\mathbb{P}\left( \mathrm{err}(\hat{G},G) >  \exp\left(-\frac{c'S^2}{2}\right) \right) &\leq \frac{c}{n^2}  +  2K   \exp \left( - \frac{n \pi_m^2}{2A_1 + 4A_2 \pi_m}\right) ,
\end{align*}

where $S^2 = \frac{n \alpha_n\pi_m D^2 }{L}$ and where $ A_1$ and $A_2$ are constants that only depend on the Markov chain $(C_i)_{i \geq 1}$ with $ \displaystyle A_1:= \frac{1+ (\lambda_+ \vee 0)}{1- (\lambda_+ \vee 0)}$ and $ \displaystyle A_2 :=  \frac13 \mathds 1_{\lambda_+ \leq 0} + \frac{5}{1-\lambda_+}\mathds 1_{\lambda_+ > 0}$. Here $1-\lambda_+$ is the right $L^2$ spectral gap of the Markov chain $(C_i)_{i \geq 1}$ (see Definition \ref{def2} in Section \ref{gap}).
\end{lmm}

\textbf{Remarks.}

\begin{itemize}

\item The fact that $\pi_m>0$ is a direct consequence of the positive recurrent property of the Markov chain.
\item The second term in the right hand side of the inequality from Lemma~\ref{lemma:1} comes from the concentration of the average number of visits of the Markov chain towards the stationary distribution of the chain.
The first term in this inequality corresponds to the 
bound from Theorem \ref{thmVerzelen} when communities have been assigned.
\\
Recalling that $\|Q\|_{\infty}$ is upper bounded by $\alpha_n L$, the condition $\alpha_n \log(n)\leq 1/L$ enforces the signal to noise ratio defined by Giraud and Verzelen $s^2 := \Delta^2/(\alpha_nL)$ (see Theorem \ref{thmVerzelen}) to be larger than $\Delta^2 . \log(n)$. Another way to interpret this condition is to say that it enforces the expected degree of all nodes of the graph to be smaller than $n/ \log(n)$. 

\item In order to get some intuition on the conditions on $n$ in the previous theorem, keep in mind that asymptotically, the size of the smallest community in the graph will be $n  \pi_m$. 

\begin{itemize}
\item The condition $\displaystyle n>  \frac{4Lc''}{\alpha_n\pi_m^2D^2}$ can be read as $(n  \pi_m) \alpha_n D^2/L >\frac{4c''}{\pi_m} =4c'' . \frac{n}{n \pi_m} $. Asymptotically,  $(n \pi_m) \alpha_n D^2/L$ provides a lower bound on the signal-to-noise ratio defined in Theorem \ref{thmVerzelen}. This shows that the condition $\displaystyle n>  \frac{4Lc''}{\alpha_n\pi_m^2D^2}$ is related to the constraint $s^2 \gtrsim n/m$ of Theorem \ref{thmVerzelen}.
\item The condition $\displaystyle n>  \frac{2}{\alpha_n L\pi_m}$ can be read as $\frac{1}{n  \pi_m} < \alpha_n L/2 $. This shows that the condition $\displaystyle n> \frac{2}{\alpha_n L\pi_m}$ is related to the constraint $1/m < \alpha_n L$ from Theorem \ref{thmVerzelen}.
\end{itemize}
\end{itemize}

The proof of Lemma \ref{lemma:1} is based on the following Lemma which is proved at the end of this subsection.

\begin{lmm}
\label{lemma:1.2}
We consider $c,c'$ and $c''$ the three numerical constants involved in Theorem \ref{thmVerzelen}.
\medskip

Let us consider $0<t<\pi_m$. Assume that $\alpha_n L \leq 1/\log(n)$. Then for any $\epsilon > 0$ and $n$ large enough such that: 

\begin{center}
\renewcommand{\arraystretch}{2}
$\begin{array}{rcl}
n  (\pi_m  -t) \geq  
\left\{
    \begin{array}{ll}
        \frac{ L \log(1/\epsilon)}{c'\alpha_n D^2}  & (i) \\
         \left(  \frac{c'' nL}{\alpha_n D^2}  \right)^{1/2}  & (ii) \\ 
         1/(\alpha_n L) & (iii) \\
    \end{array}
\right.
\end{array}$
\end{center}

it holds 
\begin{align*}
&\mathbb{P}\left( \rm{err}(\hat{G},G) >\epsilon  \right) \leq \frac{c}{n^2}  +  2K   \exp \left( - \frac{n t^2}{2(A_1/4 + A_2 t)} \right)  ,
\end{align*}

where $A_1$ and $A_2$ are constants defined in Theorem~\ref{theorem}.

 \label{lemmaVerzelen}

\end{lmm}

Note that the only constraint on $\epsilon$ is given by the condition $(i)$ which is equivalent to 
$$\epsilon \geq \exp \left(  - \frac{c' D^2 n \alpha_n (\pi_m-t)}{L}  \right)  .$$

In order to get the tighter result possible, we want to choose $ \epsilon =\exp \left(  - \frac{c' D^2 n \alpha_n (\pi_m-t)}{L}  \right)$ which leads to  

 $$ t= \pi_m- \frac{ L \log(1/\epsilon)}{c' D^2n \alpha_n }.$$

 The condition $t>0 $ is then equivalent to $$\displaystyle \pi_m > \frac{L \log(1/\epsilon)}{c' D^2n \alpha_n }  \Leftrightarrow \exp(-\pi_mn \alpha_n c'D^2/L) < \epsilon.$$

 The condition $(ii) $ is equivalent to $$\displaystyle n(\pi_m-t) = \frac{L\log(1/\epsilon)}{c' \alpha_n D^2}\geq\left(  \frac{c'' nL}{\alpha_n D^2}  \right)^{1/2} \Leftrightarrow \exp\left(- c' \sqrt{\frac{D^2c''n \alpha_n }{L}}\right) \geq \epsilon.$$
 
  The condition $(iii) $ is equivalent to $$\displaystyle n(\pi_m-t) = \frac{L\log(1/\epsilon)}{c'\alpha_n D^2}\geq (1/ \alpha_n L) \Leftrightarrow \exp\left(-\frac{c'D^2}{L^2}\right)\geq \epsilon.$$
 
 One can easily prove that for $ n \alpha_n>\max\left( \frac{4Lc''}{\pi_m^2D^2} \; ,\; \frac{2}{L\pi_m}\right)$,  $ \epsilon:= \exp\left(-\frac{\pi_mn\alpha_nc'D^2}{2L}\right) $ satisfies the three conditions above. This gives Lemma \ref{lemma:1} from Lemma~\ref{lemmaVerzelen}.

\begin{proof}[Proof of Lemma \ref{lemmaVerzelen}.]

Using Theorem~\ref{bernstein-markov} (cf. Appendix~\ref{gap} or \cite[Theorem 2]{Jiang}), we get that

\begin{equation}\forall c \in [K], \; \forall t>0, \; \; \mathbb{P}\left( \left|\frac{1}{n} \sum_{i=1}^n \mathds 1_{C_i=c} - \pi(c) \right| \geq t \right) \leq 2\exp \left( - \frac{n t^2}{2(A_1\sigma_c^2 + A_2 t)}   \right) \label{scalelinearly}\end{equation}

where $ \displaystyle A_1= \frac{1+ (\lambda_+ \vee 0)}{1- (\lambda_+ \vee 0)}$, $ \displaystyle A_2 =  \frac13 \mathds 1_{\lambda_+ \leq 0} + \frac{5}{1-\lambda_+}\mathds 1_{\lambda_+ > 0}$ and $\sigma^2_c = \pi(c)(1-\pi(c)). $

We deduce that for all $t >0$,

\begin{align*}
&\mathbb{P}\left( \bigcup_c  \left\{ \left| \frac{1}{n} \sum_{i=1}^n \mathds 1_{C_i=c} - \pi(c) \right|\geq  t \right\} \right) \leq  2K   \exp \left( - \frac{n t^2}{2(A_1\sigma^2 + A_2 t)} \right),
\end{align*}

where $\sigma^2 := \underset{c}{\max} \; \sigma^2_c \; \;( \leq 1/4).$ We define $\Omega^c := \bigcup_c  \left\{ \left|\frac{1}{n} \sum_{i=1}^n \mathds 1_{C_i=c} - \pi(c)\right| \geq  t \right\}$ and we recall $\pi_m =\underset{c}{\min} \; \pi(c) $ and $D^2= \underset{j \neq k}{\min} \sum_l ((Q_0)_{k,l}-(Q_0)_{j,l})^2$.

Suppose that $0<t<\pi_m$ and that $n$ is large enough to satisfy $(i)$, $(ii)$ and $(iii)$. Then it holds

\begin{align}
&\mathbb{P}\left( \text{err}(\hat{G},G) >\epsilon  \right) \nonumber\\
= \quad& \mathbb{P}\left( \{ \text{err}(\hat{G},G) > \epsilon \} \cap \Omega   \right)  + \mathbb{P}\left( \{ \text{err}(\hat{G},G) > \epsilon \} \cap \Omega^c   \right) \nonumber \\
  \leq \quad & \mathbb{P}\left(  \{ \text{err}(\hat{G},G) > \epsilon \} \cap \Omega  \right) +  2K   \exp \left( - \frac{n t^2}{2(A_1\sigma^2 + A_2 t)} \right) \nonumber   \\
=  \quad& \mathbb{P}\left(  \text{err}(\hat{G},G) > \epsilon \; |\; \Omega  \right) \times  \mathbb{P}\left( \Omega \right)  +  2K   \exp \left( - \frac{n t^2}{2(A_1\sigma^2 + A_2 t)} \right)  . \; \label{lem1}
\end{align}

We denote by $M$ the random variable that gives the size of the smallest cluster: $M:= \min_{k \in [K]} \; m_k$. Condition $(i)$ is equivalent to $$  \epsilon \geq \exp \left(  - c' \frac{n\alpha_n(\pi_m-t)D^2}{L} \right)  .$$

Since on the event $\Omega$ we have $n(\pi_m-t) \leq M$, we get that on $\Omega$ it holds   \begin{equation}\label{eq:epsi}\epsilon \geq \exp \left(  - c' \frac{M\alpha_nD^2}{L} \right) \geq \exp \left(  - c' s^2 \right),\end{equation}

where $s^2 = \Delta^2/(\alpha_nL)$ with $\Delta^2 = \underset{k\neq j}{\min} \; \Delta_{k,j}^2$ and $\Delta_{k,j}^2 = \sum_l m_l (Q_{k,l} - Q_{j,l})^2 .$ The last inequality comes from the fact that $\Delta^2 \geq M\alpha_n^2D^2.$ Using \eqref{lem1} we get that

\begin{align*}
\mathbb{P}\left( \text{err}(\hat{G},G) >\epsilon  \right) &\leq \mathbb{P}\left(   \text{err}(\hat{G},G) > \epsilon \; | \;\Omega  \right)   +  2K   \exp \left( - \frac{n t^2}{2(A_1\sigma^2 + A_2 t)} \right)  \\
& \leq \mathbb{P}\left(   \text{err}(\hat{G},G) > e^{-c's^2} \; |\; \Omega  \right)   +  2K   \exp \left( - \frac{n t^2}{2(A_1\sigma^2 + A_2 t)} \right)  .
\end{align*}

We note that on $\Omega$ :

\begin{itemize}
\item Condition $(ii)$ gives $$M^2 \geq  \frac{c'' nL}{\alpha_nD^2}   \; \Leftrightarrow  \frac{M\alpha_nD^2}{L} \geq c'' n/M ,$$

which implies that $s^2 = \frac{\Delta^2}{\alpha_nL} \geq c'' n/M$ since $\Delta^2 \geq M\alpha_n^2D^2.$

\item Condition $(iii)$ gives $$\frac{1}{M} \leq \alpha_nL.$$
\end{itemize}

Applying Theorem \ref{thmVerzelen} from Verzelen and Giraud, we get that

$$ \mathbb{P}\left(   \text{err}(\hat{G},G) > e^{-c's^2}  | \Omega  \right) \leq \frac{c}{n^2} .$$

Finally we obtain using Eq.\eqref{eq:epsi} that
\begin{align*}
&\mathbb{P}\left( \text{err}(\hat{G},G) >\epsilon  \right) \leq \frac{c}{n^2}  +  2K   \exp \left( - \frac{n t^2}{2(A_1\sigma^2 + A_2 t)} \right)  .
\end{align*}
\end{proof}

\subsection{Proof of Theorem~\ref{Qrecovery}}
\label{sec:Qrecovery}

We start by proving Lemma \ref{lemma:4} which enriches the statement of Theorem~\ref{Qrecovery} by giving explicitly the constants. 

\begin{lmm}
We consider $c,c'$ and $c''$ the three numerical constants involved in Theorem \ref{thmVerzelen}.

Assume that $\alpha_n \log(n)\leq \frac{1}{L}$ and that $\displaystyle n\alpha_n> \max\left( \frac{4Lc''}{\pi_m^2D^2}\;,\;  \frac{2}{L\pi_m} \; , \; \frac{2L\log(n)}{\pi_m c' D^2}\right)$. Then for all $0<t<\pi_m-\frac{1}{n}$, it holds
\begin{align*}
&\mathbb{P}\left( \|\hat{Q}-Q\|_{\infty} > t \right) \\
\leq \quad & K(K+1) \exp \left( - \frac{(n \pi_m-nt-1)^2 t^2}{\frac12+\frac23 t}   \right) +  \frac{c}{n^2}+ 2K\exp \left( - \frac{n t^2}{2(A_1/4+ A_2 t)}   \right) .
\end{align*}
\label{lemma:4}
\end{lmm}
\begin{proof}[Proof of Lemma \ref{lemma:4}.]\mbox{}

\begin{itemize}
\item \underline{Preliminary 1}

Using the standard Bernstein's inequality for independent random variables, we get that for all $ k,l \in [K]^2 \text{ with } k\neq l,$ and for all $t>0, $ it holds

$$ \mathbb{P}\left( \left|\frac{1}{|G_k|.|G_l|} \sum_{i \in G_k}\sum_{j \in G_l} X_{i,j} - Q_{k,l} \right| \geq t \right) \leq 2\exp \left( - \frac{|G_k|.|G_l| t^2}{2(Q_{k,l}(1-Q_{k,l})+t/3)}   \right)$$

and for all $ k \in [K],  t>0, $ it holds
\begin{align*} &\mathbb{P}\left( \left|\frac{1}{|G_k|.(|G_k|-1)} \sum_{i,j \in G_k \; i \neq j}X_{i,j} - Q_{k,k} \right| \geq t \right)\leq 2\exp \left( - \frac{|G_k|.(|G_k|-1) t^2}{2(Q_{k,k}(1-Q_{k,k})+t/3)}   \right).\end{align*}

\medskip
\item \underline{Preliminary 2}

We define the event $  N:=\left\{\text{err}(\hat{G},G)<\exp\left(-\frac{\pi_mn \alpha_n c'D^2}{2L}\right)\right\}$.  Note that on $N$, the partition of the clusters is correctly recovered thanks to the condition $ n \alpha_n > \frac{2L \log(n)}{\pi_mc'D^2}$.

\medskip

\item \underline{Preliminary 3}

Using Theorem~\ref{bernstein-markov} (cf. Appendix~\ref{gap} or \cite[Theorem 2]{Jiang}), we get that

$$\forall c \in [K], \; \forall t>0, \; \; \mathbb{P}\left( \left|\frac{1}{n} \sum_{i=1}^n \mathds 1_{C_i=c} - \pi(c) \right| \geq t \right) \leq 2\exp \left( - \frac{n t^2}{2(A_1/4+ A_2 t)}   \right).$$

We deduce that for all $t >0$,

\begin{align*}
&\mathbb{P}\left( \bigcup_c  \left\{ \left| \frac{1}{n} \sum_{i=1}^n \mathds 1_{C_i=c} - \pi(c) \right|\geq  t \right\} \right) \leq  2K   \exp \left( - \frac{n t^2}{2(A_1/4 + A_2 t)} \right).
\end{align*}

We define $\Omega^c := \bigcup_{c \in [K]}  \left\{ \left|\frac{1}{n} \sum_{i=1}^n \mathds 1_{C_i=c} - \pi(c)\right| \geq  t \right\}$.
\end{itemize}

Let us define for any $k,l\in[K]^2$, 
\[\widetilde Q_{k,l} =   \frac{1}{|G_k|.|G_l|} \sum_{i\in G_k} \sum_{j \in G_l} X_{i,j}.\]
Then, considering $0<t< \pi_m-\frac{1}{n}$, we have 
\begingroup
\allowdisplaybreaks
\begin{align*}
&\mathbb{P}\left( \| \hat{Q}-Q \|_{\infty} > t \right)\\
\leq \quad & \mathbb{P}\left( \underset{k,l \in [K]^2, \; k\leq l}{\bigcup} \{ |\hat{Q}_{k,l}-Q_{k,l} | > t \}  \; | \; \Omega \right) +  \mathbb{P}(\Omega^c)\\
= \quad & \mathbb{P}\left(\left[ \underset{k,l \in [K]^2, \; k\leq l}{\bigcup} \{ |\hat{Q}_{k,l}-Q_{k,l} | > t \} \right] \cap N  \; | \; \Omega \right) +  \mathbb{P}\left(\left[ \underset{k,l \in [K]^2, \; k\leq l}{\bigcup} \{ |\hat{Q}_{k,l}-Q_{k,l} | > t \} \right] \cap N^c  \; | \; \Omega \right)+\mathbb{P}(\Omega^c)\\
\leq \quad & \mathbb{P}\left( \underset{k,l \in [K]^2, \; k\leq l}{\bigcup} \left(\{ |\hat{Q}_{k,l}-Q_{k,l} | > t \} \cap N \right)  \; | \; \Omega \right) +  \mathbb{P}\left(N^c  \; | \; \Omega \right)+\mathbb{P}(\Omega^c)\\
\leq \quad & \sum_{k,l\in [K]^2, k\leq l}\mathbb{P}\left( \{ |\hat{Q}_{k,l}-Q_{k,l} | > t \} \cap N \; | \; \Omega \right)+ \mathbb{P}\left(N^c  \; | \; \Omega \right)+\mathbb{P}(\Omega^c)\\
 = \quad & \sum_{k,l\in [K]^2, k\leq l}\mathbb{E}\left( \mathds 1_{\{ |\hat{Q}_{k,l}-Q_{k,l} | > t \}} \mathds 1_{N}  \; | \; \Omega \right) +  \mathbb{P}(N^c \; |\;\Omega) +\mathbb{P}(\Omega^c).
\end{align*}
\endgroup
At this point, one can notice that it holds for any $k,l\in[K]^2$, 
\begin{equation}\label{proof:eq-review}\mathds 1_{\{ |\widehat{Q}_{k,l}-Q_{k,l} | > t \}}\mathds 1_{N}=\mathds 1_{\{ |\widetilde{Q}_{k,l}-Q_{k,l} | > t \}}\mathds 1_{N}.\end{equation} 
Indeed,
\begin{itemize}
\item either it holds $\mathds 1_N=0$ in which case the equality from Eq.\eqref{proof:eq-review} is trivially true,
\item or $\mathds 1_N=1$ (meaning that the event $N$ holds) which implies that $\hat G_s=G_s$ for all $s\in[K]$ (see Preliminary 2), leading to $\hat Q_{k,l} = \widetilde Q_{k,l}$.  
\end{itemize}
Hence for any $0<t< \pi_m-\frac{1}{n}$ we have,
\begingroup
\allowdisplaybreaks
\begin{align*}
&\mathbb{P}\left( \| \hat{Q}-Q \|_{\infty} > t \right)\\
\leq \quad & \sum_{k,l\in [K]^2, k\leq l}\mathbb{E}\left( \mathds 1_{\{ |\widetilde{Q}_{k,l}-Q_{k,l} | > t \}} \mathds 1_{N}  \; | \; \Omega \right) + \mathbb{P}(N^c \; |\;\Omega) +\mathbb{P}(\Omega^c)\\
\quad& \text{and using preliminary 3,}\\
\leq \quad & \sum_{k,l\in [K]^2, k\leq l} \mathbb{P}\left(  \{ |\widetilde{Q}_{k,l}-Q_{k,l} | > t \}  \; | \; \Omega \right) +  \mathbb{P}(N^c \; |\;\Omega) \\
\quad & + 2K\exp \left( - \frac{n t^2}{2(A_1/4 + A_2 t)}   \right)\\
\leq \quad & 2 \sum_{1\leq k \leq l \leq K}\exp \left( - \frac{n( \pi(k)-t).(n \pi(l)-nt-1) t^2}{2(Q_{k,l}(1-Q_{k,l})+t/3)}   \right) +  \frac{c}{n^2}+ \\ \quad & 2K\exp \left( - \frac{n t^2}{2(A_1/4 + A_2 t)}   \right) ,
\end{align*}
\endgroup
where we used that $\displaystyle \mathbb{P}(N^c \; |\;\Omega)\leq \frac{c}{n^2}$ (shown in the proof of Theorem~\ref{theorem}).

\end{proof}

\medskip
\begin{proof}[Proof of Theorem~\ref{Qrecovery}.]
Let us consider $\gamma >0$ and let us define $t=\frac{\gamma}{\sqrt{n}}$. Considering that $$\frac{n\alpha_n}{\log(n)} \geq a  \quad \text{ with } \quad a:= \frac{4Lc''}{c' \pi_m^2D^2} \vee \frac{2L}{c' \pi_mD^2} \vee \frac{2}{L\pi_m},$$

we ensure that $n \alpha_n$ satisfies the conditions of Lemma \ref{lemma:4}.

Now let us look into the condition $t=\frac{\gamma}{\sqrt{n}}<\pi_m-\frac{1}{n}$ of Lemma~\ref{lemma:4}. We will ask $t$ to satisfy the stronger condition \begin{equation}\label{eq:stronger}t=\frac{\gamma}{\sqrt{n}}<\frac{\pi_m}{2}-\frac{1}{n} \quad \Leftrightarrow \quad 0<\frac{\pi_m}{2} n -\gamma \sqrt{n} -1.\end{equation}

Studying the polynomial function $f:x \mapsto \frac{\pi_m}{2} x^2 - \gamma x -1 $, one can find that the zeros of $f$ are $$x_1:= \frac{\gamma - \sqrt{\gamma^2+2\pi_m}}{\pi_m}  \quad \text{ and } \quad x_2:=\frac{\gamma + \sqrt{\gamma^2+2\pi_m}}{\pi_m} \leq \frac{2\gamma+\sqrt{2\pi_m}}{\pi_m}.$$

We deduce that considering that \begin{equation} n >4\left( \frac{\gamma+1}{\pi_m}\right)^2, \label{Qrecov-ncondition}\end{equation} which implies that $\sqrt{n} > \frac{2\gamma+\sqrt{2\pi_m}}{\pi_m}$, we guarantee that $\gamma/\sqrt{n} < \pi_m-1/n$. Applying Lemma \ref{lemma:4}, we get that with probability at least $$1-\left[ (K^2+K) \exp\left( \frac{-(n\pi_m-\gamma\sqrt{n}-1)^2 \frac{\gamma^2}{n}}{\frac12+\frac23 \frac{\gamma}{\sqrt{n}}}\right) +\frac{c}{n^2} + 2K \exp \left(  \frac{-\gamma^2}{2(A_1/4+A_2 \frac{\gamma}{\sqrt{n}})} \right) \right],$$
it holds $\|\hat{Q}-Q\|_{\infty}\leq \gamma/ \sqrt{n}$.

Thanks to Eqs.\eqref{Qrecov-ncondition} and \eqref{eq:stronger}, we have $(n\pi_m-\gamma\sqrt{n}-1)^2 =n^2 (\pi_m-\gamma/\sqrt{n}-1/n)^2 \geq n^2\pi_m^2/4$ and $\gamma/\sqrt{n} \leq \pi_m / 2$. We deduce that defining 

$$b:= c + (2K(K+1)) \quad \text{ and } \quad b':= \frac{1}{2(A_1/4+A_2 \pi_m)} \wedge \frac{\pi_m^2}{2+\frac43 \pi_m},$$

it holds with probability at least $1-b(1/n^2 \vee \exp(-b'\gamma^2))$
$$ \|\hat{Q}-Q\|_{\infty} \leq \frac{\gamma}{\sqrt{n}}.$$
\end{proof}

\subsection{Proof of Theorem~\ref{pirecovery}}

\label{sec:proofmsbmpirecovery}
Lemma \ref{lemma:3} provides a more complete version of Theorem~\ref{pirecovery} by giving explicitly the constants.

\begin{lmm}
We consider $c,c'$ and $c''$ the three numerical constants involved in Theorem \ref{thmVerzelen}.

Assume that $\alpha_n \log(n)\leq 1/L$ and that $\displaystyle n \alpha_n > \max\left( \frac{4Lc''}{\pi_m^2D^2}\;,\;  \frac{2}{L\pi_m} \; , \; \frac{2L\log(n)}{\pi_m c' D^2}\right)$. Then for all $t>0$, it holds

\begin{align*}
\mathbb{P}\left(\|\hat{\pi}-\pi \|_{\infty} > t \right) \leq  2K\exp \left( - \frac{n t^2}{2(A_1/4 + A_2 t)}   \right) +\frac{c}{n^2}  +   2K   \exp \left( - \frac{n \pi_m^2}{2A_1 + 4A_2 \pi_m}\right)  .
\end{align*}
\label{lemma:3}
\end{lmm}
\begin{proof}[Proof of Lemma \ref{lemma:3}.]

Using Theorem~\ref{bernstein-markov} (cf. Appendix~\ref{gap} or \cite[Theorem 2]{Jiang}), we get that

$$\forall c \in [K], \; \forall t>0, \; \; \mathbb{P}\left( \left|\frac{1}{n} \sum_{i=1}^n \mathds 1_{C_i=c} - \pi(c) \right| \geq t \right) \leq 2\exp \left( - \frac{n t^2}{2(A_1\sigma_c^2 + A_2 t)}   \right)$$

where $ \displaystyle A_1= \frac{1+ (\lambda_+ \vee 0)}{1-( \lambda_+ \vee 0)}$, $ \displaystyle A_2 =  \frac13 \mathds 1_{\lambda_+ \leq 0} + \frac{5}{1-\lambda_+}\mathds 1_{\lambda_+ > 0}$ and $\sigma^2_c = \pi(c)(1-\pi(c)) \leq 1/4. $
\medskip

We define the event $  N:=\left\{\text{err}(\hat{G},G)<\exp\left(-\frac{\pi_mn \alpha_n c'D^2}{2L}\right)\right\}$.  Note that on $N$, the partition of the clusters is correctly recovered thanks to the condition $ n \alpha_n > \frac{2L \log(n)}{\pi_mc'D^2}$. Then,

\begin{align*}
&\mathbb{P}\left( \underset{k \in [K]}{\bigcup} \{ |\hat{\pi}(k)-\pi(k) | > t \}\right)\\  \leq \quad & \mathbb{P}\left( \underset{k \in [K]}{\bigcup} \{ |\hat{\pi}(k)-\pi(k) | > t \}  \; | \; N \right) +  \mathbb{P}(N^c)\\
=\quad & \mathbb{P}\left( \underset{k \in [K]}{\bigcup} \{ |\frac{1}{n} \sum_{i=1}^n \mathds 1_{C_i=k}-\pi(k) | > t \}  \; | \; N \right) +  \mathbb{P}(N^c)\\
\leq \quad & 2K\exp \left( - \frac{n t^2}{2(A_1/4 + A_2 t)}   \right) +\frac{c}{n^2}  +   2K   \exp \left( - \frac{n \pi_m^2}{2A_1 + 4A_2 \pi_m}\right)  ,
\end{align*}

where we apply Lemma \ref{lemma:1} in the last inequality.
\end{proof}

\subsection{Proof of Theorem~\ref{kernelrecovery}}
\medskip

We will prove a more accurate result with Lemma \ref{lemma:2}.

\begin{lmm}
\label{lemma:2}
Let us consider $\gamma> \frac{5K}{2 \pi_m^2}$.\\
Assume that $\alpha_n \log(n)\leq 1/L$, that $\displaystyle n\alpha_n> \max\left( \frac{4Lc''}{ \pi_m^2D^2}\;,\;  \frac{4}{L\pi_m}\; , \; \frac{2L\log(n)}{\pi_m c' D^2}\right)$ and that $\sqrt{n}> \frac{2}{\pi_m}(1+\pi_m^2\gamma/5)$. Then it holds
\begin{align*}
& \mathbb{P}\left( \| \hat{P} - P \|_{\infty} \geq \frac{\gamma}{\sqrt{n}}   \right)\\& \leq \quad  2K^2\exp \left( - \frac{\left(\frac{\pi_m^2\gamma}{5K}-\frac12 \right)^2}{2(B_1/4 + B_2 \frac{\pi_m^2\gamma }{5K\sqrt{n}})}\right) +\frac{c}{n^2}  +   2K   \exp \left( - \frac{n \pi_m^2}{8A_1\sigma^2 + 4A_2 \pi_m}\right)    ,
\end{align*}

where $B_1$ and $B_2$ depend only on the Markov chain and are defined by $ \displaystyle B_1:= \frac{1+ (\xi_+ \vee 0)}{1- (\xi_+ \vee 0)}$ and 
 $ \displaystyle B_2 :=  \frac13 \mathds 1_{\xi_+ \leq 0} + \frac{5}{1-\xi_+}\mathds 1_{\xi_+ > 0}$. Here $1-\xi_+$ is the right $L^2$ spectral gap of the Markov chain $(Y_i)_{i \geq 1}$ (see Definition \ref{def2} in Section \ref{gap}).

\end{lmm}

\textbf{Remarks.}

\begin{itemize}
\item  The first term in the right hand side of the inequality in Lemma~\ref{lemma:2} is due to the concentration of the average number of visits of the chain $(Y_i)_{i \geq 1}$ (defined in Section~\ref{defY} of this chapter) towards its stationary distribution. The two last terms of the inequality correspond to the bound guaranteeing the recovery of the true partitions with a direct application of Theorem~\ref{theorem}.
\item The condition $n\alpha_n >  \frac{2L\log(n)}{\pi_m c' D^2}$ ensures that $\exp\left(-\frac{\pi_mn\alpha_nc'D^2}{2L}\right)< \frac{1}{n}$. Theorem~\ref{theorem} will then guarantee that we recover perfectly the partition of the communities.
\item Expecting the accuracy $\gamma/ \sqrt{n}$, the condition  $\sqrt{n}> \frac{2}{\pi_m}(1+\pi_m^2\gamma/5)$ ensures that the Markov chain $(C_i)_{i\geq1}$ has visited enough each state $k \in [K]$ to guarantee the convergence of the average number of visits toward the stationary distribution.
\end{itemize}

\begin{proof}[Proof of Lemma \ref{lemma:2}.]\mbox{}

\underline{\textbf{I. Concentration of the average number of visits for $(Y_i)_{i \geq 1}$. }}
\medskip

We recall that $(Y_i)_{i \geq 1}$ is a Markov Chain on $[K]^2$ defined by  : $Y_i = (C_i, C_{i+1})$.

Then using again Theorem~\ref{bernstein-markov}, we get that $\forall t>0, \; \forall k,l \in [K]^2, $

$$ \mathbb{P}\left( \left| \frac{1}{n-1} \sum_{i=1}^{n-1} \mathds 1_{Y_i=(k,l)}- \pi(k)P_{k,l} \right| \geq t \right) \leq 2\exp \left( - \frac{n t^2}{2(B_1/4 + B_2 t)}   \right),$$

\underline{\textbf{II. First step toward the Theorem.}}

 We define the event $ N:=\left\{\text{err}(\hat{G},G)<\exp\left(-\frac{\pi_mn\alpha_nc'D^2}{2L}\right)\right\}.$  Note that on $N$, the partition of the clusters is correctly recovered thanks to the condition $ n\alpha_n > \frac{2L \log(n)}{\pi_mc' D^2}$. Let $\gamma>\frac{5K}{2 \pi_m^2}$ and let us define

\begin{align*}r&= \frac{\zeta}{\sqrt{n}} \; \text{ with } \zeta= \frac{ \pi_m^2 \gamma}{5K} - \frac12>0, \\ \text{and}\quad \Gamma &= \underset{k,l}{\bigcap}  \left\{ \left| \frac{1}{n-1} \sum_{i=1}^{n-1} \mathds 1_{Y_i=(k,l)} - \pi(k) P_{k,l} \right| < r \right\} .\end{align*}

Then,
\begin{align*}
 &\mathbb{P}\left(  \bigcup_{k,l} \left\{\left| \hat{P}_{k,l} - P_{k,l} \right| \geq \frac{\gamma}{\sqrt{n}} \right\}  \right)\\\leq \quad & \underbrace{\mathbb{P}\left( \bigcup_{k,l} \left\{\left| \hat{P}_{k,l} - P_{k,l} \right| \geq \frac{\gamma}{\sqrt{n}} \right\} | N, \Gamma \right)}_{\textcolor{blue}{(*)}}\mathbb{P}(N)\mathbb{P}(\Gamma|N) +  \mathbb{P}(\Gamma^c) + \mathbb{P}( N^c).
\end{align*}

Note that the condition  $\sqrt{n}> \frac{2}{\pi_m}(1+\pi_m^2\gamma/5)$ of Lemma \ref{lemma:2} implies that \begin{equation}\sqrt{n}> \frac{2}{\pi_m}(1+K \zeta). \label{majozeta}\end{equation}

\underline{\textbf{III. We prove that $\textcolor{blue}{(*)}$ is zero.}}

\medskip
In this third step of the proof, we are going to show that conditionally on the event $N \cap \Gamma$, the infinite norm between our estimate of the transition matrix $\hat{P}$ and $P$ is smaller than $\gamma/\sqrt{n}$.

  \paragraph{ \fbox{1} We split $\textcolor{blue}{(*)}$ in two terms.}

\begingroup
 \allowdisplaybreaks
\begin{align*}
&\mathbb{P}\left( \bigcup_{k,l} \left\{\left| \hat{P}_{k,l} - P_{k,l} \right| \geq \frac{\gamma}{\sqrt{n}} \right\} | N, \Gamma \right) \\= \quad & \mathbb{P}\left( \bigcup_{k,l} \left\{\left| \hat{P}_{k,l} -  \frac{\sum_{i=1}^{n-1} \mathds 1_{Y_i=(k,l)}}{(n-1)\pi(k)}  +  \frac{\sum_{i=1}^{n-1} \mathds 1_{Y_i=(k,l)}}{(n-1)\pi(k)}- P_{k,l} \right| \geq \frac{\gamma}{\sqrt{n}} \right\} |N, \Gamma \right) \\
\end{align*}
\endgroup

\paragraph{ \fbox{2} We show that on $\Gamma$:  $\displaystyle \left| \frac{1}{n}\sum_{i=1}^n \mathds 1_{C_i=k} -\pi(k)  \right| \leq \frac{1}{n} + Kr.$}

Here we show that a concentration of the average number of visits for $(Y_i)_{i \geq 1}$ gives for free a concentration result of the average number of visits for $(C_i)_{i \geq 1}$.
\medskip

Note that on the event $\Gamma$ :

\begingroup
 \allowdisplaybreaks
\begin{align*}
\bullet \;  \frac{1}{n}\sum_{i=1}^n \mathds 1_{C_i=k} &= \frac{1}{n} \sum_{l=1}^K \sum_{i=1}^{n-1} \mathds 1_{C_i=k, C_{i+1}=l} \\
&= \frac{n-1}{n} \sum_{l=1}^K \frac{1}{n-1} \sum_{i=1}^{n-1}\mathds 1_{C_i=k, C_{i+1}=l} \\
&\geq \frac{n-1}{n} \sum_{l=1}^K (\pi(k)P_{k,l}-r)\\ &=\frac{n-1}{n}(\pi(k)-Kr).  
\end{align*}
$ \text{Hence } \displaystyle \frac{1}{n}\sum_{i=1}^n \mathds 1_{C_i=k} -\pi(k) \geq -\frac{\pi(k)}{n}-\frac{n-1}{n}Kr \geq -\left(\frac{1}{n} + Kr \right) . $
\begin{align*}
\bullet \;  \frac{1}{n}\sum_{i=1}^n \mathds 1_{C_i=k} &\leq \frac{1}{n} \sum_{l=1}^K \sum_{i=1}^{n-1} \mathds 1_{C_i=k, C_{i+1}=l}+\frac{1}{n} \\
&= \frac{n-1}{n}\sum_{l=1}^K \frac{1}{n-1} \sum_{i=1}^{n-1}\mathds 1_{C_i=k, C_{i+1}=l}+\frac{1}{n} \\
&\leq \frac{n-1}{n} \sum_{l=1}^K (\pi(k)P_{k,l}+r)+\frac{1}{n}\\
&\leq \pi(k)+Kr +\frac{1}{n}.
\end{align*}
$ \text{Hence } \displaystyle \frac{1}{n}\sum_{i=1}^n \mathds 1_{C_i=k} -\pi(k) \leq \frac{1}{n} + Kr  .$
\endgroup

We deduce then that on $\Gamma$, $\displaystyle \left| \frac{1}{n}\sum_{i=1}^n \mathds 1_{C_i=k} -\pi(k)  \right| \leq \frac{1}{n} + Kr.$

  \paragraph{ \fbox{3} We show that the first term from \fbox{1} is zero.}

  In the following, we show that the definition of $\zeta$ with the condition $\sqrt{n}> \frac{2}{\pi_m}(1+\pi_m^2 \gamma/5)$ implies that the first term in \fbox{1} is zero.

 \begingroup
 \allowdisplaybreaks
\begin{align}
&\mathbb{P}\left(  \left| \hat{P}_{k,l} - \frac{1}{n-1} \frac{\sum_{i=1}^{n-1} \mathds 1_{Y_i=(k,l)}}{\pi(k)} \right| \geq \frac{\gamma}{2\sqrt{n}} \; |\; N, \Gamma \right)\nonumber\\
=\quad & \mathbb{P}\left(  \frac{1}{n-1} \sum_{i=1}^{n-1} \mathds 1_{Y_i=(k,l)} \left|  \frac{n}{\sum_{i=1}^{n}\mathds 1_{C_i=k}}-\frac{1}{\pi(k)} \right| \geq \frac{\gamma}{2\sqrt{n}}  \; | \;N, \Gamma \right)\nonumber\\
 \leq \quad & \mathbb{P}\left(  (r+\pi(k)P_{k,l}) \left|  \frac{n}{\sum_{i=1}^{n}\mathds 1_{C_i=k}}-\frac{1}{\pi(k)} \right| \geq \frac{\gamma}{2\sqrt{n}} \; | \; N,\Gamma \right) \nonumber \quad \text{(by definition of }\Gamma \text{)}\\
 =\quad & \mathbb{P}\left(  \left|  \frac{n\pi(k)-\sum_{i=1}^{n}\mathds 1_{C_i=k}}{\pi(k)\sum_{i=1}^{n}\mathds 1_{C_i=k}} \right| \geq \frac{1}{2r+2\pi(k)P_{k,l}} \cdot \frac{\gamma}{\sqrt{n}}  \; | \; N,\Gamma \right) \nonumber \text{ and using } \fbox{2},\\
  \leq \quad & \mathbb{P}\left(    \frac{\frac{1}{n} + Kr}{\pi(k) (\pi(k)-\frac{1}{n} -Kr) }  \geq \frac{1}{2r+2\pi(k)P_{k,l}}\cdot \frac{\gamma}{\sqrt{n}}  \; | \; N,\Gamma \right) \nonumber\\
   \leq \quad & \mathbb{P}\left(    \frac{\frac{1}{n} + Kr}{\pi_m (\pi_m-\frac{1}{n} -Kr) }  \geq \frac{1}{2r+2}\cdot\frac{\gamma}{\sqrt{n}}   \; | \; N,\Gamma \right)  \text{ and since } r= \frac{\zeta}{\sqrt{n}}, \nonumber\\
   = \quad & \mathbb{P}\left(    \frac{\frac{1}{n} + K\frac{\zeta}{\sqrt{n}}}{\pi_m (\pi_m-\frac{1}{n} -K\frac{\zeta}{\sqrt{n}}) }  \geq \frac{1}{2\frac{\zeta}{\sqrt{n}}+2}\cdot\frac{\gamma}{\sqrt{n}}   \; | \; N,\Gamma \right) \nonumber\\
    = \quad & \mathbb{P}\left(    \frac{(\frac{1}{n} + K\frac{\zeta}{\sqrt{n}})(2\zeta+2\sqrt{n})}{\pi_m (\pi_m-\frac{1}{n} -K\frac{\zeta}{\sqrt{n}}) }  \geq \gamma  \; | \; N,\Gamma \right) \nonumber\\
      \leq \quad & \mathbb{P}\left(    \frac{(\frac{1}{n} + K\frac{\zeta}{\sqrt{n}})(2\zeta+2\sqrt{n})}{\pi_m (\pi_m-\frac{1}{\sqrt{n}}(1+K\zeta)) }  \geq \gamma  \; | \; N,\Gamma \right). \label{proof-point3}
\end{align}
\endgroup

Since from \eqref{majozeta}, $\sqrt{n} \geq \frac{2}{\pi_m}(1+ K\zeta),$ we have $$\frac{\pi_m}{2} \leq \pi_m - \frac{1}{\sqrt{n}}(1+K \zeta), $$
which leads to 

$$\mathbb{P}\left(    \frac{(\frac{1}{n} + K\frac{\zeta}{\sqrt{n}})(2\zeta+2\sqrt{n})}{\pi_m (\pi_m-\frac{1}{\sqrt{n}}(1+K\zeta)) }  \geq \gamma  \; | \; N,\Gamma \right)
 \leq \mathbb{P}\left(    \frac{2(\frac{1}{\sqrt{n}} + K\zeta)(\frac{\zeta}{\sqrt{n}}+1)}{\pi_m^2/2 }  \geq \gamma  \; | \; N,\Gamma \right) .$$

Moreover, since from \eqref{majozeta} and the fact that $\pi_m \in (0,1)$, $\sqrt{n} \geq \frac{2}{\pi_m}(1+ K\zeta) > 2K\zeta$, it holds   $$\frac{\zeta}{\sqrt{n}} <\frac{1}{2K}<\frac14.$$

Coming back to \eqref{proof-point3}, we finally get
      
\begingroup
\allowdisplaybreaks
\begin{align*}
& \mathbb{P}\left(  \left| \hat{P}_{k,l} - \frac{1}{n-1} \frac{\sum_{i=1}^{n-1} \mathds 1_{Y_i=(k,l)}}{\pi(k)} \right| \geq \frac{\gamma}{2\sqrt{n}} \; |\; N, \Gamma \right) \\ \leq \quad &\mathbb{P}\left(    \frac{2(\frac{1}{\sqrt{n}} + K\zeta)(\frac{\zeta}{\sqrt{n}}+1)}{\pi_m^2/2 }  \geq \gamma  \; | \; N,\Gamma \right)     \\ 
      \leq \quad & \mathbb{P}\left(    \frac{5(\frac{1}{\sqrt{n}} + K\zeta)}{\pi_m^2 }  \geq \gamma  \; | \; N,\Gamma \right) \\
       = \quad & 0 .
\end{align*}
\endgroup

The last equality is due to the definition of $\zeta$. Indeed,

$$ \zeta = \frac{\gamma\pi_m^2}{5K}-\frac12 < \frac{\gamma\pi_m^2}{5K}-\frac{1}{K \sqrt{n}} \quad \text{ leading to } \quad \frac{5(\frac{1}{\sqrt{n}} + K\zeta)}{\pi_m^2 }<\gamma. $$

  \paragraph{ \fbox{4} We show that the second term from \fbox{1} is zero.}

\begin{align*}
&\mathbb{P}\left(\left| \frac{1}{n-1} \frac{\sum_{i=1}^{n-1} \mathds 1_{Y_i=(k,l)}}{\pi(k)}- P_{k,l} \right| \geq \frac{\gamma}{2\sqrt{n}} \; |\; N, \Gamma \right)\\
 = \quad & \mathbb{P}\left(\left| \frac{1}{n-1} \sum_{i=1}^{n-1} \mathds 1_{Y_i=(k,l)}- \pi(k)P_{k,l} \right| \geq \pi(k)  \frac{\gamma}{2\sqrt{n}} \; |\; N, \Gamma \right) \\
\leq \quad &  \mathbb{P}\left(\left| \frac{1}{n-1} \sum_{i=1}^{n-1} \mathds 1_{Y_i=(k,l)}- \pi(k)P_{k,l} \right| \geq \pi_m  \frac{\gamma}{2\sqrt{n}} \; |\; N, \Gamma \right) \\
=\quad & 0 ,
\end{align*}

where the last equality comes from the definition of $r=\zeta / \sqrt{n}$ and the definition of $\Gamma$ because
\begin{equation*}\zeta= \frac{\pi_m^2 \gamma}{5K} - \frac12 < \frac{\pi_m^2 \gamma}{5K} \leq \frac{\pi_m \gamma}{2}. 
\end{equation*}

\underline{\textbf{IV. Conclusion.}}

\begingroup
\allowdisplaybreaks
\begin{align*}
 &\mathbb{P}\left(  \bigcup_{k,l} \left\{\left| \hat{P}_{k,l} - P_{k,l} \right| \geq \frac{\gamma}{\sqrt{n}} \right\}  \right) \\
 \leq \quad &  \mathbb{P}\left( \bigcup_{k,l} \left\{\left| \hat{P}_{k,l} - P_{k,l} \right| \geq \frac{\gamma}{\sqrt{n}} \right\}  \; | \;  N, \Gamma \right)\mathbb{P}(N)\mathbb{P}(\Gamma \; | \; N) +  \mathbb{P}(\Gamma^c) + \mathbb{P}( N^c) \\
= \quad & \mathbb{P}(\Gamma^c) + \mathbb{P}( N^c) \\
\leq \quad & 2K^2\exp \left( - \frac{n r^2}{2(B_1/4 + B_2r )}\right) +\frac{c}{n^2}  +   2K   \exp \left( - \frac{n \pi_m^2}{8A_1\sigma^2 + 4A_2 \pi_m}\right)  \\
 \leq \quad & 2K^2\exp \left( - \frac{\left(\frac{\pi_m^2\gamma}{5K}-\frac12 \right)^2}{2(B_1/4 + B_2 \frac{\frac{\pi_m^2\gamma}{5K}-\frac12 }{\sqrt{n}})}\right) +\frac{c}{n^2}  +   2K   \exp \left( - \frac{n \pi_m^2}{8A_1\sigma^2 + 4A_2 \pi_m}\right)  ,
\end{align*}
\endgroup

where we apply Lemma \ref{lemma:1} in the last inequality.
\end{proof}

\subsection{Proof of Proposition~\ref{prop:consistance-BLP}}
\label{apdx:prop:consistance-BLP}
Let us consider $\gamma >\frac{5K}{2\pi_m^2}$. We assume that the conditions~\eqref{conditions-MSBM} of Proposition~\ref{prop:consistance-BLP} are satisfied and we deduce from Theorems~\ref{kernelrecovery} and~\ref{Qrecovery} that there exists three constants $a,b,b'>0$ and some event $\mathcal E_a$ satisfying
\[\mathbb P(\mathcal E_a)\geq 1- b \left[ 1/n^2 \vee \exp\left( - b'(\gamma-\frac{5K}{2\pi_m^2})^2 \right)  \right],\]
such that it holds on $\mathcal E_a$,\[\|\hat P - P\|_{\infty} \vee \|\hat Q - Q\|_{\infty} \leq \frac{\gamma}{\sqrt n}.\]
Taking a close look at the proof of Theorems~\ref{Qrecovery} and~\ref{kernelrecovery}, one can see that the constant $a$ is chosen so that on $\mathcal E_a$ it holds $\mathrm{err}(\hat G,G)<1/n$ meaning that the clustering algorithm has recovered correctly the partition of the nodes, i.e. $\mathbf C_{1:n}=\hat {\mathbf C}_{1:n}$. Hence we can focus only on $\left| \eta_i(\mathbf C_{1:n})-\hat \eta_i(\mathbf C_{1:n})  \right|$ to prove Proposition~\ref{prop:consistance-BLP}.

For any $i \in [n]$ we have
\begingroup
 \allowdisplaybreaks
\begin{align*}
&\left| \eta_i(\mathbf C_{1:n})-\hat \eta_i(\mathbf C_{1:n})  \right|\\
&= \left|\sum_{k \in [K]}  P_{C_i,k} Q_{C_i,k}-\sum_{k \in [K]} \hat P_{C_i,k}\hat Q_{C_i,k} \right|\\
&\leq \sum_{k \in [K]}\left|  P_{C_i,k} Q_{C_i,k}-  \hat P_{C_i,k}  Q_{C_i,k}+  \hat P_{C_i,k}  Q_{C_i,k} -\hat P_{C_i,k}\hat Q_{C_i,k} \right|\\
&\leq \sum_{k \in [K]}\left|  P_{C_i,k} -  \hat P_{C_i,k} \right| \times \left| Q_{C_i,k}\right|+ \sum_{k \in [K]} \left|   Q_{C_i,k} -\hat Q_{C_i,k} \right| \times \left| \hat P_{C_i,k}\right|\\
&\leq \|\hat P - P\|_{\infty} \sum_{k \in [K]}  Q_{C_i,k} + \|Q - \hat Q\|_{\infty} \sum_{k \in [K]}  \hat P_{C_i,k}\\
&\leq \|\hat P - P\|_{\infty} K \alpha_n L + \|Q - \hat Q\|_{\infty},
\end{align*}
\endgroup
where we used that $\|Q\|_{\infty} = \alpha_n \|Q_0\|_{\infty} \leq \alpha_n L$ and the fact that $\hat P$ is a stochastic matrix. 

We deduce that for any $i \in [n]$, it holds with probability at least $1-b \left[ 1/n^2 \vee \exp\left( - b'(\gamma-\frac{5K}{2\pi_m^2})^2 \right)  \right],$ 
\begin{align*}
 \left| \eta_i(\mathbf C_{1:n})-\hat \eta_i(\mathbf C_{1:n})  \right| \leq \frac{\gamma}{\sqrt n}\left( \alpha_n K L+1\right).
 \end{align*}
 
 Using a union bound concludes the proof.

\section{Spectral Gaps and Mixing Times for Markov Chains}
This section is largely inspired from \cite[Section 2.1]{Jiang}.
\label{gap}

\subsection{Spectral gap}

We consider a state space $E$ and a sigma-algebra $\Sigma$ on $E$ which is a standard Borel space. We denote by $(X_i)_{i\geq 1}$ a Markov chain on the state space $(E,\Sigma)$ with stationary distribution $\pi$.

For any real-valued, $\Sigma$-measurable function $h:E \rightarrow \mathbb{R}$, we define $\pi(h):= \int h(x)\pi(dx)$. The set  $$\mathcal{L}^2(E,\Sigma,\pi):=\{h:\pi(h^2)<\infty\}$$ is a Hilbert space endowed with the inner product $$\langle h_1,h_2\rangle_{\pi}=\int h_1(x)h_2(x)\pi(dx), \; \forall h_1,h_2 \in \mathcal{L}^2(E,\Sigma,\pi).$$ The map $$\| \cdot \|_{\pi}: h\in \mathcal{L}^2(E,\Sigma,\pi) \mapsto \|h\|_{\pi}=\sqrt{\langle h,h \rangle_{\pi}},$$ is a norm on $\mathcal{L}^2(E,\Sigma,\pi) $.  $\| \cdot \|_{\pi}$ naturally allows to define the norm of a linear operator $T$
 on $\mathcal{L}^2(E,\Sigma,\pi)$ as $$N_{\pi}(T)= \sup\{\|Th\|_{\pi}:\|h\|_{\pi}= 1\}.$$  To each transition probability kernel $P(x,B) $  with $x\in E$ and $B\in\Sigma$ invariant with respect to $\pi$, we can associate a bounded linear operator $h\mapsto \int h(y)P(\cdot,dy)$ on $\mathcal{L}^2(E,\Sigma,\pi)$. Denoting this operator $P$, we get $$Ph(x) =\int h(y)P(x,dy), \; \forall x \in E, \; \forall h \in \mathcal{L}^2(E,\Sigma,\pi).$$ 

 Denoting by $P^*$ the adjoint or time-reversal operator of the Markov operator $P$, we can define the self-adjoint operator $R=(P+P^*)/2$. Let $\mathcal{L}^2_0(\pi) :=\{h\in  \mathcal{L}^2(E,\Sigma, \pi) \;: \; \pi(h) = 0 \}$. The spectrum of a self-adjoint Markov operator like $R$ acting on $\mathcal{L}^2_0(\pi) $ is contained in $[-1,+1]$. The gap between $1$ and the maximum of the spectrum of $R$ is called the right $\mathcal{L}^2$-spectral gap of $P$.
 
 \begin{dfntn} (Right $\mathcal{L}^2$-spectral gap)
 A Markov operator $P$ has right $\mathcal{L}^2$-spectral gap $1-\lambda_+(R)$ if the operator $R= (P+P^*)/2$ satisfies $$\lambda_+(R) := \sup\{s:s\in \text{spectrum of }R\text{ acting on }\mathcal{L}^2_0(\pi)\}<1.$$
 \label{def2}
 \end{dfntn}

Theorem~\ref{bernstein-markov} shows that the existence of a non-zero right $\mathcal L^2$-spectral gap is of particular interest to prove concentration result for empirical processes. Theorem~\ref{bernstein-markov} is a key tool for the proofs of the theoretical results of this paper. 
\begin{thrm}
\label{bernstein-markov} \cite[Theorem 2]{Jiang}\\
Suppose that the sequence~$(X_i)_{i\geq1}$ is a Markov chain with stationary distribution~$\pi$ and non-zero right $\mathcal L^2$-spectral gap~$1-\lambda_+>0$ (see Definition~\ref{def2}). We assume further that $X_1$ is distributed according to $\pi$. Let us consider some~$n \in \mathbb N \backslash \{0\}$ and a real valued function~$f:E\to \mathbb R$ such that~$\int f(x) d\pi(x)=0$ and $\sup_{x\in E} |f(x)|<c$  for some constant $c>0$ independent of $n$. Let~$\sigma^2 =  \int f^2(x) d\pi(x)$.
Then for any~$\epsilon \geq 0$ it holds
\[\mathbb P \left( \frac{1}{n}\sum_{i=1}^n f(X_i) \geq \epsilon \right) \leq \exp\left(- \frac{n\epsilon^2}{A_2\sigma^2+A_1c\epsilon}\right) ,\]
where~$A_2 := \frac{1+\lambda_+\vee 0}{1-\lambda_+\vee0}$ and~$A_1:=\frac13 \mathbb1_{\lambda_+\leq0}+ \frac{5}{1-\lambda_+}\mathds 1_{\lambda_+ >0}$.
\end{thrm}

\subsection{Signal to noise ratio and Markov mixing times}

In this subsection, we want to briefly highlights that the SNR defined by $S^2 := \frac{n \alpha_n\pi_m D^2 }{L}$ carries information on the ergodicity of the chain through $\pi_m$. The mixing time and the spectral gap are two widely used quantities to measure how fast an ergodic Markov chain will converge to its stationary distribution $\pi$. Proposition~\ref{pi-mixing} states a direct connection between  $\pi_m$ and the latter mentioned coefficients.

\begin{prpstn}\cite[Theorem 12.3]{levin}\\ \label{pi-mixing}
In the following, we denote $\|\cdot\|_{TV}$ the total variation norm. Let $P$ be the transition matrix of a reversible, irreducible
Markov chain with state space and for $0<\epsilon<1$, let
 \[t_{\mathrm{mix}}(\epsilon):=\min \{ t>0 \; : \;  \sup_x \|P^t(x,\cdot)-\pi\|_{TV}\leq \epsilon\},\]be the mixing time of the chain. Then it holds 
 \[t_{\mathrm{mix}}(\epsilon)\leq \log( (\epsilon \pi_m)^{-1}) / (1-\lambda_+),\]where $1-\lambda_+ $ is the right $\mathcal L_2$-spectral gap of the chain from Definition~\ref{def2}.
\end{prpstn}

\section{The Baum-Welch algorithm with information gap}
\label{apdx:BW}

In Sections~\ref{sec:hmmbmlink} and~\ref{sec:colab}, we have presented a reliable approach to solve link prediction or a collaborative filtering problem when we fully observe the graph at time $n$ and when we want to perform some prediction involving future nodes. We propose to consider a more general framework considering that we fully observe the graph at time $n+\delta$ ($\delta \in \mathbb N^*$) but we consider that edges involving nodes between time $T$ (with $T <n$) and time $n$ are not reliable. Note that the simpler framework addressed in the paper is simply recovered by taking $n=T+1$. Hence, we want only to take into account the edges involving pairs of nodes in $\{1, \dots, T, n , \dots n+\delta\}.$ We denote $E_{T,n, \delta}$ this set of edges. We describe the Baum-Welch algorithm in this framework. Running the clustering algorithm on the graph $G=(\{1, \dots, T, n , \dots n+\delta\}, E_{T,n, \delta})$, we find sequences of estimates for the communities $\hat{ \mathbf{C}}_{1:T},\hat{ \mathbf{C}}_{n:n+\delta}$. In the following, we will consider by abuse of notations that for any $j \geq T+1$, the sequence $\hat{ \mathbf{C}}_{1:j}$ represents the sequence $\left( \hat{ \mathbf{C}}_{i}, i \in [j] \backslash \{T+1, \dots, n-1 \} \right).$

The Baum-Welch algorithm consists in a forward and a backward procedure followed by an update step that we describe below. Denoting $\mathbf 1_K = (1 , 1, \dots , 1)^{\top} \in \mathbb R^K$, $\theta^{(0)}= (  \mu^{(0)}, P^{(0)}, O^{(0)}  )$ is initialized as follows 
\begin{align*}
  P^{(0)} &= \frac{1}{K} \mathbf 1_K  \mathbf 1_K^{\top}, \\
\mu^{(0)} & =   \mathbf 1_K^{\top}, \\
 O ^{(0)}& = (1-\epsilon) \mathrm{Id}_K + \frac{\epsilon}{K-1} \left(  \mathbf 1_K  \mathbf 1_K^{\top} - \mathrm{Id}_K\right),
\end{align*}
where $\epsilon \in (0,1)$ (typically $\epsilon = 10^{-2}$).

\begin{itemize}
\item Forward procedure

Let us recall that we have denoted $\alpha_k^{(m)}(i) = \mathbb Q_{\theta^{(m)}}\left( \hat{\mathbf C}_{1:i}, C_i=k \;| \; \theta \right) $  the probability of seeing the observations $\hat C_1,  \dots \hat C_i$ and being in state $k$ at time $i$. This is found recursively with
\begin{align*}
\forall k \in [K], \quad \alpha_k^{(m)}(1)=& \mu_k ^{(m)}O^{(m)}_{k, \hat C_1}\\
\forall k \in [K], \; \forall i \in [n], \quad  \alpha_k^{(m)}(i)=& \left\{
    \begin{array}{ll}
            \sum_{l \in [K]} \alpha_l^{(m)}(T) \left( (P^{(m)})^{i-T}\right)_{l,k} & \mbox{if } T<i\leq n\\
       O_{k,\hat C_{i}}^{(m)} \sum_{l \in [K]} \alpha_l^{(m)}(i-1)  P^{(m)}_{l,k} & \mbox{otherwise.} 
    \end{array}
\right.
\end{align*}

\item Backward procedure

Let us recall that we have denoted $\beta_k^{(m)}(i) = \mathbb Q_{\theta^{(m)}}\left( \hat{\mathbf C}_{i+1:n+\delta} \;| \;  C_i=k, \, \theta  \right) $ the probability of the ending partial sequence $\hat {\mathbf C}_{i+1:n+\delta}$ given starting in state $k$ at time $i$. This is found recursively with
\begin{align*}
\forall k \in [K],\quad \beta^{(m)}_k(n)=& 1\\
\forall k \in [K], \; \forall i \in [n], \quad \beta^{(m)}_k(i)=& \left\{
    \begin{array}{ll}
            \sum_{l \in [K]} \beta^{ (m)}_l(n-1) \left( (P^{(m)})^{n-1-i}\right)_{k,l} &  \mbox{   if } T\leq i\leq n-2\\
      \sum_{l \in [K]} \beta^{ (m)}_l(i+1)  P^{(m)}_{k,l} O^{(m)}_{l,\hat C_{i+1}} &  \mbox{   otherwise.} 
    \end{array}
\right. 
\end{align*}

\item Update step

We can first update the temporary variables $\gamma^{(m)}$ and $\xi^{(m)}$ defined below. The probability of being in state $k$ at time $i$ given the observed sequence $\hat{\mathbf C}_{1:n+\delta}$ and the parameters $\theta^{(m)}$ is denoted $\gamma^{(m)}_k(i)$ with

\[\forall k \in [K], \; \forall i \in [n], \quad \gamma_k^{(m)}(i) = \mathbb Q_{\theta^{(m)}}(C_i=k | \hat{\mathbf C}_{1:n+\delta}) = \frac{\alpha^{(m)}_k(i) \beta^{(m)}_k(i)}{\sum_{l \in [K]} \alpha^{(m)}_l(i) \beta^{(m)}_l(i) }.\]

The probability of being in state $k$  and $l$ at times $i$ and $i+1$ respectively given the observed sequence $\hat{\mathbf C}_{1:n+\delta}$ and parameters $\theta^{(m)}$ is denoted $\xi^{(m)}_{k,l}(i)$ with for all $ k,l \in [K]$ and for all $i \in [n]$,  
\begin{align*}\xi^{(m)}_{k,l}(i) = \mathbb Q_{\theta^{(m)}}(C_{i}=k, C_{i+1}=l \; | \; \hat{\mathbf C}_{1:n+\delta}) &= \frac{\mathbb Q_{\theta^{(m)}}(C_i=k,C_{i+1}=l,\hat{\mathbf C}_{1:n+\delta} )}{\mathbb Q_{\theta^{(m)}}(\hat{\mathbf C}_{1:n+\delta}  )} .
\end{align*}

Hence,
\begin{tabular}{rll}
$\xi^{(m)}_{k,l}(i)$& $\displaystyle =
             \frac{\alpha^{(m)}_k(i)  P^{(m)}_{k,l}\beta^{(m)}_l(i+1)}{\sum_{c,b \in [K]} \alpha^{(m)}_c(i)  P^{(m)}_{c,b}\beta^{(m)}_b(i+1)}$&$  \mbox{if } T\leq i\leq n-2$\\
   $\xi^{(m)}_{k,l}(i)$ &=   $\displaystyle \frac{\alpha^{(m)}_k(i)  P^{(m)}_{k,l}\beta^{(m)}_l(i+1)O^{(m)}_{l,\hat C_{i+1}}}{\sum_{c,b \in [K]} \alpha^{(m)}_c(i)  P^{(m)}_{c,b}\beta^{(m)}_b(i+1)O^{(m)}_{b,\hat C_{i+1}}}$ &   $\mbox{otherwise.} $
\end{tabular}

The parameters of the hidden Markov model can now be updated.
\begin{align*}
\forall k \in [K], \quad  \mu^{(m+1)}_k &= \gamma^{(m)}_k(1), \\
\forall k,l \in [K], \quad  P^{(m+1)}_{k,l} &= \frac{\sum_{i=1}^{n-1} \xi^{(m)}_{k,l}(i)}{\sum_{i=1}^{n-1} \gamma^{(m)}_k(i)} ,\\
\forall k,l \in [K], \quad  O^{(m+1)}_{k,l} &= \frac{\sum_{i=1}^n \mathds 1_{\hat C_i = l} \gamma^{(m)}_k(i)}{\sum_{i=1}^n \gamma^{(m)}_k(i)}.
\end{align*}
\end{itemize}

\section{Notations}
\label{apdx:notations}

\renewcommand{\arraystretch}{2}
\begin{longtable}{c|C{14cm}}
    \multicolumn{2}{c}{\bf Standard MSBM}\\
$Q \in [0,1]^{K\times K} $ & Connectivity matrix.\\ \hline
$P \in [0,1]^{K\times K}$ & Markov kernel.\\ \hline
$X \in \{0,1\}^{n\times n}$ & Adjacency matrix of the observed graph of size $n$.\\ \hline
$\mathbf C_{1:n}$ & Sequence of hidden communities of the nodes $1$ to $n$.\\ \hline
$\hat {\mathbf C}_{1:n}$ & Sequence of estimates of the communities of the nodes $1$ to $n$ given by the clustering algorithm from a given number of communities $K$ and from the adjacency matrix $X$ .\\ \hline
$\eta_i( {\mathbf C}_{1:n}) $ &  Posterior probability of having a connection between node $i$ and the upcoming node (the node $n+1$), namely $\eta_i( {\mathbf C}_{1:n})  = \mathbb P(X_{i,n+1}=1 \, |\, \mathbf C_{1:n})$.\\\hline
$\mathbb P$ & Probability distribution (or probability mass function depending on the context) for which $\mathbf C_{1:n}$ is a Markov chain with initial distribution $\pi$ and Markov kernel $P$, and for all $i,j\in[n]$, $X_{i,j} \sim \mathrm{Bern}(Q_{C_i,C_j})$. The dependence between the random variables $\mathbf C_{1:n}$, $(X_{i,j})_{1\leq i,j\leq n}$ and $\hat {\mathbf C}_{1:n}$ is described by Figure~\ref{fig:graphicalmodel}.\\ \hline
$\hat P$ & Estimate of the Markov kernel $P$ given in Section~\ref{estimation}.\\ \hline
$\hat Q$  & Estimate of the connectivity matrix $Q$ given in Section~\ref{estimation}.\\ \hline
$\mathbb P_{(\mu',P',Q')}$ & For any probability distribution $\mu'$ on $[K]$, any Markov kernel $P' \in [0,1]^{K\times K}$ and any connectivity matrix $Q'\in [0,1]^{K\times K}$, $\mathbb P_{(\mu',P',Q')}$ is the probability distribution (or probability mass function depending on the context) for which $\mathbf C_{1:n}$ is a Markov chain with initial distribution $\mu'$ and Markov kernel $ P'$, and for all $i,j\in[n]$, $X_{i,j} \sim \mathrm{Bern}( Q'_{C_i,C_j})$. The dependence between the random variables $\mathbf C_{1:n}$, $(X_{i,j})_{1\leq i,j\leq n}$ and $\hat {\mathbf C}_{1:n}$ is described by Figure~\ref{fig:graphicalmodel}.\\ \hline
    \multicolumn{2}{c}{\bf HMM approximation}\\
\makecell{$\mathbb Q$ \\~\\ $O_{k,l}$}& $\mathbb Q$ is the probability distribution (or probability mass function depending on the context) for which $\mathbf C_{1:n}$ is a Markov chain with initial distribution $\pi$ and Markov kernel $P$, and for all $i,j\in[n]$, $X_{i,j} \sim \mathrm{Bern}( Q_{C_i,C_j})$. Moreover, the joint distribution of $(\mathbf C_{1:n},\hat {\mathbf C}_{1:n})$ factorizes according to the graph of a homogeneous HMM with emission probabilities $O_{k,l}=\mathbb P(\hat C_1=l \, |\, C_1=l)$ for $k,l\in [K]$.\\ \hline
$\alpha_k(i)$ & For any $k\in [K]$ and $i\in [n]$, $\alpha_k(i) = \mathbb Q(\hat {\mathbf C}_{1:i},C_i=k)$.\\\hline
$\beta_k(j)$ & For any $k\in [K]$ and $j\in [n]$, $\beta_k(j) = \mathbb Q(\hat {\mathbf C}_{j+1:n} \, |\, C_j=k)$.\\\hline
$\chi^{(i,j)}_{k,l}$ & For any $k,l\in [K]$ and $i,j\in [n]$ with $i<j$, $\chi^{(i,j)}_{k,l} = \mathbb Q(C_j=l,\hat {\mathbf C}_{i+1:j} \, |\, C_i=k)$.\\\hline
    \multicolumn{2}{c}{\bf HMM approximation \& Estimates from the Baum-Welch algorithm}\\
$\hat \alpha_k(i)$ & For any $k\in [K]$ and $i\in [n]$, $\hat \alpha_k(i)$ is the estimate of $\alpha_k(i)$ given by the Baum-Welch algorithm.\\\hline
$\hat \beta_k(i)$ & For any $k\in [K]$ and $i\in [n]$, $\hat \beta_k(i)$ is the estimate of $\beta_k(i)$ given by the Baum-Welch algorithm.\\\hline
$\hat \chi^{(i,j)}_{k,l}$ & For any $k,l\in [K]$ and $i,j\in [n]$ with $i<j$, $\hat \chi^{(i,j)}_{k,l}$ is the estimate of $\chi^{(i,j)}_{k,l}$ computed using Eq.\eqref{eq:chihat}.\\\hline
$\hat \mu$ & Estimate of the initial distribution of the Markov chain $\mathbf C_{1:n}$ obtained using the Baum-Welch algorithm.\\ \hline
$\hat O_{k,l}$ & Estimate of the emission probability $O_{k,l}$ given by the Baum-Welch algorithm.\\ \hline
$\hat {\mathbb Q}$ & Probability distribution (or probability mass function depending on the context) for which $\mathbf C_{1:n}$ is a Markov chain with initial distribution $\hat \mu$ and Markov kernel $\hat P$, and for all $i,j\in[n]$, $X_{i,j} \sim \mathrm{Bern}(\hat Q_{C_i,C_j})$. Moreover, the joint distribution of $(\mathbf C_{1:n},\hat {\mathbf C}_{1:n})$ factorizes according to the graph of a homogeneous HMM with emission probabilities $\hat O_{k,l}$.\\\hline
$\hat \eta_i^R( \hat {\mathbf C}_{1:n}) $ &  Estimate of the posterior probability $\eta_i(\mathbf C_{1:n})$. $\hat \eta_i^R( \hat {\mathbf C}_{1:n}) $  is defined in Eq.\eqref{eq:etaR}.\\\hline
\end{longtable}

\end{document}